\documentclass[a4paper,11pt]{article}
\pdfoutput=1 
\usepackage{jcappub}
\usepackage[T1]{fontenc} 
\usepackage{hyperref}
\usepackage{amsmath}
\usepackage{amsfonts}
\usepackage{mathrsfs}
\usepackage{amssymb}
\usepackage{geometry}
\usepackage{physics}
\usepackage{cancel}
\usepackage{slashed}
\usepackage{scalerel}
\usepackage{comment}
\usepackage{caption}
\usepackage{subcaption}
\usepackage[most]{tcolorbox}
\usepackage{xargs}
\usepackage{color}
\usepackage{duckuments}
\usepackage[normalem]{ulem}

\newcommand{\GeV}{\text{GeV}}
\newcommand{\TeV}{\text{TeV}}
\newcommand{\Gtpi}{\Gamma^{2\text{PI}}}
\newcommand{\mc}[1]{\mathcal{#1}}

\newcommand{\Pc}{\mathcal{P}}
\newcommand{\Kc}{\mathcal{K}}

\newcommand{\Uc}{\mathcal{U}}

\newcommand{\dm}{\text{DM}}

\newcommand{\DeA}{\Delta^{\mathcal{A}}}
\newcommand{\PiA}{\Pi^{\mathcal{A}}}

\newcommand{\SsA}{\slashed{S}^{\mathcal{A}}}

\newcommand{\SiSh}{\slashed{\Sigma}}

\newcommand{\SsH}{ \slashed{S}^{\mathcal{H}}}

\newcommand{\dslS}{\dsl{ S}}
\newcommand{\dslSigma}{{\slashed{\Sigma}}}
\newcommand{\SigmaH}{\Sigma^{\mc{H}}}
\newcommand{\SigmaA}{\Sigma^{\mc{A}}}
\newcommand{\Deltas}{{\Delta}_s}

\newcommand{\omdm}{\Omega_\text{DM}h^2}
\newcommand{\sgn}[1]{\text{sign}\big({#1}\big)}

\newcommand{\mdm}{m_\text{DM}}
\newcommand{\ydm}{y_\text{DM}}
\newcommand{\Ydm}{Y_\text{DM}}
\newcommand{\Mp}{M_\text{Pl}}
\newcommand{\gs}{g_\star}
\newcommand{\gss}{g_{\star s}}
\newcommand{\dsl}[1]{\slashed{#1}}
\newcommand{\dfour}[1]{\dfrac{\dd^4 {#1}}{(2\pi)^4}}

\newcommand{\Li}[2]{\text{Li}_{#1}\left({#2}\right)}

\newcommand{\rp}[1]{\big({#1}\big)}

\newcommand{\kvec}{\lvert\vec{k}\rvert}
\newcommand{\qvec}{\lvert\vec{q}\rvert}
\newcommand{\pvec}{\lvert \vec{p} \rvert}

\title{\boldmath Dark matter freeze-in from non-equilibrium QFT: towards a consistent treatment of thermal effects
}

\author[a]{Mathias Becker,}
\author[a]{Emanuele Copello,}
\author[a]{Julia Harz,}
\author[a]{and Carlos Tamarit}

\affiliation[a]{$PRISMA^+$ Cluster of Excellence \& Mainz Institute for Theoretical Physics, FB 08 - Physics, Mathematics and Computer
Science, Johannes Gutenberg-Universit\"{a}t Mainz, 55099 Mainz, Germany}

\emailAdd{bmathias@uni-mainz.de}
\emailAdd{ecopello@uni-mainz.de}
\emailAdd{julia.harz@uni-mainz.de}
\emailAdd{ctamarit@uni-mainz.de}

\abstract{

We study thermal corrections to a model of real scalar dark matter (DM) interacting feebly with a SM fermion and a gauge-charged vector-like fermion mediator. We employ the Closed-Time-Path (CTP) formalism for our calculation and go beyond previous works by including the full dependence on the relevant mass scales as opposed to using (non)relativistic approximations. In particular, we calculate the DM production rate by employing 1PI-resummed propagators constructed from the leading order term in the loop expansion of the 2PI effective action, beyond the Hard-Thermal-Loop (HTL) approximation. We compare our findings to commonly used approximation schemes, including solving the Boltzmann equation using momentum-independent thermal masses in decay processes and as regulators for $t$-channel divergences. We also compare with the result when employing HTL propagators and their tree-level limit. We find that the DM relic abundance when using thermal masses in the Boltzmann approach deviates between $-10\%$ and $+30\%$ from our calculation, where the size and sign strongly depend on the mass splitting between the DM candidate and the gauge-charged mediator. The HTL-approximated result is more accurate at small gauge couplings, only deviating by a few percent at large mass splittings, whereas it overestimates the relic density up to $25\%$ for small mass splittings. Calculations using tree-level propagators in the CTP formalism or semiclassical Boltzmann equations without scatterings underestimate the dark matter abundance and can lead to deviations of up to $-100\%$ from the 1PI-resummed result.}

\begin{document}

\begin{flushright}
    MITP-23-085\\
    December 2023
\end{flushright}

\maketitle
\flushbottom

\newpage
\section{\label{sec:Intro}Introduction}

One of the most puzzling questions of physics is the origin and nature of Dark Matter (DM), a non-luminous, gravitationally interacting, and (preferably) non-relativistic substance permeating the Universe at every cosmological scale \cite{Planck:2018parameters,Bertone:2016nfn}.
If DM has a particle nature, it cannot be accommodated in the Standard Model (SM) of particle physics so that new physics beyond the Standard Model (BSM) needs to be involved.
Among the various possibilities, one of the most studied scenarios is that of weakly interacting massive particles (WIMPs). 
However, the lack of signatures relatable to such particles has challenged standard WIMP models \cite{Arcadi:2017kky,Roszkowski:2017nbc}.
Even though this could be due to a less-trivial structure of the dark sector, perhaps involving a mass-compressed spectrum \cite{PhysRevD.43.3191,Becker:2022iso}, another possibility is that DM features very small interactions with the visible sector, rendering its detection more challenging.

This is the case for DM being a feebly interacting massive particle (FIMP) produced via the \textit{freeze-in} mechanism \cite{McDonald:2001vt,Hall:2009bx,Bernal:2017kxu}.
In this framework, DM is assumed to have been created in the early universe from a negligible initial abundance via $1~\rightarrow~2$ decays or $2~\rightarrow~2$ annihilations possibly involving heavier dark mediators and SM particles.
The crucial assumption is that DM interacts with these particles through very small couplings, in such a way that it can never reach thermal equilibrium.
As a consequence, contrary to freeze-out, backreactions are negligible and the comoving number density of FIMP DM grows from zero until it freezes in to the present-day value around $T\sim M$, when its production rate becomes inefficient in comparison to the expansion rate of the Universe.
This means that the production dynamics is sensitive to the high-temperature regime in which DM is relativistic.
In this regime, if the dark mediators are part of the thermal plasma, their physical properties get corrected by finite-temperature effects and, in return, the production of DM is also affected.
This is in contrast to freeze-out, where the production dynamics peak in the non-relativistic regime.

Such corrections have recently received major attention.
Firstly, accounting for the full relativistic quantum statistics (Fermi-Dirac or Bose-Einstein) in the freeze-in collision operators instead of Maxwell-Boltzmann distributions can lead up to $\mc{O}(10\%)$ differences in the DM abundance \cite{Lebedev:2019ton,Belanger:2018ccd,Belanger:2020npe,Bringmann:2021sth}.
Secondly, thermal corrections in the form of thermal masses have been included in decay and scattering processes in order to regulate infrared divergences affecting $t$-channel diagrams involving massless states.
Depending on the model considered, thermal masses open up a broader range of kinematically allowed regimes, enabling decay and inverse decay processes that would be forbidden at zero temperature, and, moreover, they make the solution of the Boltzmann equation infrared-finite (see, for example, Refs.~\cite{Garny:2017rxs,Baker:2017zwx,No:2019gvl,Belanger:2020npe,Decant:2021mhj,Calibbi:2021fld,Bringmann:2021sth,Alguero:2023zol}).

Despite the inclusion of relativistic statistics and thermal masses, the previous works relied on a semiclassical Boltzmann approach, where production rates originate from integrating a cross section with thermal masses added as an \emph{ad-hoc} correction.
The cross-section itself is derived from an \emph{in-out} S-matrix element for asymptotically-free states.
However, in a finite density medium, the concept of free particles and asymptotic states breaks down.
Here, a consistent treatment of finite-density effects is necessary, requiring a full quantum-field-theoretical approach at finite-temperature \cite{le_bellac_1996,Kapusta:2006pm}, going beyond the simple addition of thermal masses.
This can be done essentially in two ways, namely within the \emph{imaginary-time} (or Matsubara) formalism \cite{Matsubara:1955ws,Kapusta:2006pm,Laine:2016hma}, which only applies in thermal equilibrium, and within the \emph{real-time} (also known as closed-time-path, or Keldysh-Schwinger) formalism \cite{Keldysh:1964ud,Schwinger:1960qe,Cornwall:1974vz,Chou:1984es,Cutkosky:1960sp,Kobes:1984vb,Kobes:1985kc,Kobes:1986za,Kobes:1986_Cutkosky,Landsman:1986uw,Bedaque:1996af,Landshoff:1996ta,Lundberg:2020mwu}.
This is also the standard theoretical framework to study non-equilibrium processes \cite{Calzetta:1986cq,Calzetta:2008iqa}.

Recently, a major step towards a consistent treatment of DM freeze-in at finite temperature that does not rely on the semi-empirical Boltzmann approach has been taken by Ref.~\cite{Biondini:2020ric}, where the authors rely on the imaginary-time formalism to compute the production rate of Majorana DM fermion particles from three-body and four-body reactions involving a heavier gauge charged scalar mediator and a SM fermion in thermal equilibrium (for related results in the freeze-out context see also \cite{Biondini:2017ufr,Biondini:2018pwp}).
The scatterings are computed in the ultrarelativistic regime by exploiting the large momentum limit of the hard thermal loop (HTL) resummed propagators of the equilibrated fields \cite{Pisarski:1988vd,Braaten:1989mz,Kraemmer:2003gd}.
Decays are corrected by partial resummation including effects of vacuum masses, and by adding on top reactions enhanced by multiple soft scatterings with the plasma, efficient at high temperatures.
Such quantum phenomenon, named the Landau-Pomeranchuk-Migdal (LPM) effect \cite{Landau:1953um,Migdal:1956tc}, is calculated for massless particles and switched off by a phenomenological prescription in the non-relativistic regime $T\ll M$, where the Born rate is assumed to hold.
Such an interpolation allows the characterization of particle production in every temperature regime, although relying on a high-temperature approximation of the production rate around the bulk of freeze-in production ($T\sim M$).
Even though improved interpolations between LPM and Born rates are possible \cite{Ghiglieri:2021vcq}, whether accounting for a massive mediator and DM candidate would sensibly change the production rate is an open question due to the highly complex calculations needed to correctly and smoothly switch from one regime to another.

In our work, we choose to follow a different and complementary approach.
It relies on deriving the evolution of the DM abundance from quantum kinetic Kadanoff-Baym equations (see \cite{kadanoff2018quantum}) in the Closed-Time-Path (CTP) formalism of non-equilibrium quantum field theory.
This corresponds to following the time evolution of two-point correlators defined on a complex time contour via Schwinger-Dyson equations derived as the stationary points of a two-particle-irreducible (2PI) effective action \cite{Cornwall:1974vz,Chou:1984es,Calzetta:1986cq,Berges:2004pu,Berges:2004yj,Calzetta:2008iqa}.
There are several advantages to this method. 
First, the CTP is constructed to track the temporal evolution of expectation values of $n$-point correlation functions in a fully interacting background, in contrast to the S-matrix formalism employed in the semiclassical Boltzmann approach (it is also called \emph{in-in} formalism, in opposition to \emph{in-out}).
Moreover, since the expectation values of two-point functions are directly related to statistical properties of fields, the CTP formalism is well-suited to describe particle production in the early Universe and the evolution equations for the correlators (which involve exact propagators and self-energies constructed from them) can be turned into rate equations for particle number densities.
Additionally, 2PI effective actions provide a framework that can systematically account for quantum and thermal corrections by resumming a larger set of diagrams compared to more standard 1PI approaches, better probing the field configuration space.
Importantly, this approach ultimately leads to a Boltzmann-like time-evolution equation for the DM number density. 
However, in contrast to conventional approaches, the collision term in these equations is derived rigorously from first principles, as described above. 
This ensures that for instance momentum-dependent corrections beyond simple thermal masses are consistently included.
The chosen formalism has been extensively employed in the context of leptogenesis \cite{Riotto:1995hh,Buchmuller:2000nd,Giudice:2003jh,Cirigliano:2009yt,Salvio:2011sf,Prokopec:2003pj,Prokopec:2004ic,Garbrecht:2008cb,Garny:2009qn,Garny:2009rv,Garny:2010nj,Garny:2011hg,Beneke:2010dz,Beneke:2010wd,Fidler:2011yq,Frossard:2012pc,Garbrecht:2011xw,Garbrecht:2013gd,Garbrecht:2013bia,Garbrecht:2014aga,Garbrecht:2014kda,Garbrecht:2015cla,Drewes:2016gmt,Garbrecht:2018mrp,Garny:2018ali,Garbrecht:2019zaa,Depta:2020zmy,Kainulainen:2021oqs}, with fewer concrete applications to DM models \cite{Beneke:2014gla,Binder:2018znk,Binder:2020efn,Ai:2023qnr}.

In order to quantify our results, we choose a model where DM is a real scalar singlet interacting via a feeble Yukawa coupling with an exotic vectorlike fermion and a SM fermion. Gauge invariance enforces the BSM fermion to be charged under the SM gauge interactions, which keep the two fermions  in thermal equilibrium.
This model has also been considered in freeze-out \cite{Giacchino:2013bta,Giacchino:2015hvk,Colucci:2018vxz,Belanger:2018sti,Arina:2020udz}
and freeze-in contexts~\cite{Calibbi:2021fld,Belanger:2018sti,Becker:2023tvd}.
In practice, in order to arrive at a tractable rate equation for DM one needs  to choose a truncation of the loop expansion for the self-energies, and select an approximation for  the exact propagators. 
In our implementation we truncate the 2PI effective action at leading order (LO), which results in one-loop DM self-energies. 
We approximate the exact propagators by resumming 1PI-diagrams. 
In principle, next-to-leading order (NLO) contributions in the self-energies can be relevant in certain kinematic regimes; 
for example, the aforementioned LPM effect can be captured by resumming an infinite series of 2PI ``ladder'' diagrams. We will however omit such terms because of the aforementioned theoretical uncertainties and complexities when several mass scales are involved besides the temperature.  
Nevertheless, we will comment on the omitted effect and we will provide an estimate of its impact based on existing literature, while we plan to address it in a follow-up work.

With the DM production rate estimated as above, the resulting DM relic density can be obtained by integrating over time. 
In our work we compare our results with the DM abundances obtained with the following prescriptions:
\begin{enumerate}
    \item Using the conventional Boltzmann equation approach and accounting only for decays, neglecting thermal corrections to the masses.
    \item As before but including thermal masses.
    \item Using Boltzmann equations accounting for decays and scatterings and incorporating thermal masses.
    \item Using the CTP formalism with tree-level-like, simplified HTL propagators.
    \item Employing the CTP formalism with full HTL resummed propagators.
\end{enumerate}
To summarize our results, we find that the accuracy of each method strongly depends on the mass splitting of the two vacuum mass scales involved.
More specifically:
\begin{itemize}
    \item The relic density obtained from a Boltzmann approach considering only decay contributions fails to accurately describe the DM production.
    Especially for small mass splittings, the relic density can be underestimated by up to $100\%$ with respect to the CTP result with one-loop resummed propagators. 
    Remarkably, the inclusion of thermal masses worsens the accuracy of the result when \emph{only} decays are considered.

    \item Adding DM production from scatterings with propagators regulated by thermal masses can accidentally cancel the overestimated production from decays, deviating from $-10\%$ to $+30\%$ when increasing the mass splitting.
    The discrepancies can be cut in half when considering the appropriate equilibrium distribution functions instead of Boltzmann statistics, which soften DM production because of the Pauli blocking associated with the accompanying final state fermion.
    The accuracy of this method only mildly depends on the effective gauge coupling of the parent particle.

    \item The HTL-approximated propagators lead to percent-level accurate results with respect to one-loop resummed propagators, as long as the two vacuum mass scales are not of the same order. 
    For smaller mass splittings, however, they can overestimate the relic density by roughly $25 \%$. 
    Additionally, the size of this deviation grows with larger effective gauge couplings of the interacting fermions.
\end{itemize}

The paper is organized as follows: In Sec.~\ref{sec:ModelBEQ}, we introduce the example DM model setup and we briefly review the conventional freeze-in production mechanism with a semiclassical Boltzmann equation approach.
In Sec.~\ref{sec:FIfromCTP}, we first recall some elements of the closed-time-path (CTP) formalism and we show how the rate equations for DM are obtained (we refer to Appendices~\ref{sec:AppA}, \ref{sec:AppB} for a more detailed discussion).
Then, we derive the DM production rates for 1PI-resummed, HTL-resummed and tree-level fermionic propagators.
In Sec.~\ref{sec:results}, we present the results of our analysis and discuss the deviations of each method from the relic abundance obtained with one-loop resummed propagators. 
Finally, we conclude in Sec.~\ref{sec:conclusions}.

\section{\label{sec:ModelBEQ}Example model setup and semiclassical Boltzmann approach}
In this section we introduce the class of freeze-in models for which we will study the impact of a more accurate treatment of finite-temperature effects. Furthermore, we will sketch the calculation of the relic density for freeze-in DM in the well-established semiclassical Boltzmann approach. 
\subsection{Vectorlike portal FIMP models \label{subsec:Model}}
We will focus on a concrete class of models where DM is a gauge-singlet scalar field $s$ featuring a Yukawa interaction with a heavier vectorlike fermion $F$ (which will be referred to as the ``mediator'') and a SM fermion $f$.
This model has been sometimes named \emph{vectorlike portal} and has been considered in the context of freeze-out~\cite{Giacchino:2013bta,Giacchino:2015hvk,Colucci:2018vxz,Arina:2020udz}
and freeze-in~\cite{Calibbi:2021fld,Belanger:2018sti,Becker:2023tvd} without thermal effects.
Its Lagrangian density reads as
\begin{align}
    \mc{L} = \mc{L}_{\text{SM}} + \frac{1}{2}(\partial_\mu s)^2-\frac{1}{2}m_s^2 s^2 - V(s,H) + \Bar{F}\left(i\dsl{D}-m_F\right)F - \left[\ydm \Bar{F} P_{L/R} f s +h.c. \right]\,,
    \label{eq:ModelLagrangian}
\end{align}
where $V(s,H)$ is the scalar potential of the DM including its possible interaction with the Higgs field, $D_\mu$ is the covariant derivative, $\ydm$ is the portal Yukawa coupling, and $P_{L/R}=(1\mp \gamma_5)/2$ are the chiral projectors.
Since we want to focus on the freeze-in regime, this portal coupling is assumed to be $\ydm\ll 1$.
The mediator and the DM scalar are stabilized by a $Z_2$-symmetry under which they are odd, while the SM fields remain even.
This allows DM to be the lightest stable dark sector particle.

In this work we will not consider the influence of DM self-interactions and of the interaction terms with the Higgs in the scalar potential.
This allows us to focus on a reduced set of parameters, namely the gauge couplings, the Yukawa coupling $\ydm$, and the masses $m_s = \mdm$ and $m_F$.
In particular, we require the self-coupling and the coupling to the Higgs to be sufficiently small in order not to equilibrate the scalar singlet with itself or with the SM bath (see, e.g., Refs.~\cite{Yaguna:2011qn,Bernal:2015ova,Belanger:2018sti,Bringmann:2021sth}).

    To preserve gauge invariance, the heavy mediators $F$ must belong to the same representation of the SM gauge groups as the SM fermion $f$ they interact with, and hence they are also gauge charged. 
In total, there are five model realizations corresponding to the different representations of the SM fermions.
In this work, we parameterize the gauge interaction of the vectorlike fermion $F$ and of the SM $f$ via an effective gauge coupling $G$ given by
\begin{align}
    G= Y^2 g_1^2 + C_2 \left( \mc{R}_2 \right) g_2^2 + C_2 \left( \mc{R}_3 \right) g_3^2 \, ,
    \label{eq:G_def}
\end{align}
where $g_1$, $g_2$ and $g_3$ are the SM gauge couplings for $U(1)_Y$, $SU(2)_L$ and $SU(3)_C$, respectively, $Y$ is the  weak hypercharge of $f/F$, while $\mc{R}_2$ and $\mc{R}_3$ are the corresponding representations under $SU(2)_L$ and $SU(3)_C$, whose Casimir invariants are denoted as $C_2({\cal R_X})$.
We state some typical values for the effective gauge coupling $G$ for the five different realizations of this model in Table \ref{tab:G_val}. 
{
\setlength{\tabcolsep}{6pt}
\renewcommand{\arraystretch}{2}
\begin{table}[!t]
    \centering
    \begin{tabular}{c|c@{\hspace{2mm}}c@{\hspace{2mm}}c| c | c@{\hspace{3mm}} c@{\hspace{3mm}} c@{\hspace{3mm}} c}
    \hline
         & $Y$ & $SU(2)$ & $SU(3)$ & $G$ & $\mu=M_Z$ & $10^{4}\,\GeV$ & $10^{7}\,\GeV$ & $10^{10}\,\GeV$\\
    \hline
       $e_L$  & $-1/2$ & $\mathbf{2}$ & $\mathbf{1}$ & $\dfrac{g_1^2}{4}+\dfrac{3g_2^2}{4}$ & 0.37 & 0.35 & 0.33 & 0.32\\
      $q_L$  & $+1/6$ & $\mathbf{2}$ & $\mathbf{3}$ & $\dfrac{g_1^2}{36}+\dfrac{3g_2^2}{4}+\dfrac{4g_3^2}{3}$ & 2.3 & 1.6 & 1.2 & 1.0\\
       $e_R$  & $-1$ & $\mathbf{1}$ & $\mathbf{1}$ & $g_1^2$ & 0.21 & 0.22 & 0.24 & 0.26\\
       $u_R$  & $+2/3$ & $\mathbf{1}$ & $\mathbf{3}$ & $\dfrac{4g_1^2}{9}+\dfrac{4g_3^2}{3}$ & 2.1 & 1.3 & 0.9 & 0.7\\
       $d_R$  & $-1/3$ & $\mathbf{1}$ & $\mathbf{3}$ & $\dfrac{g_1^2}{9}+\dfrac{4g_3^2}{3}$ & 2.0 & 1.2 & 0.8 & 0.6
    \end{tabular}
    \caption{The effective gauge coupling $G$ (cf. Eq.\eqref{eq:G_def}) in various models, classified according to the gauge quantum numbers of the SM fermion and the BSM vectorlike fermion involved.
    We use the 1-loop beta functions of the SM gauge couplings to evaluate $G$ at several renormalization scales $\mu$ encompassing some of the values of $G$ considered in this work.}
    \label{tab:G_val}
\end{table}
}

The gauge interactions are sufficiently large to safely assume that the mediators, as well as the SM fermions, interact quickly enough with the thermal bath to reach a state of thermal equilibrium, so that we do not need to study their kinetic evolution.
Furthermore, we parameterize the mass splitting in the dark sector by the dimensionless variable
\begin{align}
    \delta = \frac{m_F - \mdm}{\mdm} \, .
\end{align}
With these definitions, the four parameters of our  class of models are the portal Yukawa $\ydm$, the effective gauge coupling $G$, the mass splitting in the dark sector $\delta$, and the parent particle mass $m_F$.

\subsection{\label{sec:Boltzmann}Semiclassical Boltzmann approach}
In this section, we summarize the most standard treatment of DM freeze-in production in the context of the semiclassical Boltzmann equations, to which we will later compare our work. The evolution of the DM number density $n_\text{DM}$ is given by
\begin{align}
    \dot{n}_\text{DM} + 3 H n_\text{DM} =  \sum_\text{processes} \gamma^\text{eq}_\mathrm{DM} \left( \mathcal{I} \rightarrow \mathcal{F} \right) \left( \prod_{i \in \mathcal{I}} \frac{n_i}{n_i^\text{eq}} - \prod_{f \in \mathcal{F}} \frac{n_f}{n_f^\text{eq}} \right) \equiv \gamma_\text{DM} \, , \label{eq:BoltzmannEquation}
\end{align}
with
\begin{align}\begin{aligned}
   \gamma^\text{eq}_\mathrm{DM} \left( \mathcal{I} \rightarrow \mathcal{F} \right) = &\, \int g_{{\rm DM},{\cal F}} \left( \prod_{i \in \mathcal{F} \cup \mathcal{I}} \frac{d^3\vec{ p}_i}{\left( 2 \pi \right)^3 E_i} \right)\left| \mathcal{M} \left( \mathcal{I} \rightarrow F \right) \right|^2 \times\\
   &\, \times(2\pi)^4\delta \left( \sum_{i \in \mathcal{I}} p_i - \sum_{j \in \mathcal{F}} p_j  \right) \prod_{i \in \mathcal{I}} f_i^\text{eq} \,.\label{eq:CollisionBoltzmann}
\end{aligned}\end{align}
Above, $\mathcal{I} \rightarrow \mathcal{F}$ describes a process with an initial state $\mathcal{I}$ and a final state $\mathcal{F}$ that contains $g_{{\rm DM},{\cal F}}\geq1$ DM degrees of freedom.
$\mathcal{M}$ is the matrix element of the process, while $n_i$ and $n_i^\text{eq}$ are the actual and the equilibrium number densities of particle species $i$. 
Finally, $f_i^\text{eq}$ is the equilibrium phase-space distribution function of particle species $i$ and $H$ is the Hubble parameter encoding the expansion of the Universe.
For the case of freeze-in, the Boltzmann equation simplifies due to the fact that DM never reaches thermal equilibrium, such that $n_\text{DM} \ll n_\text{DM}^\text{eq}$ at all times relevant to the production of DM. 
This directly implies that the last term in the brackets of Eq.~\eqref{eq:BoltzmannEquation} is subleading and can be set to zero, which corresponds to neglecting the effect of backreactions.  
The expansion of the Universe, whose effect is captured by the term $3 H n_\text{DM}$ in Eq.~\eqref{eq:BoltzmannEquation}, can be conveniently treated by expressing number densities through the yields or comoving number densities $Y = n/s$, where $s$ is the entropy density. 
In terms of the yield and by employing a dimensionless time variable $z = m_F/T$, the evolution of the DM yield is given by
\begin{align}
    z H s \frac{\dd Y_\text{DM}}{dz} = \gamma_\text{DM} \, .
\end{align}
Due to the negligible backreactions, this evolution equation can be easily integrated to obtain
\begin{align}
    \Ydm(z)=\int_0^z \dfrac{\dd z'}{z'}\dfrac{\gamma_\text{DM}(G,\delta;z')}{s(z')H(z')}=\dfrac{\ydm^2\,\Mp}{m_F}\dfrac{135\sqrt{10}}{2\pi^3\gss(m_F)\gs(m_F)} \mc{I}(G,\delta;z) ,
\end{align}
where we chose to parameterize the result in terms of a dimensionless integral $\mc{I}(G,\delta;z) = \int_0^z \dd z' \gamma_\text{DM}/(T^4 \ydm^2)$, while $\Mp~=~2.4~\times~10^{18}~\GeV$ is the reduced Planck mass, and $\gss$ and $\gs$ denote the entropy and energy effective number of relativistic degrees of freedom in the primordial plasma, respectively, defined by expressing the entropy and energy densities as $s=2\pi^2/45\, \gss(T) T^3$ and $\rho=\pi^2/30\, \gs T^4$.
The DM relic abundance is related to the comoving number density via the definition
\begin{align}
\label{eq:OmegaDM_def}
    \Omega_\mathrm{DM}h^2=\dfrac{\rho_\mathrm{DM}}{\rho_c/h^2}=\dfrac{s_0}{\rho_c/h^2}\mdm\Ydm^{\infty}=0.12\left(\dfrac{\mdm}{\mathrm{TeV}}\right)\left(\dfrac{\Ydm^\infty}{10^{-12}}\right)\,,
\end{align}
where $\rho_c~\simeq~10^{-5} h^2\,\mathrm{GeV}\,\mathrm{cm}^{-3}$ is the critical density of the Universe, $s_0~\simeq~2891\,\mathrm{cm}^{-3}$ the entropy density today, $h~=~H_0/(100\,\mathrm{km}\,\text{s}^{-1}\,\text{Mpc}^{-1})$ with $H_0$ the Hubble parameter today, and $\Ydm^\infty~=~\Ydm(z=\infty)$ the comoving DM yield today.
Therefore, by also expressing $m_F=(1+\delta)\,\mdm$, the relic abundance at the time $z$ reads as
\begin{align}
    \omdm (z) = 0.12\, \left(\dfrac{\ydm}{2.46\times 10^{-13}}\right)^2\,\dfrac{\mc{I}(G,\delta;z)}{1+\delta}\,.
    \label{eq:OmegaDM_final}
\end{align}
In the following, we sketch the derivation of the DM interaction rate density $\gamma_\text{DM}$ in the following cases: when including only DM production from decays, with either vacuum or thermally corrected masses, and when accounting for production from both decays and scatterings  while including thermal masses.\\
\newline
\textbf{Boltzmann approach considering decays with in-vacuum masses:}
the simplest way to treat freeze-in is to consider only leading-order contributions to DM production and to approximate all distribution functions with Maxwell-Boltzmann statistic, independent of the spin of the considered particle.
In the class of models discussed in this article, the leading order freeze-in process is the parent particle decay into DM and a SM fermion, $F \rightarrow f + s$, for which the production rate simply reads as
\begin{align}
    \left( \gamma_\mathrm{DM} \right)_\text{dec} = \dfrac{\ydm^2 m_F}{16\pi^3}\left(1+\frac{m_f^2}{m_F^2}-\frac{\mdm^2}{m_F^2}\right)\sqrt{\lambda(m_F^2,m_f^2,\mdm^2)}\,T\,K_1(m_F/T)\,,
    \label{eq:DecayRate}
\end{align}
where $\lambda(a,b,c)=(a-b-c)^2-4bc$ is the K\"{a}llén function, $K_1$ the first modified Bessel function of the second kind, and $m_i$ is the vacuum mass of the particle species $i$. This approach was chosen e.g. in \cite{Hall:2009bx,Becker:2023tvd,Calibbi:2021fld}. \\ \newline
\textbf{Boltzmann approach considering decays with thermal masses:}
Sometimes, for instance, in Ref.~\cite{Chakrabarty:2022bcn,No:2019gvl}, thermal effects are simply included at the leading order of the perturbative expansion in vacuum by also considering momentum-independent thermal masses instead of only vacuum masses. 
The DM production rate density is still given by Eq.~\eqref{eq:DecayRate}, but the vacuum mass $m_i$ for the $i$-th species is replaced by $m_i^2 \rightarrow M_i^2 = m_i^2 + m_{i,\text{th}}^2$, where $m_{i,\text{th}}$ is its thermal mass. 
In our model, the thermal masses for the fermions $f$ and $F$ are simply given by $m_{i,\text{th}}^2 = G T^2/4$.
This correction finds its rigorous motivation, as discussed later, in the large-momentum limit of the quasi-particle dispersion relation of the HTL-resummed propagator, as described by Eq.~\eqref{eq:Thermal_Mass}.
The DM thermal mass is neglected since it is suppressed by two powers of the feeble coupling $\ydm$.
The interaction rate reads now as
\begin{align}
    \left( \gamma_\mathrm{DM} \right)_\text{dec,th} = \dfrac{\ydm^2 M_F}{16\pi^3}\left(1+\frac{M_f^2}{M_F^2}-\frac{\mdm^2}{M_F^2}\right)\sqrt{\lambda(M_F^2,M_f^2,\mdm^2)}\,T\,K_1(M_F/T)\,.
    \label{eq:DecayRateTh}
\end{align}
 \newline
\textbf{Boltzmann approach considering decays and scatterings with thermal masses:}
\begin{figure}[!t]
    \centering
\begin{subfigure}{.2\textheight}
\includegraphics[width=0.95\textwidth]{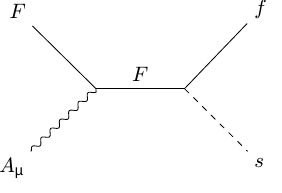}
\end{subfigure}
\hspace{8pt}
\begin{subfigure}{.2\textheight}
\includegraphics[width=0.95\textwidth]{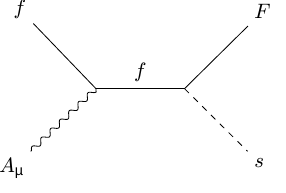}
\end{subfigure}
\\
\begin{subfigure}{.2\textwidth}
\includegraphics[width=0.95\textwidth]{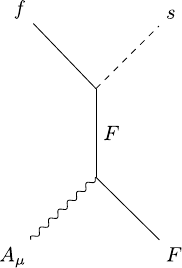}
\end{subfigure}
\hspace{8pt}
\begin{subfigure}{.2\textwidth}
    \includegraphics[width=0.99\textwidth]{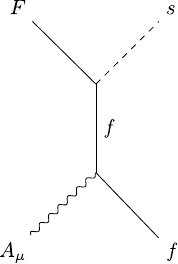}
\end{subfigure}
\hspace{8pt}
\begin{subfigure}{.2\textwidth}
    \includegraphics[width=0.99\textwidth]{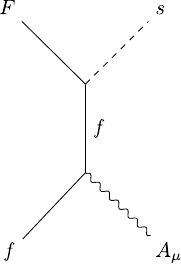}
\end{subfigure}
    \hspace{8pt}
\begin{subfigure}{.2\textwidth}
    \includegraphics[width=0.99\textwidth]{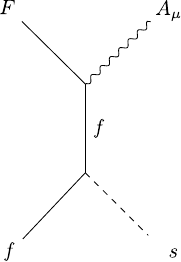}
\end{subfigure}
    \caption{$s$- and $t$-channel scattering diagrams for single particle production of the scalar singlet DM particle $s$. The double production diagram is suppressed by $\ydm^2/G$ and not taken into account. We have omitted the arrows corresponding of the orientation of the fermion flow; one has to consider both allowed orientations for each diagram.}
    \label{fig:DecScatDiagrams}
\end{figure}
Additionally, $2 \leftrightarrow 2$ scatterings can be relevant at high temperatures, as their thermally averaged rates are proportional to the square of the typical number density $n^\text{eq}$ in the plasma -- in contrast to a single power for decay rates -- and $n^\text{eq}\sim T^3$ grows with  with $T$.
Scatterings are especially important if the decay contribution is suppressed, for instance in the case of a small mass splitting $\delta$ between the vectorlike fermion $F$ and DM. 
These scatterings, however, can be plagued by infrared singularities appearing when (almost) massless states of mass $m \ll T$ are involved in $t$-channel exchanges (cf. Fig.~\ref{fig:DecScatDiagrams}), resulting in a logarithmic divergence $\sim \log \left(m / T \right)$ in the cross section. 
A common method to deal with these divergences is to introduce thermal masses to regulate the $t$-channel propagators.
Since the thermal mass itself grows linearly with $T$, the infrared divergence is automatically cured. 
For example, this method was chosen in the Refs.~\cite{Baker:2017zwx,Goudelis:2021qla}.
In this paper, we will restrict DM production from scatterings in the Boltzmann approach to the diagrams shown in Fig.~\ref{fig:DecScatDiagrams}, namely to the processes 
$F + f \rightarrow s + A$ and $f/F + A \rightarrow F/f + s$, with $A$ being a gauge boson and $t$-channel divergences regulated by a thermal mass. 
Importantly, when squaring the matrix elements, we only include the squared $t$- and $s$-channel terms and neglect the interference terms. 
Although this might seem like an unjustified choice, it is a necessary one to guarantee the comparability of the semi-classical Boltzmann equation approach with the calculation on the CTP relying on the interaction rate obtained from the 2PI effective action truncated at LO. 
As will be explained in Sec.~\ref{sec:rate}, by perturbatively expanding the effective action, one realizes that the interference terms only emerge starting at NLO, whereas the squared elements are already appearing at LO.
While the former can be easily incorporated in the semiclassical Boltzmann approach, which relies on perturbative quantum field theory in vacuum, extending the computation beyond LO in the effective action would require evaluating two-loop contributions to the DM self-energy.
Such an extension is beyond the scope of this paper and, therefore, is left for future work.
Furthermore, we have verified that in the Boltzmann formalism the interference between $s$- and $t$-channel scatterings yields a sub-leading contribution of $\mathcal{O} \left( 10 \% \right)$ to the squared matrix elements.
To this end, we also expect the omitted NLO terms in the effective action to contribute to a similar order.
As a consequence, we omit the interference terms from the computation using the Boltzmann equation to be able to more consistently quantify the differences between both approaches.
With these comments at hand, the DM production rate from scatterings reads as
\begin{align}
    \left( \gamma_\text{DM} \right)_\text{scat}=\gamma_{\scaleto{F f \rightarrow s A}{6pt}} + \gamma_{\scaleto{F A \rightarrow s f}{6pt}} + \gamma_{\scaleto{f A \rightarrow s F}{6pt}}\,,
    \label{eq:gammaTOT_scatt}
\end{align}
where, for each process $i$, the rate is given by
\begin{align}
    \gamma_i=\dfrac{T}{64\pi^4}\int_{s_\text{max}}^{\infty}\dd s\, \hat{\sigma}_i(s)\,\sqrt{s}\,K_1(\sqrt{s}/T)\,,
    \label{eq:gamma_scatt}
\end{align}
with $s_\text{max}=\max\{(m_1+m_2)^2,(m_3+m_4)^2\}$, and with $\hat{\sigma}_i(s)$ being the reduced cross section for the $i$-th scattering process, defined as
\begin{align}
    \hat{\sigma}_i(s)=2s\,\lambda\left(1,\frac{m_1^2}{s},\frac{m_2^2}{s}\right)\,\sigma_i(s)\, ,
    \label{eq:reduced_sigma}
\end{align}
where the masses $m_1$($m_3$) and $m_2$($m_4$) are the masses of the initial (final) state particles. 
The expressions for the various cross sections are lengthy and thus not displayed in this article.
The total DM interaction rate is then simply given by the sum of the decay and scattering contribution
\begin{align}
    (\gamma_\text{DM})_\text{dec+scat,th} = \left( \gamma_\text{DM} \right)_\text{dec,th} + \left( \gamma_\text{DM} \right)_\text{scat,th} \, . \label{eq:Rate_scatdec}
\end{align}
where the subscript $\text{th}$ indicates that the vacuum mass $m_i$ of a given particle is replaced by $M_i$ as defined above.  
\subsection{\label{sec:SoA}State-of-the art and goals}
The three interaction rate densities reviewed in the last section represent three methods commonly used in the literature to calculate the relic abundance from freeze-in DM. 
Note, however, that there exist more accurate treatments of the freeze-in production of DM, even in the context of the semiclassical Boltzmann approach.
For instance, Refs.~\cite{Belanger:2018ccd,Lebedev:2019ton,Arcadi:2017kky,Bringmann:2021sth} include the effects of considering the correct quantum statistics, namely Fermi-Dirac and Bose-Einstein statistics, for the bath particles producing DM, which lead to corrections of order few percent on the relic abundance. 

In recent work \cite{Biondini:2020ric}, a notable step has been made towards a consistent treatment of DM freeze-in at finite temperature, going beyond the Boltzmann approach.
The authors employed the imaginary-time formalism of thermal field theory to compute the production rate of Majorana DM fermion particles through decays and scatterings. 
Interaction rates for scatterings were computed in the ultrarelativistic regime relying on HTL approximated resummed propagators, while decay contributions were obtained employing momentum-dependent thermal masses.
Additionally, reactions enhanced by multiple soft scatterings with the plasma were included, a phenomenon commonly referred to as the LPM effect. 
The quantities calculated in the ultrarelativistic limit are switched off smoothly by hand when approaching the non-relativistic regime.
While this treatment allows for a description of the DM production rate at all temperatures, it might be inaccurate at temperatures around the largest vacuum mass scale of the model $T \sim m_F$ due to incorporating the effects of the vacuum mass of the parent particle by a phenomenological prescription rather than an ab-initio calculation. However, this is the temperature regime where the freeze-in dynamics are most relevant. 

In what follows, we address this issue by taking a different and complementary approach with respect to Ref.~\cite{Biondini:2020ric}. 
As detailed in the next section and in appendix \ref{sec:AppA}, we rely on the DM relic density calculated from from Kadanoff-Baym equations in the closed-time-path formalism (CTP) of thermal field theory.
The DM production rate, in resemblance to the optical theorem, is directly related to the imaginary part of the retarded DM self-energy, which itself can be calculated from one-loop resummed propagators. 
Crucially, we do not employ any approximations to simplify the form of these resummed propagators -- such as the HTL approximation that in the relevant regime ($T \sim m_F$) loses the information about the vacuum masses in the thermal corrections. 
Hence our approach is valid in the relevant temperature regime and we expect our calculation to capture the dynamics of freeze-in for temperatures around the largest vacuum mass scale more accurately.  
Importantly, we account for both vacuum mass scales $m_F$ and $\mdm$ at all levels of our calculation, effectively extending existing works in the context of leptogenesis \cite{Garbrecht:2019zaa,Garbrecht:2013bia} that only involve a single vacuum mass scale, the heavy neutrino mass. 

This paper aims to accurately describe the freeze-in DM production rate at all temperatures, most importantly including those around the largest vacuum mass scale of the model, without relying on the HTL approximation. 
Moreover, we want to assess how accurately the different commonly used approximation schemes -- as discussed above -- capture the dynamics of freeze-in production of DM for different parameter regimes in this model class, and provide guidance on which approach yields the most accurate results. 
We provide a comparison of our results obtained with the CTP formalism using 1PI-resummed propagators for 
\begin{itemize}
    \item the Boltzmann approach including only decays with vacuum masses. The interaction rate density for this approach is given in Eq.~\eqref{eq:DecayRate}; 
    \item the Boltzmann approach including only decays but with thermal masses. The interaction rate density for this approach is also given in Eq.~\eqref{eq:DecayRateTh} but with vacuum masses replaced by thermal masses;
    \item the Boltzmann approach including both decays and scatterings where the decays include thermal masses and IR singularities in scatterings are regulated by thermal masses. The interaction rate density for this approach is given by Eq.~\eqref{eq:Rate_scatdec};
    \item the interaction rate density calculated from the CTP formalism but relying on tree-level propagators instead of resummed propagators. This approach reproduces results from the Boltzmann approach only considering decays but with appropriate quantum statistics. The interaction rate density for this approach is given in Eq.~\eqref{eq:numbdensEQ} evaluated with Eq.~\eqref{eq:InRate_tree2};
    \item the interaction rate density calculated from the CTP formalism but relying on HTL-approximated 1PI-resummed propagators. The interaction rate density for this approach is given in Eq.~\eqref{eq:numbdensEQ} evaluated with Eq.~\eqref{eq:SelfEnergy_HTLTOT}. Note that this approach does not coincide with the aforementioned treatment of Ref.~\cite{Biondini:2020ric} but simply serves as a measure to evaluate the deviations induced by HTL-approximated propagators.
\end{itemize}
\section{\label{sec:FIfromCTP}Freeze-in in the CTP formalism}

The CTP formalism is a powerful framework to systematically study initial-value problems in non-equilibrium quantum field theory \cite{Keldysh:1964ud,Schwinger:1960qe,Calzetta:1986cq,Calzetta:2008iqa,Millington:2013isa}.
Differently from the \emph{in-out} S-matrix formalism, which is suited for the computation of transition amplitudes between an initial and a final state, the CTP is tailored to follow the time-dependence of \emph{expectation values} of quantum field operators, like $n$-point Green's functions, when the final state is not known a priori, while the initial state can be arbitrary. 
The CTP is appropriate to study the dynamics in the early Universe. 
In this case, one is not interested in the \emph{probability} that an initial state in the infinite past interacts at a given time and transitions to a final state in the infinite future, where both states are assumed to be asymptotically free, but rather in the evolution of a statistical ensemble (such as the thermal bath of the primordial plasma), where the fields continuously interact.
This situation clearly departs from the particle picture employed in the S-matrix formalism of QFT in vacuum.

In the following, we will outline the essential ingredients to understand how to obtain a freeze-in rate equation for DM in the CTP formalism and how it allows to include thermal corrections from first principles of QFT in a manifest way.
More details are left for the reader in Appendix~\ref{sec:AppA}.
\subsection{\label{sec:CTP}Basics of the CTP formalism}
\begin{figure}[!t]
\vspace{-3cm}
    \centering
    \includegraphics[width=\textwidth]{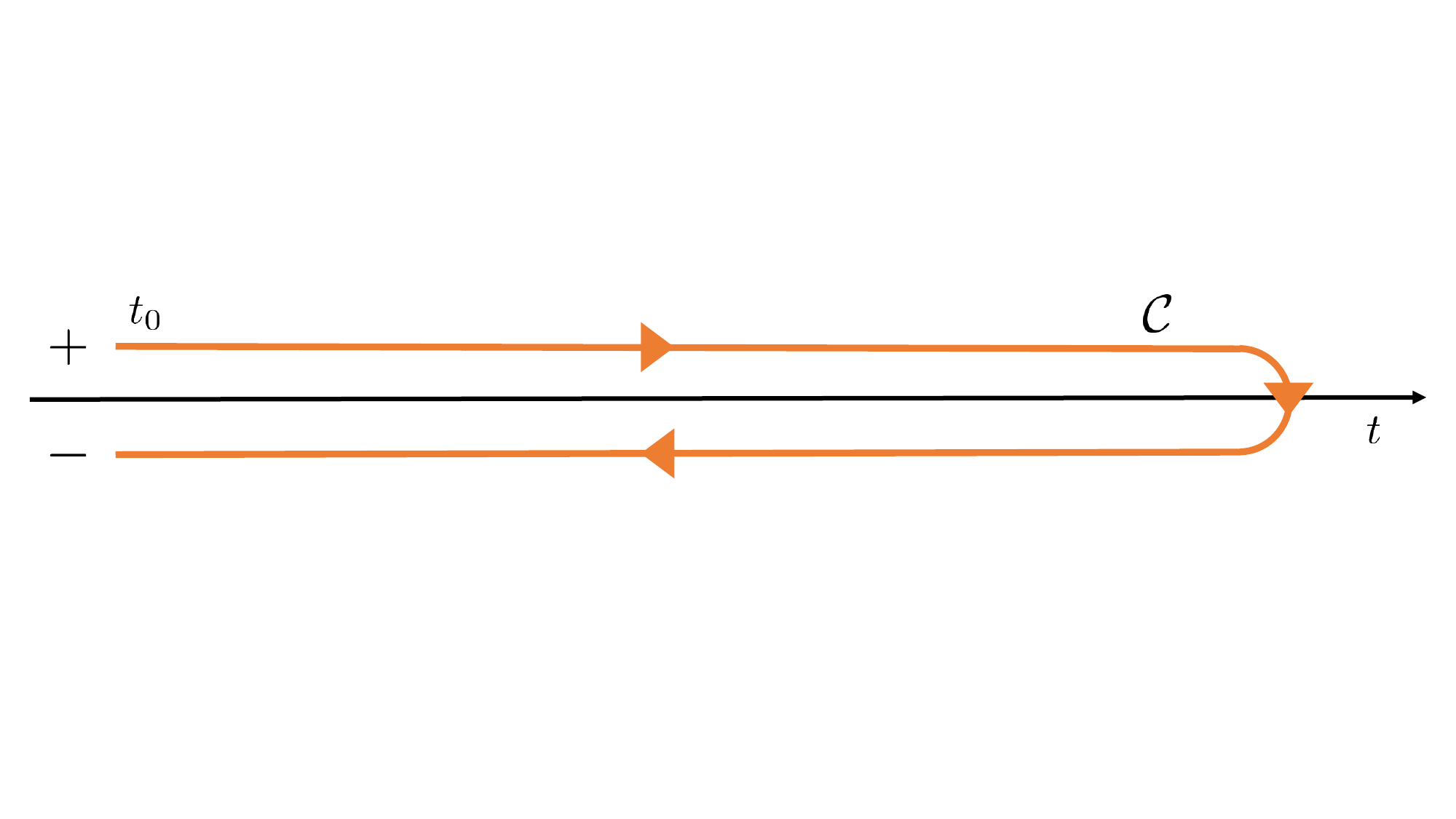}
\vspace{-3.5cm}
    \caption{The CTP contour $\mc{C}$ in the complex plane of the time variable $t$ with the upper $+$ and the lower $-$ branches starting at the initial time $t_0$.}
    \label{fig:CTP_contour}
\end{figure}
The CTP formalism is aimed at computing time-dependent expectation values $\langle \mc{O}(t)\rangle_\rho$ for observables $\mc{O}$ in an arbitrary state $\rho$, which, in our case, will be represented by the bath of particles in the early Universe. The correlators $\langle \mc{O}(t)\rangle_\rho$  can be obtained from a path integral in which the time integration contour $\mc{C}$ has two branches $\mc{C}_+$ and $\mc{C}_-$ going from $-\infty$ to $t$, and from $t$ to $-\infty$, respectively, as illustrated in Fig.~\ref{fig:CTP_contour}. 
As a consequence of the existence of these two branches, the time-ordered two-point correlators of fields along $\mc{C}$ can be labeled 
with two indices $a,b\in \{+,-\}$  which denote the branches at which each of the fields in the correlator is evaluated. 
Representing generic fields of arbitrary spin as $\phi$ and their conjugates (Dirac conjugates in the case of fermions) as $\bar\phi$, Green's functions can be written as follows,
\begin{align}
    i G^{ab}(x,y)=\big\langle T_{\mc{C}}\,\phi(x^a)\bar\phi(y^b)\big\rangle.
\end{align}
We will use $G=\Delta$ for scalars and $G=S$ for fermions.

As for traditional S-matrix calculations, it is useful to express Green's functions in momentum space by performing a Fourier transformation with respect to the relative coordinate $r\equiv x-y$. 
However, in an out-of-equilibrium state, translation invariance is not guaranteed, so the functions $iG^{ab}(x,y)$ do not necessarily only depend on $x-y$, but may also vary under changes of the average coordinate $(x+y)/2$. 
Hence, a more suited type of transformation that we will use is the Wigner transform, defined by
\begin{align}
 G^{ab}(k,x)={\int {\rm d}^4 r}\,e^{ik\cdot r}G^{ab}\left(x+\frac{r}{2},x-\frac{r}{2}\right)\,.
    \label{eq:Wigner_transform}
\end{align}
Among the four possible combinations of $G^{ab}$, the $G^{++}$ corresponds to the usual time-ordered Feynman propagators, while $G^{--}$ is a propagator with an inverted time ordering, with fields sitting on the $-$ branch in Fig.~\ref{fig:CTP_contour}.
With $G^{++}$ and $G^{--}$ one defines the Hermitian propagator as
\begin{align}
    G^\mc{H}\equiv\frac{1}{2}\left(G^{++}-G^{--}\right),
\end{align}
which can be shown to be the Hermitian part of the retarded propagator.
The propagators with mixed branch indices
\begin{align}
    i G^<\equiv &\,iG^{+-}, & i G^>\equiv &\,iG^{-+},
\end{align}
are also called ``Wightman functions'' and are of particular importance because they are related to the so-called spectral (anti-Hermitian) propagator
\begin{align}\label{eq:spectral_props}
    G^{\mc{A}}\equiv \frac{i}{2}\left(iG^>-iG^<\right),
\end{align}
which is connected to the phase-space density of propagating degrees of freedom in the chosen state $\rho$ and contains information about their number densities and energy shells.
For example, in a free theory with a scalar field of mass $m$, one finds (both in the vacuum and in a thermal state)
\begin{align}\label{eq:DeltaA_eq}
    \Delta_{0}^{\rm{eq},\mc{A}}(p) = \pi\,\mathrm{sign}(p^0)\delta(p^2-m^2), 
\end{align}
which gives a nonzero contribution only on the mass-shell $p^2=m^2$ --- and hence is related to the density of states in the phase space, as anticipated before --- while the Wightman functions also involve the particle distribution functions in equilibrium. 
Writing the bosonic/fermionic equilibrium distribution functions as
\begin{align}\label{eq:BEFD_distr}
    f_\mp(p^0) = \frac{1}{e^{p^0/T}\mp 1},
\end{align}
with $-$ corresponding to bosons, and $+$ to fermions, the free scalar Wightman functions in a state of thermal equilibrium at a temperature $T$, read
\begin{align}\label{eq:Delta_Wightman_eq}
    i\Delta_{0}^{\rm{eq},<}(p) = &\,2\Delta_{0}^{\rm{eq},\mc{A}}(p) \,f_-(p^0), &    i\Delta_{0}^{\rm{eq},>}(p) = &\,2\Delta_{0}^{\rm{eq},\mc{A}}(p)\,\left(1+ f_-(p^0)\right).
\end{align}
The subscript ``$0$'' indicates that the free propagators are nothing but tree-level propagators, of order zero in the loop expansion of the path-integral with time contour $\mc{C}$, which can be obtained by inverting the quadratic terms of the classical action along $\mc{C}$ with appropriate boundary conditions. 
Once one goes beyond free fields and beyond thermal equilibrium, the full propagators differ from the results above.
For example, particle distribution functions might not follow their equilibrium values, and there might be corrections to the on-shell relations.

In any case, the distribution functions, and hence the number densities, can still be related to the Wightman functions. 
For the DM scalar field $s$ with a particle distribution function $f_s(\vec{p})$ and number density $n_s$, one has
\begin{align}
    f_s(\vec{p}) &=\int_0^\infty \frac{\dd p^0}{\pi}\, p^0 i\Delta_s^<(p) \, , \\ n_s &= \int \frac{d^3p}{(2\pi)^3}\,f_s(\vec{p})=\int \frac{d^3p}{(2\pi)^3}\int_0^\infty \frac{\dd p^0}{\pi}\, p^0 i\Delta_s^<(p)\,. \label{eq:n_s}
\end{align}
Note that, in the equilibrium case, this follows immediately from Eqs.~\eqref{eq:DeltaA_eq} and \eqref{eq:Delta_Wightman_eq}. 
As a consequence of the relation between number densities and Wightman functions, one can obtain the evolution equation for the DM number density from appropriate dynamical equations for the Wightman functions. 
In the CTP formalism, such dynamical equations can be derived from the Schwinger-Dyson equations for the two-point functions, which in position space can be expressed compactly in terms of the tree-level propagators $i \Delta^{ab}_{0}$, $i \slashed{S}^{ab}_{0}$ and bosonic/fermionic self-energies $\Pi^{ab},\slashed{\Sigma}^{ab}$, as follows:
\begin{align}
    &i\Delta_{0}^{ab^{-1}}(x,y)=i\Delta^{ab^{-1}}(x,y)+i\Pi^{ab}(x,y)\,,\label{eq:EoM_Delta}\\
    &i\slashed{S}_{0}^{ab^{-1}}(x,y)=i \slashed{S}^{ab^{-1}}(x,y)+i \slashed{\Sigma}^{ab}(x,y)\, ,\label{eq:EoM_S}
\end{align}
with $\slashed{S}^{ab} = \gamma^{\mu} S^{ab}_\mu$ and similarly for other quantities. 
The self-energies encode interactions between fields and can be formally defined in terms of functional derivatives with respect to the corresponding propagators of 2PI vacuum diagrams constructed with full propagators $iG^{ab}$.
These vacuum diagrams are part of a generating functional known as the \textit{2PI-effective action} (cf. \ref{sec:2PIEA}) \cite{Cornwall:1974vz,Chou:1984es,Calzetta:1986cq,Berges:2004pu,Berges:2004yj}. 
For the dark matter model considered in this paper, the vacuum diagrams of the 2PI-effective action and the corresponding self-energies are given in Eq.~\ref{eq:2PIEA_model} and Fig.~\ref{fig:DMselfenergy}, respectively.
As the self-energies correspond to loop diagrams with full propagators, the Schwinger-Dyson equations \eqref{eq:EoM_Delta} and \eqref{eq:EoM_S} constitute a very complicated system of infinitely many coupled integro-differential equations, which can only be solved if an approximation is performed, for example by truncating the loop expansion to a fixed order. 
We limit ourselves to leading order in the expansion of the 2PI-effective action and take into account 1PI-resummed fermionic propagators. 

The interactions in the plasma captured by the self-energies can introduce decays of a given field into plasma constituents, and can also modify the dispersion relation for the propagating degrees of freedom. 
Similarly to the case for the propagators, one can introduce Wightman self-energies as
\begin{align}
    \Pi^{<}\equiv&\, \Pi^{+-}, & \Pi^{>}\equiv&\, \Pi^{-+}, \label{eq:Def_Wightman}
\end{align}
as well as spectral and Hermitian self-energies as
\begin{align}\label{eq:PiAH}
    \Pi^{\mc{A}}\equiv &\,\frac{i}{2}\left(\Pi^>-\Pi^<\right),& \Pi^\mc{H}\equiv&\,\frac{1}{2}\left(\Pi^{++}-\Pi^{--}\right).
\end{align}
Analogous relations hold for $\slashed{\Sigma}^{<},\slashed{\Sigma}^{>}, \slashed{\Sigma}^{\mc{A}}, \slashed{\Sigma}^{\mc{H}}$.
In regards to the physical interpretation of these quantities,  $\Pi^{\mc{A}}/\slashed{\Sigma}^{\mc{A}}$ can be related to decay widths or more general interaction rates, while  $\Pi^{\mc{H}}/\slashed{\Sigma}^{\mc{H}}$ yield a modification in the fields' dispersion relations (e.g. changes in the effective pole masses).
This connection can be made explicit in the case of a state without spatial and temporal inhomogeneities, as happens in thermal equilibrium. In this case, the Schwinger-Dyson equations \eqref{eq:EoM_Delta} for the full propagators can be solved algebraically in Wigner space, leading to so-called ``resummed propagators''. For a scalar field, the result is
\begin{align}
 {\Delta}^{\mc{A}}(k)=&\,\dfrac{{\Pi}^{\mc{A}}}{\left(k^2-m^2-{\Pi}^{\mc{H}}\right)^2+\left({\Pi}^{\mc{A}}\right)^2}\,,
\end{align}
where in the denominator one can recognize a structure leading to a Lorentzian resonance in which the mass in the dispersion relation is shifted by $\Pi^{\mc{H}}$, while $\Pi^\mc{A}$ enters as a width for the resonance. 
Similarly, in the fermionic case, one has
\begin{align}
     \slashed{S}^{\mc{A}}(k)=&\,\left(\dsl{k}+m-\slashed{\Sigma}^{\mc{H}}\right)\dfrac{\Gamma}{\Omega^2+\Gamma^2}-\slashed{\Sigma}^{\mc{A}}\dfrac{\Omega}{\Omega^2+\Gamma^2}\,,\label{eq:Resum_S_AntiH_0}
\end{align}
%
and
\begin{align}\begin{aligned}
    &\Gamma = 2 \left(k-\Sigma^{\mc{H}}\right)\cdot\Sigma^{\mc{A}},\\
    &\Omega = \left(k-\Sigma^{\mc{H}}\right)^2-\rp{\Sigma^{\mc{A}}}^2-m^2\,.\label{eq:Omega_Fer_0}
\end{aligned}\end{align}
Again, one still has a Lorentzian pole structure with a width proportional to $\Sigma^\mc{A}$, and with $\Sigma^\mc{H}$ modifying the dispersion relation.  In the ensuing calculations, we will use resummed propagators for the fermions $F$ and $f$ in thermal equilibrium, which is well justified as both carry SM charges.

To finalize, we shall mention that in the case of thermal equilibrium, the propagators and self-energies satisfy additional constraints, known as the Kubo-Martin-Schwinger (KMS) relations. For  propagators in Wigner space, these are
\begin{align}
  i  { \Delta}^{\mathrm{eq},>}(k)=&\,e^{\beta k^0}i{ \Delta}^{\mathrm{eq},<}(k)\,, & 
   i \dslS^{\mathrm{eq},>}(k)=&\,- e^{\beta k^0}i{\dslS^{\mathrm{eq},<}}(k),
\end{align}
where $\beta = T^{-1}$.
With the anti-hermitian propagators defined as in Eq.~\eqref{eq:spectral_props}, the previous relations can be used to generalize the result of \eqref{eq:Delta_Wightman_eq} beyond tree-level and beyond scalar fields, giving
\begin{align}\label{eq:KMS_Delta_S}\begin{aligned}
 i\Delta^{\mathrm{eq},<}(k)=&\,2\Delta^{\mathrm{eq},\mc{A}}(k)f_-(k_0)\,,   &i\dslS^{\mathrm{eq},<}(k)=&\,-2\dslS^{\mathrm{eq},\mc{A}}(k)f_+(k_0)\,,\\
   i\Delta^{\mathrm{eq},>}(k)=&\,2\Delta^{\mathrm{eq},\mc{A}}(k)(1+f_-(k_0))\,, &i\dslS^{\mathrm{eq},>}(k)=&\,2\dslS^{\mathrm{eq},\mc{A}}_F(k)(1-f_+(k_0))\,.
\end{aligned}\end{align}
Analogous properties hold for self-energies in equilibrium. In particular,
for the DM self-energy arising from the equilibrated fields, one arrives at
\begin{align}\label{eq:Pi_s_KMS}
   i{\Pi}^{<}_s(k)=&\,2f_{-}(k_0) {\Pi}^{\mc{A}}_s(k), & i{\Pi}^{>}_s(k)=&\,2(1+f_{-}(k_0) ){\Pi}^{\mc{A}}_s(k)
\end{align}

\subsection{\label{sec:FluidEqFI}Fluid equation for Freeze-in}
As summarized in the previous section, the Wightman functions in the CTP formalism contain information about particle number densities, c.f.~\eqref{eq:n_s}.
Hence, the Schwinger-Dyson equations for the DM two-point functions $\Delta_s^<$ associated with the DM scalar field $s$ in our model (cf. Sec.~\ref{sec:ModelBEQ}) can be turned into an equation for the rate of DM production. 
While a detailed derivation based on the CTP formalism (outlined with more details in Appendix~\ref{sec:AppA}) is left to Appendix~\ref{sec:AppB}, here we provide a summary of the strategy and the final equation for the DM production rate.  
The starting point are the Schwinger-Dyson equations \eqref{eq:EoM_Delta} for $i\Delta_s^<$, which is related to the number density via Eq.~\eqref{eq:n_s}. 
The equations are translated into Wigner space, and separated into real and imaginary parts. Assuming spatial homogeneity, one can ignore spatial gradients. Furthermore, as the scalar self-energies $\Pi^{ab}$ are dominated by the contributions of the fields in thermal equilibrium, they are time-translation invariant, so that one can neglect their time derivatives, see Eq.~\eqref{eq:KinEq_scalar}. Taking this into account, the imaginary part of the Schwinger-Dyson equation for $i\Delta_s^<(p,x)$ in Minkowski space (from now on we use $p$ for the DM momentum) becomes
\begin{align}
    p^0 \partial_t \left( i \Delta_s^{<,>} \right) &=- \frac{1}{2} \left( i {\Pi_s}^{>}\,i \Delta_s^{<} -  i {\Pi_s}^{<} \, i \Delta_s^{>} \right).
    \label{eq:KinEq_scalar_0}
\end{align}
We obtain an equation for the DM production rate, following Eq.~\eqref{eq:n_s}, by integrating over positive $p^0$, leading to
\begin{align}
 \partial_t f_s(t,\pvec)&=\int_0^\infty \frac{\dd p^0}{\pi}\,\frac{1}{2}\left[ \left( i {\Pi}_s^< \right) \left( i \Deltas^> \right) - \left( i {\Pi}_s^> \right) \left( i \Deltas^< \right)\right].
\end{align}
Furthermore, one can use the KMS relations \eqref{eq:KMS_Delta_S} and \eqref{eq:Pi_s_KMS} for the DM self-energy. Additionally, we assume that the DM spectral propagator has the form of a tree-level propagator as in Eq.~\eqref{eq:DeltaA_eq}, as expected for a weakly coupled scalar field. Integrating over spatial momenta and accounting for the expansion of the Universe, the final result can be written as
\begin{tcolorbox}[frame empty, sharp corners, ams align]
      \dot{n}_s+3 Hn_s = \gamma_\mathrm{DM}\equiv \dfrac{1}{2\pi^2}\int \dd|\vec{p}|\dfrac{|\vec{p}|^2}{\omega_p}{\Pi}_s^{\mc{A}}(\omega_p,|\vec{p}|)f_-(\omega_p)\,.
     \label{eq:numbdensEQ}
\end{tcolorbox}
\noindent In the equation above, $\PiA_s(\omega_p,|\vec{p}|)=\PiA_s(p^0,|\vec{p}|)$ is the spectral self-energy of the DM field $s$ in Wigner space, evaluated on the DM momentum shell with $p^0$ substituted by the on-shell energy
\begin{align}\label{eq:omega_p}
    \omega_p = \sqrt{|\vec{p}|^2+m^2_s}\,.
\end{align}
The self-energy is dominated by the effects of fields assumed to be in thermal equilibrium --- the mediator $F$, the SM fermion $f$, and gauge fields --- and as a consequence, it is translation invariant, so that one can omit the dependence on the average coordinate in the Wigner transform. 
Finally, $f_-(\omega_p)$ is a bosonic equilibrium abundance, c.f. Eq.~\eqref{eq:BEFD_distr}.
Eq.~\eqref{eq:numbdensEQ} corresponds to the DM rate equation we aim to study, where the change in time of the DM number density (l.h.s) is controlled by the interaction rate $\gamma_\mathrm{DM}$ (r.h.s.).
Here, we can appreciate that DM interactions are --- as expected --- determined by the DM spectral self-energy.
Additionally, since $\PiA_s$ can be related to the imaginary part of a retarded self-energy, Eq.~\eqref{eq:numbdensEQ} can be unerstood as analogous to the optical theorem for the S-matrix, which relates transition rates to the imaginary part of appropriate amplitudes.
We dedicate the following section to the computation and analysis of the DM interaction rate density $\gamma_\mathrm{DM}$.
\subsection{\label{sec:rate}Production rate and DM self-energy}
\begin{figure}[!t]
    \centering
    \includegraphics[width=0.75\textwidth]{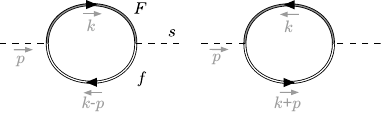}
    \caption{The DM self energy diagrams corresponding to Eq.~\ref{eq:Pi_DM_ab}.
    The double lines in the fermion loop indicate a full propagator and we account for contributions coming for both fermion flow orientations.}
\label{fig:DMselfenergy}
\end{figure}

In this section, we specify the form of the collision term (the interaction rate density $\gamma_\mathrm{DM}$) in the DM rate equation (cf. Eq.~\ref{eq:numbdensEQ}).
We evaluate the required DM self-energy using two levels of approximation:
\begin{enumerate}
    \item The loop order at which the effective action (or, equivalently, the expansion of the DM self-energy) is truncated.
    \item The level of approximation for the exact propagators appearing in the DM self-energy.
\end{enumerate}
To address the first point, we use the 2PI effective action for our model, which is diagrammatically given by
\begin{equation}
\centering
   \Gamma_2 = -i \vcenter{\hbox{\includegraphics[height=18ex]{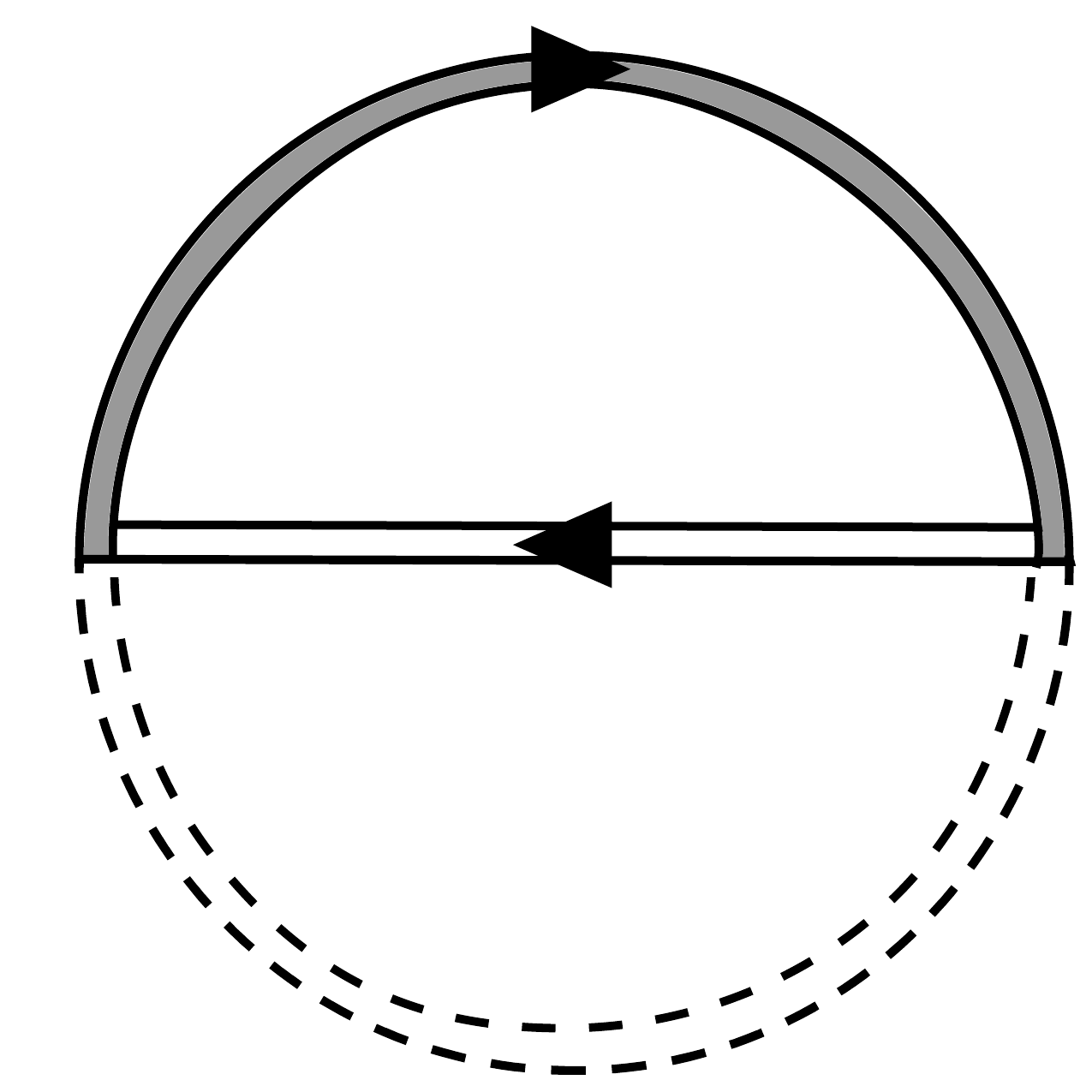}}}\;-i \,\,\vcenter{\hbox{\includegraphics[height=17ex]{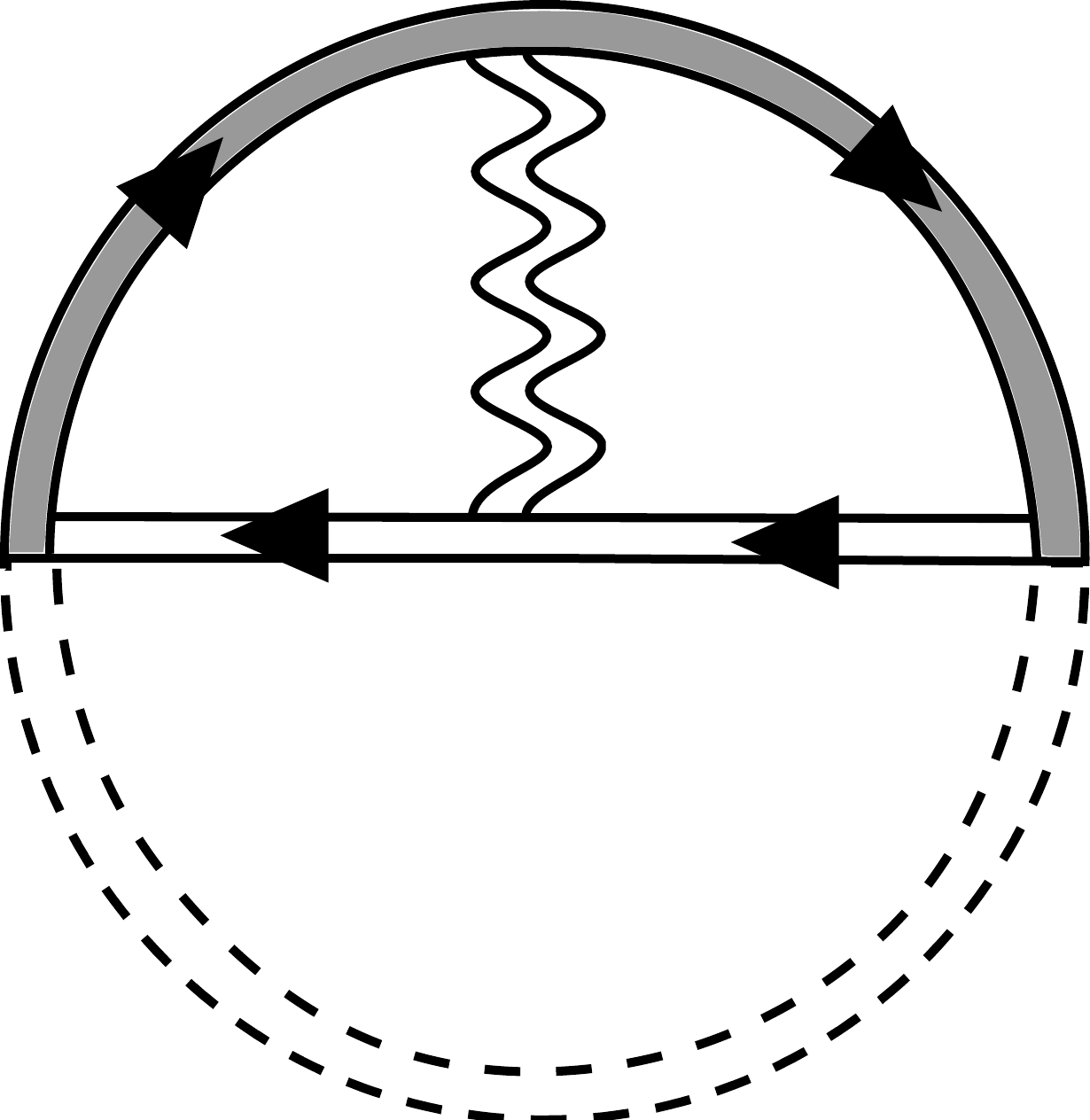}}}\;+\dots \, ,
   \label{eq:2PIEA_model}
\end{equation}
where we used double lines to represent full propagators. 
The DM scalar lines are dashed, the $F$ and $f$ fermion propagators are solid with a gray and white filling, respectively, while the wavy lines represent gauge bosons.
The $\dots$ indicate contributions proportional to further powers of $\ydm$, or higher loops.
The self-energies $\Pi_s$ are obtained by taking functional derivatives of $\Gamma_2$, corresponding to cutting the diagrams in Eq.~\eqref{eq:2PIEA_model}.
For instance, let us consider the DM-self-energy at 1-loop level, corresponding to 2-loop contributions in the effective action, namely the first term in Eq.~\eqref{eq:2PIEA_model}.
The resulting DM self-energy is obtained by cutting the DM dashed propagator line in Eq.~\eqref{eq:2PIEA_model} and taking the resulting diagram with external lines amputated, resulting in Fig.~\ref{fig:DMselfenergy}.
Here, we also account for both orientations of the fermion loop, corresponding to particle and antiparticle contributions to DM production.
On the CTP contour, the 1-loop DM self-energy can be written as
\begin{align}
    -i{\Pi}^{ab}_s(p) = y_{\dm}^2\int\dfour{k}\tr{P_L\,i\,\dslS^{ab}_F(k)\,P_R\,i\,\dslS^{ba}_f(k-p)+P_L\,i\,\dslS^{ba}_F(k)\,P_R\,i\,\dslS^{ab}_f(k+p)}
    \label{eq:Pi_DM_ab}\,,
\end{align}
where $ab$ are the CTP indices, $\ydm$ is the DM coupling, $p$ is the external DM momentum, $k$ is the loop momentum, and where the trace is performed over the Clifford algebra. 
Using \eqref{eq:Pi_DM_ab} together with Eqs.~\eqref{eq:Def_Wightman},\eqref{eq:PiAH},\eqref{eq:Pi_s_KMS}, one can write the spectral self-energy as
\begin{align}\begin{aligned}
    {\Pi}^{\mc{A}}_s(p)=-\frac{y_{\dm}^2}{2}f^{-1}_{-}(p_0)\int\dfour{k}{\rm tr}\left\{P_L\,i\,\dslS^{<}_F(k)\,P_R\,i\,\dslS^{>}_f(k-p)\right.\\
    \left.+P_L\,i\,\dslS^{>}_F(k)\,P_R\,i\,\dslS^{<}_f(k+p)\right\}\, .
\end{aligned}\end{align}
Substituting the KMS relations for the fermionic propagators in Eq.~\eqref{eq:KMS_Delta_S} leads to
\begin{align}\begin{aligned}
{\Pi}^{\mc{A}}_s(p)=-2y_{\dm}^2f^{-1}_{-}(p_0)\int\dfour{k}\left[\tr{P_L\dslS^{\mc{A}}_F(k)P_R\dslS^{\mc{A}}_f(k-p)}f_+(k_0)(1-f_+(k_0-p_0))]\right.\\
\left.+\tr{P_L\dslS^{\mc{A}}_F(k)P_R\dslS^{\mc{A}}_f(k+p)}f_+(k_0+p_0)(1-f_+(k_0))\right]\,,
\end{aligned}\end{align}
which finally becomes
\begin{tcolorbox}[frame empty, sharp corners, ams align]
\nonumber{\Pi}^{\mc{A}}_s(p)=&\,-2\,y_{\dm}^2\int\dfour{k}\tr{P_L\dslS^{\mc{A}}_F(k)P_R\dslS^{\mc{A}}_f(k-p)}\left[1-f_+(k_0)-f_+(p_0-k_0)\right]\\
&\,-(p\leftrightarrow-p)\,,
\label{eq:PiA_master}
\end{tcolorbox}
\noindent where we used the following identities
\begin{align}\begin{aligned}
    f_{-}^{-1}(p_0)f_+(k_0)(1-f_+(k_0-p_0))=&\,1-f_+(k_0)-f_+(p_0-k_0),\\
     f_{-}^{-1}(p_0)f_+(k_0+p_0)(1-f_+(k_0))=&\,-(1-f_+(k_0)-f_+(-p_0-k_0)).
\end{aligned}\end{align}
Eq.~\eqref{eq:PiA_master} is the central object we need to calculate. 
We will later integrate over the DM momentum $\dd\pvec$ to arrive at the DM interaction rate density $\gamma_\mathrm{DM}$ in Eq.~\eqref{eq:numbdensEQ}.
The resulting integral is
\begin{tcolorbox}[frame empty, sharp corners, ams align]
    \gamma_\text{DM} &= \frac{\ydm^2}{4 \pi^5} \int \dd \pvec \, \dd k^0 \, \dd\kvec \, \dd\cos\theta \, \frac{\kvec^2 \pvec^2}{\omega_p} \tr{P_L\dslS^{\mc{A}}_F(k)P_R\dslS^{\mc{A}}_f(k-p)} \nonumber\\
    &\hspace{4.5cm}\times f_-(\omega_p) \left[ 1 - f_+ (k^0) - f_+ (\omega_p - k^0)  \right] \, , 
    \label{eq:RateDensityResummed}
\end{tcolorbox}
\noindent where $\theta$ refers to the angle between $\Vec{p}$ and $\Vec{k}$ and where $\omega_p$ is the DM energy of Eq.~\eqref{eq:omega_p}.
Notice that the azimuthal angular integration in $\dd^4 k$ can be performed trivially.
The remaining four-dimensional integral has to be evaluated numerically (cf. Sec.~\ref{sec:results}).
Before proceeding further, we comment on the accuracy of our chosen approximations in the context of power-counting in terms of the gauge coupling $G$.
The power-counting scheme for freeze-in is
temperature dependent, as for different temperatures the full propagators have different leading
dependence on the coupling constant for the relevant regions of momenta:
(i) At small temperatures $T \ll m_F$, the leading-order interaction rate is dominated by decays and thus independent of the gauge coupling.
This behavior is successfully captured by our truncation of the DM self-energy and additional gauge boson rungs would introduce increasing powers of $G$.
(ii) For large temperatures $T \gg m_F$ resummations are necessary and 1PI-resummation (as discussed in the next subsection) lead to a well-defined power counting~\cite{Braaten:1989mz}. 
However, higher loop contributions to the DM self-energy can contribute at leading order in $G$. 
Our truncation of the DM self-energy successfully captures all contributions scaling as $\sim G \ln G$, as shown in \cite{Arnold:2001ba, Kapusta:1991qp, Baier:1991em}.
However, it does not include all contributions scaling as $\sim G$.
One such missed contribution corresponds to interference between $t$- and $s$-channel $2 \leftrightarrow 2$ scattering diagrams.
We have estimated these terms to be subleading, contributing at most \( \mathcal{O}(10\%) \) (see discussion above Eq.~\eqref{eq:gammaTOT_scatt}).
A second class of missed contributions involves additional gauge boson exchanges between the fermion lines in the DM self-energy, corresponding to multiple soft gauge boson scatterings in the plasma; this is the so-called Landau-Pomeranchuk-Migdal (LPM) effect  \cite{Landau:1953um,Migdal:1956tc}.
As we explain in the following, so far there exists no prescription of the LPM effect that remains accurate at $T \sim m_F$, the regime most relevant to freeze-in. 
Hence, we leave a consistent treatment for future work and will limit ourselves to assess the resulting error that we estimate to lie between $1 \%$ to $30 \%$ (see Sec.~\ref{sec:results}).   \\
{
The LPM effect arises from so-called "ladder" diagrams (as in the diamond diagram of Eq.~\eqref{eq:2PIEA_model} but with arbitrary numbers of gauge boson exchanges between the fermion lines)  and is relevant in certain kinematic regimes involving lightcone momenta and emissions of multiple light particles from the fermion legs.
At high temperatures, all vacuum mass scales are irrelevant and the wavepacket of one of these emitted particles can interfere with the wavepackets from subsequent emissions. 
The consequences of such effect are well-known in hot gauge theories, where a resummation of ladder diagrams allows to obtain a leading-order interaction rate at the hard lightcone scale \cite{Aurenche:2000gf,Arnold:2001ms,Aurenche:2002wq,Arnold:2002ja,Arnold:2002zm}.
The first works to investigate the LPM effect for particle production in the early Universe were performed in the context of leptogenesis \cite{Anisimov:2010gy,Besak:2012qm,Garbrecht:2013bia,Ghiglieri:2016xye}, where the right-handed neutrino production rate was found to be enhanced up to $\mc{O}(25\%)$ by multiple scatterings with soft gauge bosons in the primordial plasma.
Such calculations were performed in the ultra-relativistic regime, where $T\gg M$, $M$ being the mass of the heavy neutrino.
However, when dealing with the expanding Universe, we are interested in the evolution of production rates across several temperature regimes including the temperature regime $T \sim M$, where typically the largest contribution to a frozen-in abundance arises.
Here, the LPM resummation as in \cite{Ghisoiu:2014ena,Ghiglieri:2021vcq} becomes less accurate because it relies on a high-temperature expansion of $n$-point correlation functions (the HTL approximation, cf. Sec.~\ref{sec:HTL}) without taking into account in-vacuum masses.
Thus, phenomenological prescriptions that interpolate between the high and low-temperature regimes have to be employed \cite{Ghisoiu:2014ena,Ghiglieri:2021vcq}\footnote{The LPM effect has also been employed for studying thermalization processes in the early Universe and their consequences on particle production \cite{Mukaida:2015ria,Drees:2021lbm,Drees:2022vvn,Mukaida:2022bbo}.}.

In our context, the situation is even more complicated than in leptogenesis, as two distinct scales are involved: the mediator and the DM in-vacuum masses.
For instance, in Ref.~\cite{Biondini:2020ric}, the authors analyze a similar scenario to ours (with swapped statistics for the dark sector particles, a scalar mediator, and fermionic DM) and address the impact of the LPM effect on the production of DM.
Similarly to the case in Ref.~\cite{Ghisoiu:2014ena}, the rate is calculated relying on an ultrarelativistic expansion and connected to the low-temperature regime via a phenomenological switch-off prescription.
Employing a consistent resummation that would allow for an accurate transition between regimes is a challenging task and still not fully addressed in the literature (improvements in this direction were provided in Ref.~\cite{Ghiglieri:2021vcq}).

Acknowledging that the LPM effect might be relevant especially for small mass splittings, the fact that there are still several theoretical uncertainties to be understood when multiple scales interfere led us to postpone the inclusion of this for future work. 
In particular, the state-of-the-art approach to include the LPM effect, as described above, would spoil our motivation to improve the accuracy in the dominant freeze-in production regime $T \sim m_F$. 
Nevertheless, in Sec.~\ref{sec:results} we will provide an estimate of the impact of the LPM effect on our results based on the results of Ref.~\cite{Biondini:2020ric}.
}
\subsection{Spectral densities and pole structure}
Having understood these premises, we now turn to analyzing the structure of the resummed fermion spectral densities $\slashed{S}_F^A,\slashed{S}_f^A$ entering the DM self-energy.
As we have seen in Section~\ref{sec:CTP} (cf. also Appendix~\ref{sec:AppA}), the two-point functions derived as stationary points of the 2PI effective action are formally exact and can be determined by solving their Schwinger-Dyson equations.
In practice, however, these correspond to an infinite tower of equations that needs to be truncated and, hence, the two-point correlators need to be approximated.
In fact, the full theory is not known and the self-energies can only be computed perturbatively.
Thus, we need to expand the propagators into an infinite series of loop corrections to the tree-level contribution.
This series is eventually partially resummed into dressed propagators (cf. Sec.~\ref{sec:resum_prop} and Sec.~\ref{sec:1PI}).
In what follows, we discuss three levels of approximation:
\begin{enumerate}
    \item 1PI-resummed propagators,
    \item HTL-resummed propagators,
    \item tree-level propagators.
\end{enumerate}
In the first approach (cf. Sec.~\ref{sec:1PI}) we resum one-loop self-energy insertions.
The second approach, to be better defined in Sec.~\ref{sec:HTL}, relies on a high-temperature (and small coupling) expansion of the one-loop self-energy insertions known as Hard-Thermal-Loop (HTL) resummation \cite{Braaten:1989mz,Pisarski:1988vd,Frenkel:1989br,Taylor:1990ia,Braaten:1991gm,Braaten:1992gd,Pisarski:1993rf}.
Thirdly, we further simplify the HTL propagators to the tree-level form by taking the limit with negligible thermal widths and momentum-independent thermal masses (cf. Sec.~\ref{sec:Tree}). 
In this way, the spectral densities in the DM self-energy simplify to $\delta$-functions.
\subsubsection{\label{sec:1PI}1PI-resummed spectral densities}
To determine the 1PI-resummed spectral densities entering Eq.~\eqref{eq:PiA_master}, we need to compute the anti-Hermitian part of the retarded 1PI-resummed propagators, given by Eq.~\eqref{eq:Resum_S_AntiH_0}, for both the fermion fields $F$ and $f$, yielding
\begin{align}
    &\dslS_F^{\mc{A}}(k)=\left(\dsl{k}-\dslSigma^{\mc{H}}_F(k)+m_F\right)\dfrac{\Gamma_F(k)}{\Omega_F^2(k)+\Gamma_F^2(k)}-\dslSigma_F^{\mc{A}}(k)\dfrac{\Omega_F(k)}{\Omega_F^2(k)+\Gamma_F^2(k)}\,, \label{eq:SFresummed}\\ 
    &\dslS_f^{\mc{A}}(q)=\left(\dsl{q}-\dslSigma^{\mc{H}}_f(q)\right)\dfrac{\Gamma_f(q)}{\Omega_f^2(q)+\Gamma_f^2(q)}-\dslSigma_f^{\mc{A}}(q)\dfrac{\Omega_f(q)}{\Omega_f^2(q)+\Gamma_f^2(q)}\,,
    \label{eq:Sfresummed}
\end{align}
where 
\begin{align}
    &\Omega_F(k)\equiv\left(k-\SigmaH_{F}(k)\right)^2-\left(\SigmaA_{F}(k)\right)^2-m_F^2\,,\label{eq:OmegaF} \\ \
    &\Omega_f(q)\equiv\left(q-\SigmaH_{f}(q)\right)^2-\left(\SigmaA_{f}(q)\right)^2\,, \label{eq:Omegaf}\\
    &\Gamma_F(k)\equiv 2 \left(k-\SigmaH_{F}(k)\right) \cdot \Sigma_F^{\mc{A}}(k)\,, \label{eq:GammaF}\\
    &\Gamma_f(q)\equiv 2 \left(q-\SigmaH_{f}(q)\right) \cdot \Sigma_f^{\mc{A}}(q)\, , \label{eq:Gammaf}
\end{align}
and $q = k-p$. Here, $\dsl{\Sigma}^{\mc{H}}$ and $\dsl{\Sigma}^{\mc{A}}$ are the Hermitian and anti-Hermitian part of the retarded fermionic self-energy.
The functions $\Omega(k)$ encapsulate the information regarding the poles of the propagators.
If $\Gamma(k)\ll\Omega(k)$ the spectral densities sharply peak around $\Omega(k)$, which can therefore be interpreted as describing propagating degrees of freedom with well-defined dispersion relations.
The functions $\Gamma_{F,f}$ are therefore regarded as thermal widths, responsible for broadening the density of states.

{
As indicated in Eqs.~\eqref{eq:OmegaF}-\eqref{eq:Gammaf}, the resummed spectral propagators require knowledge about the spectral (anti-Hermitian) and Hermitian self-energies, given by $\SiSh^{\mc{A}}(p)$ and $\SiSh^{\mc{H}}(p)$, respectively.
\begin{figure}[!t]
\centering
\includegraphics[width=0.4\textwidth]{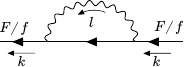}
    \caption{\small Gauge contribution to the fermion self energy at one-loop. For right-handed SM fermions and the corresponding vector-like dark fermion $F$ in Eq.~\eqref{eq:ModelLagrangian}, the wavy line corresponds to $U(1)$ gauge bosons or gluons (if color-charged). For fermions with left-handed gauge quantum numbers, the wavy line can also represent the $SU(2)$ gauge bosons.}
\label{fig:fermionSelfEn}
\end{figure}
We calculate the fermion self-energy at one-loop by exclusively taking into account the gauge corrections from the sunset diagram of Fig.~\ref{fig:fermionSelfEn}, implying that we neglect loop diagrams involving a Higgs boson exchange and Yukawa vertices for SM fermions.\footnote{Notice that, in models where DM couples to top quarks, Yukawa interactions are relevant and can give the largest contribution to the thermal mass in the dispersion relation.
Moreover, since the vectorlike fermions do not receive mass corrections beyond those arising from gauge interactions, decays of top quarks into DM and a vectorlike fermion are kinematically allowed in the ultrarelativistic regime.
This is in contrast the case in which the SM Yukawa couplings are negligible.
At lower temperatures, decays are again kinematically forbidden until the in-vacuum mass of the vector-like fermion is the dominant scale.
Such effects are not captured in this study such that our results apply for DM couplings to any SM fermion but top quarks.}
Furthermore the SM fermions' Yukawa coupling to $F$ is very small, such that we can neglect its contribution.
\\
On the CTP, the fermionic self-energy (cf. Fig.~\ref{fig:fermionSelfEn}) reads as
\begin{align}
    -i \slashed{\Sigma}^{ab}(k)_{L/R}=-G_{L/R}\int\dfour{\ell}\gamma^\mu P_{L/R}i\dslS_{L/R}^{ab}(k-\ell)P_{L/R}\gamma^\nu i\Delta^{ab}_{\mu\nu}(\ell) \, .
\end{align}
Here, $a,b$ are CTP indices, indicating on which branch of the contour the propagators are evaluated, $P_{L/R}$ are chiral projectors, and the factor $G_{L/R}$ summarizes the gauge interactions of the chiral fermion considered.
For SM interactions only, it is given by Eq.~\eqref{eq:G_def}.
Furthermore, $\Delta^{ab}_{\mu\nu}(\ell)$ is the gauge boson propagator which in Feynman gauge\footnote{We comment on the gauge dependence of our result at end of Sec.~\ref{sec:results}.} is obtained from the scalar propagator by $\Delta^{ab}_{\mu\nu}(\ell) = -g_{\mu \nu} \Delta^{ab}(\ell)$.

The Hermitian and anti-Hermitian self-energies, needed to evaluate the Eqs.~\eqref{eq:OmegaF}-\eqref{eq:Gammaf}, follow from
\begin{align}
    \SiSh^\mc{H} \left( k \right) = \frac{1}{2} \left[ \SiSh^{++} \left( k \right) - \SiSh^{--} \left( k \right) \right] \, , \quad
    \SiSh^\mc{A} \left( k \right) = \frac{1}{2} \left[ \SiSh^{-+} \left( k \right) - \SiSh^{+-} \left( k \right) \right] \, .
\end{align}
Next, we need to determine the Lorentz-vector $\Sigma^\mu_{L/R}$ (where we have suppressed its CTP-indices), which is defined via $\SiSh_{L/R} = P_{R/L} \gamma_\mu \Sigma^\mu_{L/R}$ and can be expressed in terms of the Lorentz vectors it depends on, namely $p^\mu$, the momentum of the particle, and $u^\mu$, the plasma reference vector with $u^2 =1$ and $u^0 > 0$. 
In what follows, we choose to work in the rest frame of the plasma, implying $u^{\mu}=(1,\vec{0})$. 
The self-energy can be decomposed as
\begin{align}
    &\Sigma^{\mc{H}/\mc{A},\mu}(k)=\frac{1}{2|\Vec{k}|^2}[(k^0 u^\mu- k^\mu) \mc{K}^{\mc{H}/\mc{A}}+(k^0 k^\mu-k^2 u^\mu) \mc{U}^{\mc{H}/\mc{A}}] \, ,
    \label{eq:Sigma^mu}\\
    &\mc{K}^{\mc{H}/\mc{A}}\equiv\left( k \cdot \Sigma^{\mc{H}/\mc{A}} \right), \qquad \mc{U}^{\mc{H}/\mc{A}}\equiv\left( u \cdot \Sigma^{\mc{H}/\mc{A}} \right)\,,
    \label{eq:PU}
\end{align}
as can be shown by contracting the $\Sigma^\mu(k)$ once with $k^\mu$ and once with $u^{\mu}$. 
The quantities above, $\mc{K}^{\mc{H}/\mc{A}}$ and $\mc{U}^{\mc{H}/\mc{A}}$, have to be determined for both the Hermitian and anti-Hermitian self-energies. 
The complete derivation of $\mathcal{K}^{\mc{H}/\mc{A}}$ and $\mathcal{U}^{\mc{H}/\mc{A}}$ is long and tedious, and parts of them need to be determined numerically.
Importantly, we keep track of the mass of the vector-like fermion $F$, which complicates the analytical structure of our results. On the contrary, the SM fermion $f$ can be regarded as massless since we restrict ourselves to DM production in the unbroken electroweak symmetry phase.
We present the full expressions in Appendix~\ref{sec:AppC}.

The quantities $\mc{K}^{\mc{H}/\mc{A}}$ and $\mc{U}^{\mc{H}/\mc{A}}$ can be decomposed into a vacuum ($T=0$) and a thermal part ($T\neq0$). 
The vacuum part is given by the full self-energy evaluated in the $T \rightarrow 0$ limit of all distribution functions involved in the calculation, while thermal part contains the rest.
The vacuum part of the Hermitian self-energy involves both ultraviolet and infrared divergences.
The first ones need to be renormalized as in quantum field theory at zero temperature.
The resulting observables (e.g., spectral densities) expressed in terms of renormalized quantities conserve the same analytical form. 
Importantly, finite-$T$ corrections do not affect the renormalization conditions and do not contain additional UV divergences since there is a natural cutoff at short distances given by the correlation length of thermal fluctuations.
Our results are to a good approximation independent of the vacuum part and hence we can safely assume that our observables have been renormalized appropiately. 
Regarding IR divergences, the Bloch-Nordsiek \cite{Bloch:1937pw} and the Kinoshita-Lee-Nauenberg \cite{Kinoshita:1962ur,Lee:1964is} theorems ensures that, at zero-temperature, every observable is free of any soft and collinear divergences.
The thermal part of the Hermitian self-energy, however, as well as the vacuum and thermal part of the anti-Hermitian self-energy is free of divergences. 

}

{
\begin{figure}[!t]
    \centering
\begin{subfigure}{.485\textwidth}
\includegraphics[width=0.98\textwidth]{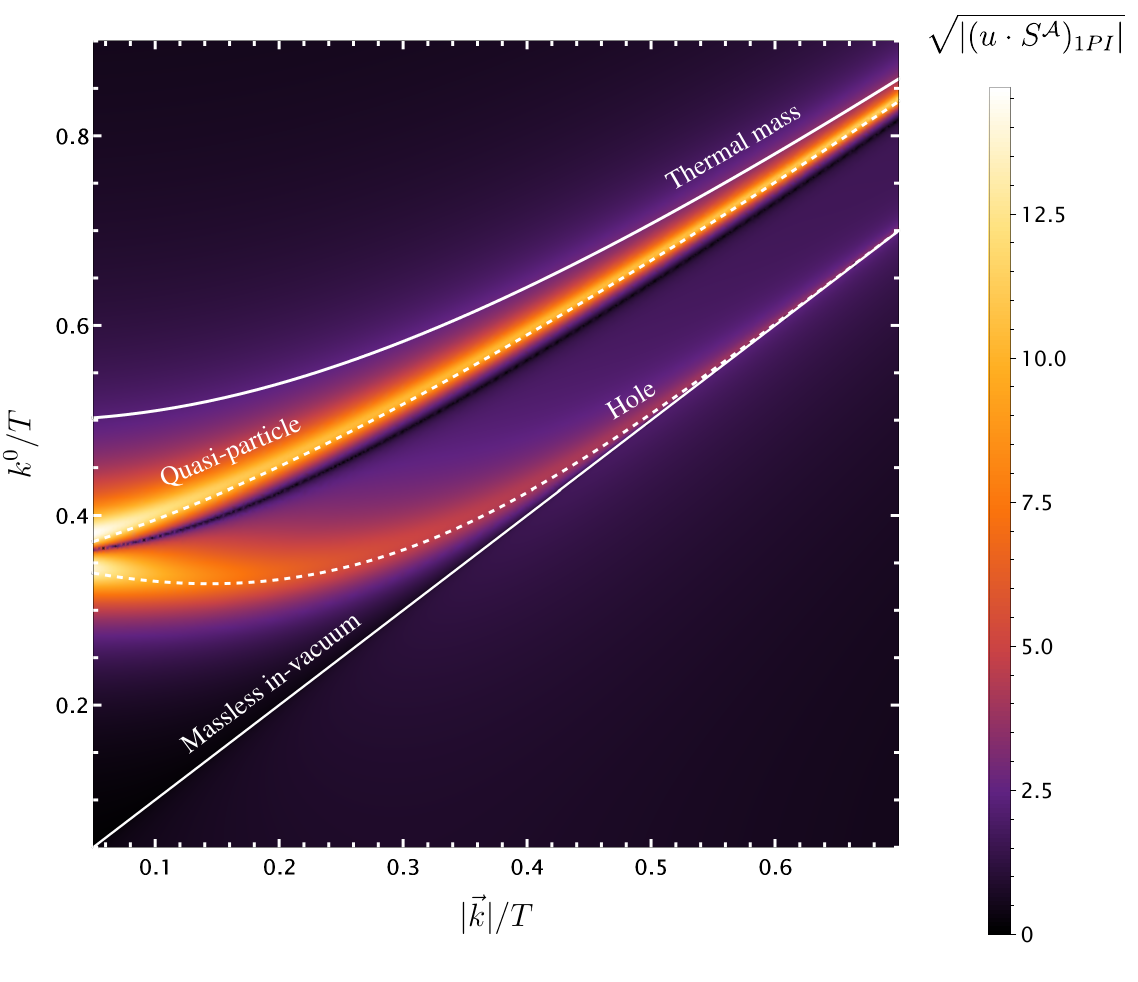}
\end{subfigure}
\begin{subfigure}{.485\textwidth}
    \includegraphics[width=0.9\textwidth]{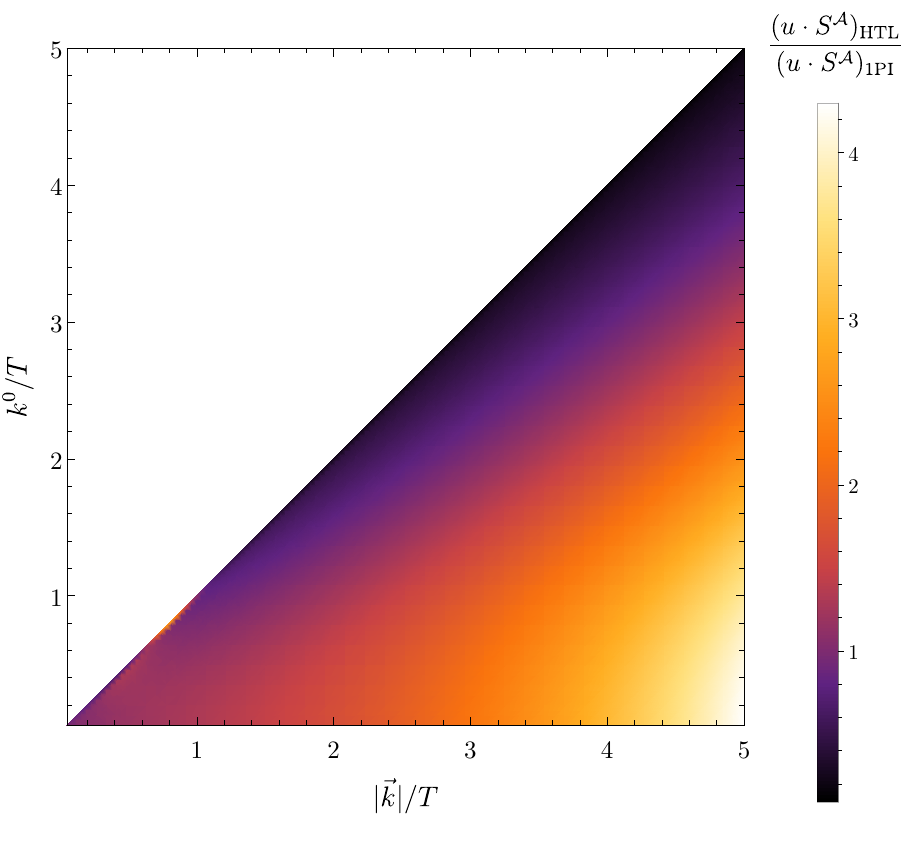}
\end{subfigure}
\caption{\emph{Left panel:} the spectral density $(u\cdot S^{\mc{A}})_{\mathrm{1PI}}$ in linear scale for a massless fermion as a function of the energy $k^0/T$ and spatial components  $|\vec{k}|/T$ of its momentum, normalized to the plasma temperature.
Brighter colors indicate larger values. The largest values correspond to the quasi-particle and hole peaks smeared out by the thermal width.
The holes fade away at large momenta when they decouple from the plasma.
The white dashed lines mark the quasi-particle and hole branches in the HTL approximation (cf. Eqs.~\eqref{eq:omega+} and \eqref{eq:omega-}), while the white solid line corresponds to the dispersion relation with a momentum-independent thermal mass in Eq.~\eqref{eq:Thermal_Mass}.
Note that in the dark regions $(u\cdot S^{\mc{A}})_{\mathrm{1PI}}$ is small but not zero. 
\emph{Right panel:} ratio of $u\cdot S^{\mc{A}}$ in the HTL approximation and using 1PI-resummed self energies for a broader kinematic range.
This ratio is only defined below the light cone, where the HTL spectral density becomes larger than the 1PI-resummed spectral density for hard momenta, as the latter captures an exponential suppression when $k\gtrsim T$.}
    \label{fig:Resummed_Fermion}
\end{figure}
The (pole) structure of the Eqs.~\eqref{eq:SFresummed} and \eqref{eq:Sfresummed} can be very convoluted and, unless a high-temperature (HTL) expansion is performed (cf. Sec.~\ref{sec:HTL}), its precise form needs some numerical input (cf. Appendix~\ref{sec:AppC}).
In particular, at finite-$T$ there could be several poles representing propagating modes and collective excitations\cite{le_bellac_1996}.
For fermions in a hot plasma, the two existing poles are usually referred to as screened \emph{quasi-particle} if they have a positive helicity over chirality ratio, or \emph{hole} modes if such ratio is negative.
These modes can be clearly distinguished in the low-momentum kinematic region, where, as shown in the left panel of Fig.~\ref{fig:Resummed_Fermion}, the brighter contours illustrate where the spectral density sharply peaks in correspondence with the poles. 
In this region, each of the two modes approximately accounts for half the density of states.
At large momenta, instead, hole excitations become massless and decouple from the plasma, while the screened quasi-particle modes only reach the in-vacuum fermionic dispersion relation asymptotically for momenta $|\vec{k}| \gg T$.
Outside the ``mass-shell'', there are also non-zero contributions which can be referred to as ``virtual'' or ``off-shell'' modes, where there is energy transfer between the fermion quanta and the thermal bath even if they would be strictly forbidden in vacuum.
The contribution in this kinematic regime is proportional to the thermal width $\Gamma$ and, although suppressed, it accounts for the irreducible \emph{Landau damping} by the medium.
Compared to its HTL-approximated version (cf. Sec.~\ref{sec:HTL}), the spectral densities exhibit two major differences. Firstly, unlike for HTL spectral densities, the thermal width $\Gamma$ does not vanish for time-like momentum, allowing for the existence of off-shell contributions away from the precise position of the poles.  Below the light cone, the 1PI-resummed spectral densities are more damped, as can be seen in the right panel of Fig.~\ref{fig:Resummed_Fermion} \footnote{For a deeper discussion on collective excitations in the plasma, we also refer to earlier works \cite{Klimov:1981ka,Weldon:1982bn,Weldon:1982aq} and more recent applications in the context of early Universe particle production \cite{Drewes:2010pf,Drewes:2013iaa}.}.
}

{
The DM interaction rate in Eq.~\eqref{eq:RateDensityResummed} receives contributions from four distinct kinematic regimes of the spectral densities, depending on whether the fermion momenta ($k$ for the BSM field $F$ and $q$ for the SM field $f$) are above or below their light cone.
\\
\newline
$\mathbf{k^2>0},$ $\mathbf{q^2>0:}$
both fermions have timelike momenta and their spectral densities are peaked at the poles, where a particle-like interpretation of the interacting fermionic states applies if the thermal width is sufficiently narrow.
DM production from hole peaks, which completely decouple at large momenta, is negligible compared to the DM production from quasi-particle peaks.
As a result, the DM interaction rate in Eq.~\eqref{eq:RateDensityResummed} is dominated by the kinematics that allow the quasi-particle peaks of the two fermionic spectral densities to overlap.
In a particle-like interpretation, these contributions can be seen as \textbf{\textit{decay processes}} of the type $F\rightarrow f + s$ with masses given approximately by $M_i^2=m_i^2+m_{\text{th},i}^2$, being $m_i$ the in-vacuum masses and $m_{\text{th},i}=\sqrt{G}\, T/2$ the thermal masses of the fermions\footnote{Decays of the type $f\rightarrow F + s$ as well as inverse decays $F+f\rightarrow s$ are always kinematically forbidden since we neglect Yukawa interactions and in-vacuum masses for the SM fermions.}.
Since at high temperatures the dispersion relations of the two fermions are identical (they receive the same gauge thermal corrections), decays are not kinematically allowed at early ``times'' $z=m_F/T$, or, in other words, the contribution to DM production from 1PI-resummed spectral densities is suppressed.
The kinematic window opens whenever the in-vacuum mass $m_F$ starts to be relevant, namely as soon as
\begin{align}
    z > \sqrt{G} \frac{\mdm m_F}{m_F^2 - \mdm^2} \, .\label{eq:Estimate_kinematic_threshold}
\end{align} 
This relation is exact if tree-level-like or HTL propagators\footnote{The HTL resummed propagators feature momentum dependent thermal masses such that the relation would develop a (mild) momentum dependence} are considered. 
For 1PI-resummed propagators, on the other hand, the non-zero thermal width relaxes the kinematic constraints such that the kinematic cutoff is softend.\\
\newline
$\mathbf{k^2>0}$ \textbf{and} $\mathbf{q^2<0}$, $\mathbf{k^2<0}$ \textbf{and} $\mathbf{q^2>0:}$
this regime constitutes the dominant contribution to the DM interaction rate when the kinematics prevent both fields from being simultaneously on the timelike peaks, as estimated above.
The spacelike fermion can be interpreted as a mediator of interactions with the thermal plasma, such that these regimes are seen as \textbf{\textit{$\mathbf{2 \leftrightarrow 2}$ scattering processes}} producing DM quanta.
\\
\newline
$\mathbf{k^2<0,\,q^2<0:}$
when both momenta are spacelike, both spectral densities are in the continuum and thus none of them is enhanced from being nearly ``on-shell''. 
Thus, this regime gives the smallest contribution to the DM-interaction rate.
In a particle interpretation, they can be viewed as two exchanges of ``off-shell'' virtual fermion quanta with the thermal plasma, corresponding to \textbf{\textit{$\mathbf{2 \leftrightarrow 3}$ scattering processes}}.
}

\subsubsection{\label{sec:HTL}Resummed propagators in the HTL approximation}
The spectral densities obtained in the previous section can be simplified by assuming that the relevant loop momenta $\ell$ in the fermion self-energies (see Fig.~\ref{fig:fermionSelfEn}) are much larger than any other mass scale involved ($k, m_F, \mdm$) and as if a net separation between hard scales $\mc{O}(\pi T)$ and soft scales $\mc{O}(gT)$ applies.
This is, in essence, what is referred to as Hard Thermal Loop (HTL) effective perturbation theory, where the self-energies resum the leading contributions from thermal fluctuations with hard virtual momenta \cite{Braaten:1989mz,Pisarski:1988vd,Frenkel:1989br,Taylor:1990ia,Braaten:1991gm,Braaten:1992gd,Pisarski:1993rf}.
Practically speaking, we can obtain the HTL-approximated self-energies by operating an expansion in $k/\ell\sim g/\pi \sim \sqrt{G}/T$ and considering only the leading terms.
Thus, we expect the expansion to be reliable if $\sqrt{G}\ll \pi$   .

It is now important to examine whether such a framework, which is notably simpler to work with and often employed in leptogenesis and DM scenarios, can be regarded as reliable for the freeze-in production of DM.
The fact that the relevant freeze-in dynamics take place at temperatures $T\sim m_F$ suggests that the momentum of the heavy fermion $F$ in the DM self-energy is at least of order $k\sim m_F\sim T$. 
Consequently, the expansion parameter $k/\ell$ is of $\mc{O} (1)$, when calculating the self-energy of $F$, indicating that the HTL approximation is not well-justified in the region where the bulk of DM production takes place.
Nevertheless, in Sec.~\ref{sec:results}, we will compute the DM interaction rate from HTL-resummed propagators to provide the deviation of this result from the one obtained from the full form of the 1PI-resummed propagator.

To understand the differences between the HTL and 1PI-resummed method, in the following we summarize the structure of the HTL self-energy, spectral densities, and production rates.
First, we need the HTL limit of the gauge-charged fermion self-energies $\Sigma_{F/f}$, which coincide in the HTL approximation.
The HTL version of $\mc{K}_{\mathrm{HTL}}=(k\cdot \Sigma_\mathrm{HTL})$ and $\mc{U}_{\mathrm{HTL}}=(u\cdot \Sigma_\mathrm{HTL})$ reads as
\begin{align}
    \mathcal{K}_\text{HTL}^\mc{H} &= \frac{1}{4} G T^2,   &&\mathcal{U}_\text{HTL}^\mc{H} = \frac{G}{8 \kvec} T^2 \log \left| \frac{k^0 + \kvec}{k^0 - \kvec} \right|, \label{eq:PUHerHTL} \\
    \mathcal{K}_\text{HTL}^\mc{A} &= 0,   &&\mathcal{U}_\text{HTL}^\mc{A} = \theta \left( - k^2 \right) \frac{G} \pi{8 \kvec} T^2 \, . 
    \label{eq:PUantiHerHTL}
\end{align}
Note that the functions in Eqs.~\eqref{eq:PUHerHTL} and \eqref{eq:PUantiHerHTL} only depend on $T$ and the external momentum, while any in-vacuum mass scale has been neglected in accordance with the high-temperature expansion.
We now insert these expressions into Eq.~\eqref{eq:Sigma^mu} and then into the functions $\Omega_{F/f}$ and $\Gamma_{F/f}$ in Eqs.~\eqref{eq:OmegaF}-\eqref{eq:Gammaf}, leading to the HTL-limit of the spectral densities from Eqs.~\eqref{eq:SFresummed} and \eqref{eq:Sfresummed}.

The thermal widths $\Gamma_{F/f}^{\mathrm{HTL}}$, directly proportional to $\mc{U}^{\mc{A}}_\mathrm{HTL}$, are non-zero only for spacelike momenta $k^2<0$.
In this kinematic regime, we recover the so-called \emph{continuum}.
Analogously to 1PI-resummed spectral densities, the continuum describes Landau damping, with ``off-shell'' states exchanged with the plasma (notice that a particle picture cannot be established without a properly defined dispersion relation).
However, the HTL continuum parts are less suppressed at large momenta because they do not capture an exponential suppression as soon as $k\gtrsim T$ as we quantify for the $u \cdot S^{\mc{A}}$ component in the right panel of Fig.~\ref{fig:Resummed_Fermion}.
We refer to Appendix~\ref{sec:AppD} for more details about the analytical structure of the HTL continuum.

For timelike momenta $k^2>0$, the HTL widths are vanishing, $\Gamma_{F/f}^{\mathrm{HTL}} = 0$, and the HTL spectral densities simplify to $\delta$-functions
\begin{align}
    \SsA_{F/f} (k) &= \pi\, \sgn{k^0} \left( \slashed{k} + m_{F/f} -\SiSh^{\mc{H},\mathrm{HTL}}_{F/f}(k) \right) \delta \left( \left[ k - \Sigma^{\mc{H},\mathrm{HTL}}_{F/f} (k) \right]^2 - m_{F/f}^2 \right) \, ,
    \label{eq:HTL_prop}
\end{align}
where $m_{F/f}$ is the in-vacuum mass.
Here, we recover a situation similar to tree-level in-vacuum spectral densities, with a particle-like behavior featuring a well-defined dispersion relation given by 
\begin{align}
    \left[ k - \Sigma^{\mc{H},\mathrm{HTL}}_{F/f} (k) \right]^2 - m_{F/f}^2 = 0 \, . \label{eq:HTL_Dispersion_Eq}
\end{align}
For a vanishing in-vacuum mass, this equation can be analytically solved by (see, for instance, \cite{Kiessig:2010pr})
\begin{align}
    &\left( k^0 \right)^2 = \omega_{+}^2(\kvec)=\kvec^2\left[\dfrac{W_{-1}\left(-\exp(-\alpha \kvec^2-1)\right)-1}{W_{-1}\left(-\exp(-\alpha \kvec^2-1)\right)+1}\right]^2\,,\label{eq:omega+}\\
    &\left( k^0 \right)^2 = \omega_{-}^2(\kvec)=\kvec^2\left[\dfrac{W_{0}\left(-\exp(-\alpha \kvec^2-1)\right)-1}{W_{0}\left(-\exp(-\alpha \kvec^2-1)\right)+1}\right]^2\,,\label{eq:omega-}
\end{align}
where $\alpha = 16/ (G T^2)$ and where $W_0$ and $W_{-1}$ are Lambert-W functions. 
These solutions physically correspond to the same \textit{quasi-particle} and \textit{hole} poles, where the 1PI-resummed spectral densities peak, as described in the previous section (cf. left panel of Fig.~\ref{fig:Resummed_Fermion}).
For a fermion with an in-vacuum mass, an analytic form does not exist and Eq.~\eqref{eq:HTL_Dispersion_Eq} has to be solved numerically. 
Note that massive fermions also have two pole branches with the same qualitative dispersion relation.
Notably, in the large momentum limit, $\kvec \gg T$, the quasi-particles have a dispersion relation of the form
\begin{align}
    \left( k^0 \right)^2 = \kvec^2 + \frac{G}{4} T^2 \equiv \kvec^2 + m_\text{th}^2 \, , \label{eq:Thermal_Mass}
\end{align}
where $m_\text{th}$ is the \textit{thermal mass} of the field and it corresponds to the momentum-independent part of the thermal Hermitian self-energy.
For fermions interacting with gauge bosons in the plasma, the thermal mass reads
\begin{align}
    m_\mathrm{th}^2=\frac{G}{4}T^2\,. \label{eq:ThermalMass_pindependent}
\end{align}
Inserting the HTL-resummed spectral densities into the DM self-energy in Eq.~\eqref{eq:PiA_master}, we can identify the same four kinematic regions and DM production processes discussed at the end of Sec.~\ref{sec:1PI}, corresponding to whether the fermionic four-momenta $k$ and $q$ are spacelike or timelike.
The DM self-energy is then given by the sum of the four regimes
\begin{align}
    \Pi_{s,\text{HTL}}^\mc{A} (p) = \Pi_{s,\text{TT}}^\mc{A} (p) + \Pi_{s,\text{TS}}^\mc{A} (p) +\Pi_{s,\text{ST}}^\mc{A} (p) +\Pi_{s,\text{SS}}^\mc{A} (p) . \label{eq:SelfEnergy_HTLTOT}
\end{align}
The subscripts TT, TS, ST, and SS indicate if the momenta $k$ and $q$ are timelike (T) or spacelike (S).
The same interpretations that were outlined for 1PI-resummed spectral densities, also apply in the HTL case, with some adjustments.
For instance, above the light cone, the vanishing thermal width implies that we do not have resonant peaks anymore, but $\delta$-functions that strongly constrain which momenta give a non-zero DM production rate.
If ``decay'' processes dominate, we should expect 1PI-resummed and HTL spectral densities to yield a similar DM abundance for small gauge couplings (the 1PI-resummed peaks are quite narrow), but be dissimilar as soon as the widths become wider and the HTL approximation less reliable at larger $G$ (which is indeed the case, as outlined in Sec.~\ref{sec:results}).
Moreover, as already discussed, HTL spectral densities are less suppressed compared to 1PI-resummed spectral densities at spacelike momenta because the in-vacuum mass of the vectorlike fermion $F$ is neglected (cf. right panel of Fig.~\ref{fig:Resummed_Fermion}).
Therefore, HTL propagators are expected to lead to larger DM production from scattering processes compared to 1PI-resummed propagators.
As we will show in Sec.~\ref{sec:results}, such differences will have the largest impact on the DM relic abundance precisely when decay production is kinematically forbidden and scattering processes are the only available channels.
\subsubsection{\label{sec:Tree}Tree-level propagators}
We now take the tree-level limit of the HTL fermionic spectral densities, considering only the leading-order correction in $gT$ to the HTL fermionic self energies.
This implies that the thermal widths are identically zero and that, in the dispersion relations, we only retain the momentum-independent thermal masses (cf. Eq.~\ref{eq:ThermalMass_pindependent}). 
In this way, the spectral densities in the DM self-energy loop simplify to delta functions defined above the light cone, yielding
\begin{align}
    &\dslS^{\mc{A}}_F(k)=\pi\delta(k^2-M_F^2)\rp{\dsl{k}+M_F}\,\sgn{k^0}\,,\\
    &\dslS^{\mc{A}}_f(k-p)=\pi\delta\rp{(k-p)^2-M_f^2}\rp{\dsl{k}-\dsl{p}+M_f}\,\sgn{k^0-p^0}\,,
\end{align}
where the masses are determined by the in-vacuum masses and the (momentum-independent) thermal masses, according to $M_i^2 = m_i^2 + m_{i,\mathrm{th}}^2$.
In analogy to what we have discussed in the previous sections, such a configuration corresponds to only accounting for DM production from decays of on-shell fermionic states with in-vacuum and thermal masses as in Eq.~\eqref{eq:Thermal_Mass}.
To obtain the tree-level limit of the DM self-energy, we insert these expressions into Eq.~\eqref{eq:PiA_master}, yielding
\begin{align}\label{eq:PiA_tree0}
    &\Pi^{\mc{A}}_s(p)=4\pi^2\ydm^2\int\dfour{k}\delta(k^2-m_F^2)\delta\rp{(k-p)^2-m_f^2}\rp{k^2-k\cdot p}\nonumber\\
    &\hspace{4cm}\times\left[1-f_+(k_0)-f_+(p_0-k_0)\right]\,\sgn{k_0}\sgn{k_0-p_0}\,.
\end{align}
This integral can be analytically resolved.
For instance, by transforming the integration measure with
\begin{align}
    \dd^4 k\longrightarrow\frac{1}{2\pvec}\dd\phi\,\dd k_0\,\dd(p\cdot k)\,\dd(k^2)\,,
\end{align}
it is easier to perform the integration over the delta functions.
Moreover, since the particles in the loop are forced to be on-shell, one can show that
\begin{align}
    \sgn{k_0}\,\sgn{k_0-p_0}=-\sgn{p^2-m_F^2-m_f^2}\,.
\end{align}
With these observations at hand, we can turn Eq.~\eqref{eq:PiA_tree0} into
\begin{align}\label{eq:PiA_tree1}
    \Pi^{\mc{A}}_s(p)=\dfrac{\ydm^2}{16\pi\,|\vec{p}|}\Big|p^2-m_F^2-m_f^2\Big|\int_{\mc{B}}\dd k_0\,\Big[1-f_+(k_0)-f_+(p_0-k_0)\Big]\,,
\end{align}
where the integration boundaries of the integral $\mc{I}$ are such that
\begin{align}
    \mc{I}\Big|_\mc{B}=\mc{I}\Big|_{k_{-}^{0}}^{k_{+}^{0}} \,\theta\rp{p^2} + \mc{I}\Big|_{-\infty}^{+\infty} \,\theta\rp{-p^2}\,,
\end{align}
with
\begin{align}
    k_{\pm}^0=\frac{p_0}{2p^2}\left(p^2+m_F^2-m_f^2\right)\pm\frac{\pvec}{2p^2}\sqrt{\lambda\left(p^2,m_F^2,m_f^2\right)},\\
    \lambda\left(p^2,m_F^2,m_f^2\right)=\left(p^2-(m_F+m_f)^2\right)\left(p^2-(m_F-m_f)^2\right)\,.
\end{align}
By integrating the expression in Eq.~\eqref{eq:PiA_tree1}, accounting for the full distribution functions at finite-temperature, we obtain
\begin{align}
    \Pi^{\mc{A}}_s(p)=\dfrac{\ydm^2}{16\pi\pvec}\Big|p^2-m_F^2-m_f^2\Big|\left(p_0\,\theta(-p^2)+T\ln\left|\dfrac{\cosh\rp{(p_0-k^0_{-})/2T}}{\cosh\rp{(p_0-k^0_{+})/2T}}\,\dfrac{\cosh\rp{k^0_{+}/2T}}{\cosh\rp{k^0_{-}/2T}}\right|\right)\,.
\end{align}
Finally, the integrand in Eq.~\eqref{eq:numbdensEQ} with $p^2=\mdm^2>0$ and $p_0=\omega_p=\sqrt{\mdm^2+\pvec^2}$ is proportional to
\begin{align}
    \dfrac{\Pi^{\mc{A}}_s(p)}{\omega_p}=\dfrac{\ydm^2\,T}{16\pi\,\omega_p\,\pvec}\Big|\mdm^2-m_F^2-m_f^2\Big|\,\ln\left|\dfrac{\cosh\rp{(\omega_p-k^0_{-})/2T}}{\cosh\rp{(\omega_p-k^0_{+})/2T}}\,\dfrac{\cosh\rp{k^0_{+}/2T}}{\cosh\rp{k^0_{-}/2T}}\right|\,.
\label{eq:InRate_tree2}
\end{align}
Notice that, the anti-hermitian DM self-energy in principle describes all kinematically viable (inverse) decay processes.
In our setup, the only mass hierarchy that can be realized at any temperature is $M_F>M_f+\mdm$ (cf. Eq.\eqref{eq:Estimate_kinematic_threshold}), and thus kinematics would select only $F~\rightarrow~f~+~s$ decays. 
The expression can be derived from a Boltzmann equation approach by only accounting for decays without approximating the equilibrium distribution functions by Maxwell-Boltzmann distributions \cite{Lebedev:2019ton,Belanger:2018ccd,Bringmann:2021sth}.

The total DM production rate $\gamma_\mathrm{DM}$ from decays, including relativistic quantum statistics, can be obtained simply by integrating Eq.~\ref{eq:InRate_tree2} over $\dd \pvec$.
Similar to the calculation involving HTL-propagators, when compared to 1PI-resummed spectral densities, we shall expect tree-level-like propagators to yield the largest discrepancy at the level of the DM relic abundance whenever decay processes are suppressed (small mass splitting between $F$ and DM), and when a high-temperature and small-coupling expansion is not reliable anymore.
Compared to the HTL case, this effect is enhanced since tree-level-like propagators completely miss scattering contributions.
The results of the different methods discussed in this section are compared to each other on the level of the total DM interaction rate density and the level of the relic density in the next section.   
\section{\label{sec:results}Comparing the 1PI-resummed calculation with other approaches}
In the following, we discuss the interaction rate densities $\gamma_\text{DM}$, and the relic abundance of DM, defined by Eq.~\eqref{eq:OmegaDM_final}, obtained in the various approaches previously described.
In particular, we want to compare the results obtained based on the 1PI-resummed propagators in the CTP formalism (cf. Sec.~\ref{sec:1PI}) with the other, in the literature well-established approaches:
\begin{itemize}
    \item the Boltzmann approach including decays with in-vacuum mass, where the DM interaction rate density is given in Eq.~\eqref{eq:DecayRate}. 
    \item the Boltzmann approach including decays with in-vacuum \textit{and} thermal masses, where the DM interaction rate density is given in Eq.~\eqref{eq:DecayRateTh}.
    \item the Boltzmann approach including both decays with in-vacuum \textit{and} thermal masses, and scatterings with propagators regulated by thermal masses, where the interaction rate density is given in Eq.~\eqref{eq:Rate_scatdec}.
    \item the CTP approach with tree-level propagators, where the DM interaction rate density is given in Eq.~\eqref{eq:numbdensEQ} evaluated with Eq.~\eqref{eq:InRate_tree2}.
    \item the CTP approach with HTL-approximated 1PI-resummed propagators, where the DM interaction rate density is given in Eq.~\eqref{eq:numbdensEQ} evaluated with Eq.~\eqref{eq:SelfEnergy_HTLTOT}.
\end{itemize}
Notice that, as previously discussed, for the methods relying on the CTP, we take into account the full relativistic quantum statistics, while in the three scenarios using the Boltzmann equation, we employ non-relativistic Maxwell-Boltzmann distribution functions.

\subsection{\label{sec:IntRateDens}Interaction rate densities}
We start by comparing the interaction rate densities derived in the previous section.
This allows us to identify where and how the various approximations deviate from the 1PI-resummed result using the CTP formalism -- the central result of this work.
\subsubsection{Numerical procedure}
We first normalize each dimensionful parameter to the plasma temperature such that we can operate with dimensionless variables.
Then, we fix the free parameters
\begin{align}
    G,\quad \delta=\frac{m_F-\mdm}{m_F},\quad z\equiv m_F/T,
    \label{eq:params}
\end{align}
where $\delta$ is the relative mass splitting between the mediator and DM, while $G$ is the effective gauge coupling defined in Eq.~\eqref{eq:G_def}.
In particular, we scan over $G \in \left[ 0.4,1.6 \right]$ in steps of $0.1$ and $\log_{10}\delta \in \left[-1, 1 \right]$ in steps of $2/9$, for $32$ values of $z$ between $z=0.01$ and $z=10$, resulting in $13\times10\times32~=~4160$ parameter points. 
The interval for $G$ is motivated by some of the typical values the running gauge couplings assume at one-loop order when considering different gauge quantum number assignments for the parent particle at several energy scales $\mu$ (cf. Tab.~\ref{tab:G_val}).
This energy scale is identified with the energy scale of DM production, which can be identified by the maximum of the parent particle mass $m_F$ and the temperature $T$, such that the effective gauge couplings run when $z<1$. 
We estimated the changes in the relic density induced by running effects when $G (\mu=\text{max}\left[ m_F, T \right])$ compared to a constant effective gauge coupling defined via $G(\mu=m_F)$ by assuming a color-dominated effective gauge coupling.
We employ $\beta$-functions at one-loop order and we only consider contributions from scatterings (which dominate for $z<1$) for this estimate. 
We find that corrections are at most $\mc{O}(1\%)$, as expected from the logarithmic dependence of the running on $z$.
Thus, in the following, we assume a constant effective gauge coupling, evaluated at the energy scale of the vacuum mass $m_F$ of the heavy vectorlike fermion.
Moreover, since what we have discussed so far only applies in the unbroken phase of the electroweak (EW) symmetry, our computation is reliable if freeze-in happens before its breaking, namely if $T_\text{fi}\simeq m_F/5 \gtrsim T_\text{EW}$, where $T_\text{EW}\simeq 160\,\GeV$ \cite{DOnofrio:2015gop}.
Thus, for masses above $m_F \gtrsim \TeV$, we can safely assume that DM production lies in the EW unbroken phase.
\begin{figure}[!t]
    \centering
    \begin{subfigure}{.49\textwidth}
        \centering
        \includegraphics[width=.99\textwidth]{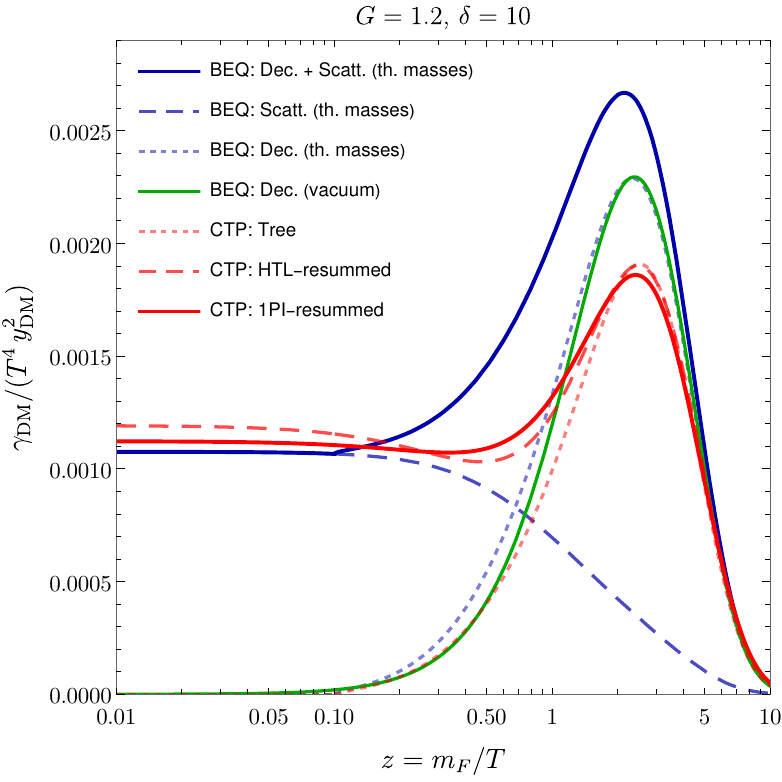}
    \end{subfigure}
    %
    %
    \begin{subfigure}{.49\textwidth}
        \centering
        \includegraphics[width=.99\textwidth]{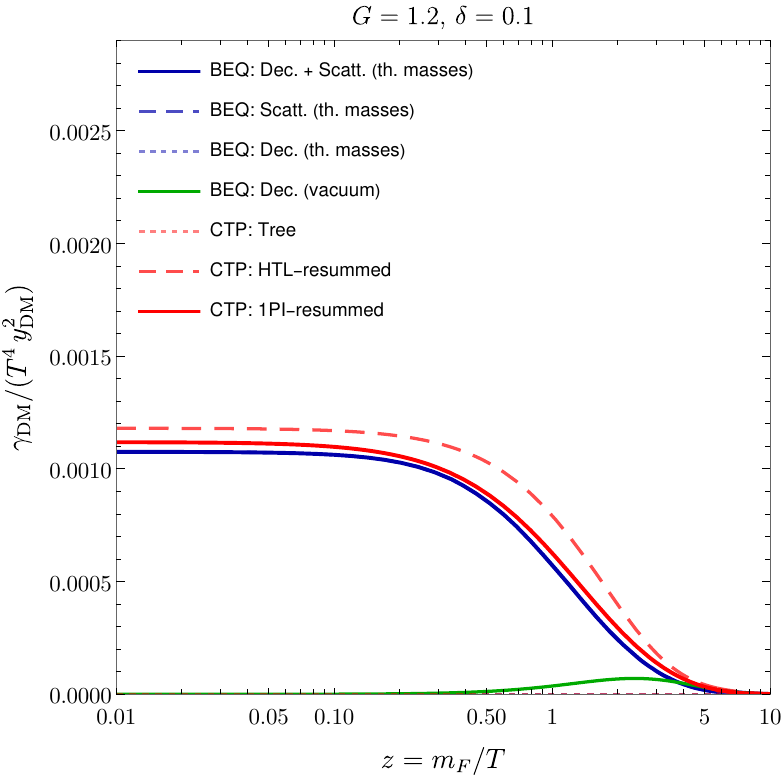}
    \end{subfigure}\\
    \vspace{4mm}
    \begin{subfigure}{.49\textwidth}
        \centering
        \includegraphics[width=.99\textwidth]{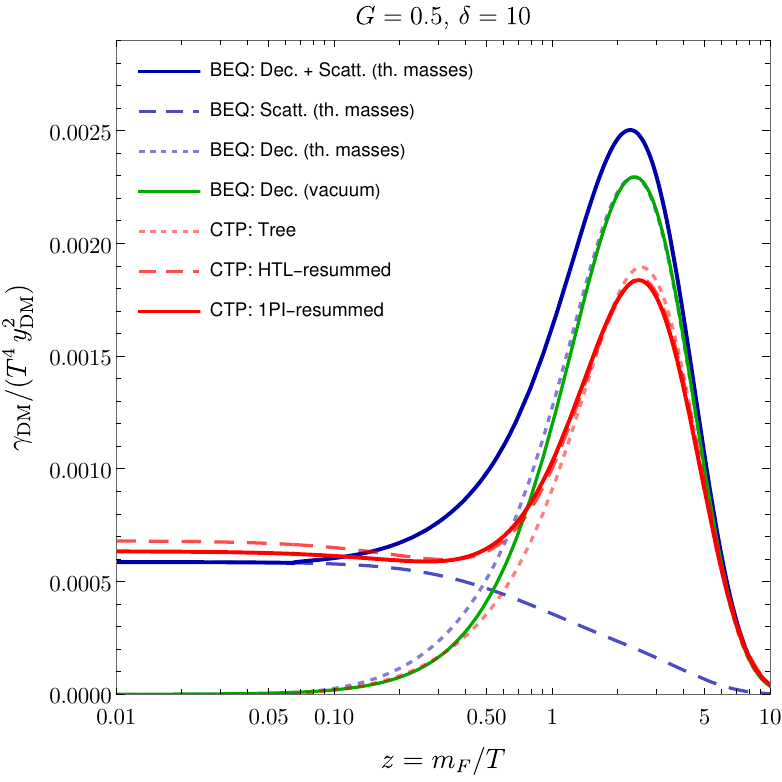}
    \end{subfigure}
    %
    %
    \begin{subfigure}{.49\textwidth}
        \centering
        \includegraphics[width=.99\textwidth]{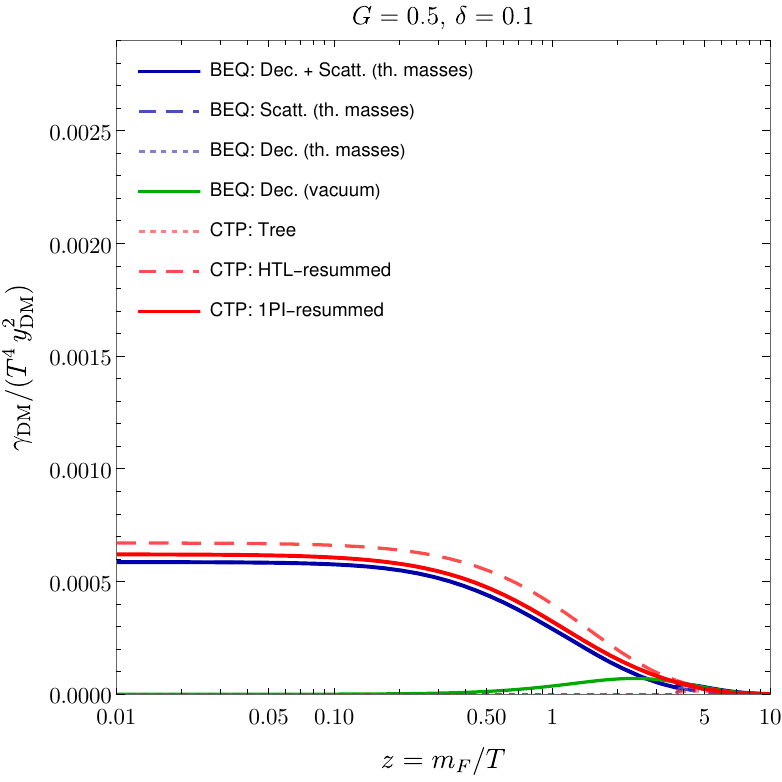}
    \end{subfigure}
    \caption{Evolution of the interaction rate density in terms of the time variable $z = m_F / T$ for a calculation using: the Boltzmann equation (BEQ) including decays with thermal masses (blue and short dashing), scatterings (blue and long dashing) with thermal masses and their sum (solid blue).
    The rate from decays with in-vacuum masses is shown with solid green lines.
    In the CTP formalism, the rates are derived with 1PI-resummed propagators (red, solid), using HTL propagators (red, long dashing), and tree-level propagators (red, short dashing).
    The latter also corresponds to decays with thermal masses and full quantum relativistic statistics.
    }
    \label{fig:InteractionRateComp} 
\end{figure}

The integration in Eq.~\eqref{eq:RateDensityResummed} is performed numerically for each of the considered $4160$ parameter points, by employing the VEGAS adaptive Monte Carlo integration algorithm \cite{Lepage:1977sw} of the GNU Scientific Library \cite{GSL}.
We make sure that the results of the integration feature an estimated error of at most $ 1\%$.
In Fig.~\ref{fig:InteractionRateComp}, we present the time evolution of the interaction rate densities for the various methods described in the previous sections for four different benchmark parameter choices, $(G, \delta)=  \left \lbrace (1.2, 10), (1.2, 0.1), (0.5, 10), (0.5, 0.1) \right \rbrace$. 

\begin{figure}[!t]
    \centering
    \includegraphics[width=0.6\textwidth]{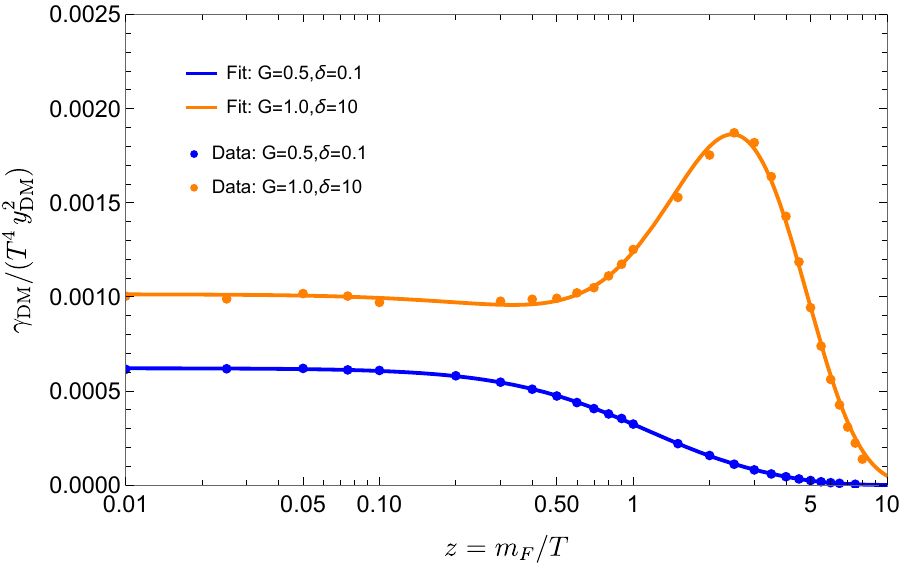}
    \caption{
    We display the numerically obtained data points for the DM interaction rate density using 1PI-resummed propagators, as well as our fit to the data for two choices of $(G,\delta)$. }
    \label{fig:Quality_Fit}
\end{figure}

For the case of 1PI-resummed and HTL-resummed propagators, we fit the values of $\gamma_\text{DM}$ obtained with the numerical integration by using the following function
\begin{align}
    \frac{\gamma_\text{DM}(z)}{\ydm^2 T^4} = A z^3 K_1 \left( a z \right) + B z K_1 \left( b z \right) \, , 
    \label{eq:FitFunction}
\end{align}
where $A$, $a$, $B$, and $b$ are the parameters of the fit, which depend on the parameters $G$ and $\delta$, and where $K_1(z)$ is a modified Bessel function of the second kind. 
We have verified that the relic density obtained with the fitted function does not deviate more than $2\%$ from the relic density obtained from a linear interpolation of our data points. 
In Fig.~\ref{fig:Quality_Fit}, we show the numerically obtained data points using 1PI-resummed propagators as well as the fit using Eq.~\eqref{eq:FitFunction} for two exemplary data points.

The results are shown with solid (long-dashed) red lines in Fig.~\ref{fig:InteractionRateComp} for the 1PI-resummed (HTL-resummed) propagators.
The form of this function is motivated by the behavior of the interaction rate in the high- and low-temperature regimes:\\
\\
\textbf{At high-temperatures} (small $z$), decays are suppressed and, eventually, kinematically forbidden (cf. Eq.~\eqref{eq:Estimate_kinematic_threshold}), while scatterings dominate due to the enhanced number density of potential scattering partners ($n^\text{eq} \sim T^3$).
The interaction rate for scatterings in vacuum typically behaves as $\sim z K_1 \left( z \right)$, motivating the second term in Eq.~\eqref{eq:FitFunction}.
Notice that, if $z\ll1$, this term is approximately constant in $z$, explaining the plateau in Fig.~\ref{fig:InteractionRateComp}.
Since at high temperatures the mass splitting between the parent particle and the DM candidate becomes negligible, the strength of the scattering contribution is solely regulated by its direct proportionality to the effective gauge coupling $G$, explaining why the plateau at small $z$ is higher for larger $G$.
Thus, we expect the ratio of the fitting parameters $\frac{B}{b}$ to exclusively depend on $G$, since
\begin{align}
A z^3 K_1 \left( a z \right) + B z K_1 \left( b z \right) \overset{z\rightarrow 0}{\sim} \frac{B}{b} \, .    
\end{align}
%
We can estimate a functional dependence of the fitting parameters based on the assumption that at high-$T$ the interaction rate is dominated by scatterings involving one gauge boson vertex ($\propto G$) and featuring fermions exchanged in the $t$-channel ($\propto G \ln G$) \cite{Garbrecht:2013bia}.
For the spectral densities from 1PI-resummed propagators, we find from our fit
\begin{align}
    \frac{B}{b} = 1 \cdot 10^{-3}\,G - 3.32 \cdot 10^{-4}\,G \ln G \, . \label{eq:UVFit}
\end{align}
\textbf{At low-temperatures} ($z \gtrsim 1$), parent particle decays constitute the dominant contribution to the interaction rate density, as long as the mass splitting is large enough to kinematically allow for them (cf.  Eq.~\eqref{eq:Estimate_kinematic_threshold}) while the parent particles are sufficiently abundant. 
In this case, the interaction rate typically behaves as $\sim z^3 K_1 \left( z \right)$, explaining the first term in Eq.~\eqref{eq:FitFunction} and the peaks observed in Fig.~\ref{fig:InteractionRateComp} for $\delta=10$.
Up to small corrections in $G$, the height of the peak is mainly controlled by the mass splitting $\delta$. 
Note that, by lowering $\delta$, the contribution from decays becomes increasingly negligible due to the suppression of the available phase space, until they disappear for degenerate masses.
For large $z$ values we can identify the following asymptotic behavior
\begin{align}
  A z^3 K_1 \left( a z \right) + B z K_1 \left( b z \right) \overset{z\rightarrow \infty}{\sim} \frac{A}{\sqrt{a}} \exp (- a z) z^\frac{5}{2}  \, .
\end{align}
For decay-dominated freeze-in, one typically can identify $A/\sqrt{a}=\Gamma_F/ m_F^{2}$, with $\Gamma_F$ being the in-vacuum decay width of the decaying $F$ parent particle.
Since this identification is only valid in the non-relativistic expansion, we can simply use in-vacuum masses to evaluate $\Gamma_F/ m_F^{2}$ and thus obtain a parametric dependence only on the mass splitting of the form
\begin{align}
    \frac{A}{\sqrt{a}} = 0.067 \frac{\delta^2 \left( \delta + 2 \right)^2}{16 \pi \left( 1 + \delta \right)^4} \, . \label{eq:IRFit}
\end{align}
\subsubsection{Interpretation}
Let us now compare the solutions depicted in Fig.~\ref{fig:InteractionRateComp} obtained using HTL-resummed (long-dashed red lines) and tree-level propagators (short-dashed red lines) to our main result obtained using 1PI-resummed propagators (solid red lines) within the CTP formalism. 
\\ 
\newline
\textbf{CTP with HTL resummed propagators:}
Overall, the HTL-resummed solution has a larger contribution from scatterings, while the rates are similar in the decay-dominated regime (large $z$).
In the transition region around $z \sim 1$, the HTL solution underestimates the interaction rate density. 
These effects can be understood as follows:
for small $z$, where scatterings dominate, the leading contribution comes from momentum configurations where one of the two fermionic propagators has a spacelike momentum. 
Since the HTL approximation neglects the in-vacuum mass of the parent, it fails to capture an exponential suppression at large spacelike momenta, instead present in the 1PI-resummed propagator (cf. Fig.~\ref{fig:Resummed_Fermion}), causing the larger rate in this regime.
On the other hand, decays start to become relevant as soon as the dispersion relations of the $F$ and $f$ fermions allow them to take place.
This strictly applies for HTL-resummed propagators, which, for timelike momenta, are $\delta$-functions.
On the contrary, 1PI-resummed propagators have finite widths even for timelike momenta such that the ``smeared'' spectral densities allow to capture the decay contribution for a wider range of momenta, also at smaller $z$ values (explaining why the red solid lines rise earlier in Fig.~\ref{fig:InteractionRateComp}).
This effect is enhanced for large values of $G$.
As soon as the decays are fully accessible, HTL rates overestimate the decay contribution for the same reason: the finite width of the 1PI-resummed propagators smear out the quasi-particle peak, leading to a slight reduction of the rate when the quasi-particle solutions are kinematically accessible. 
\\ 
\newline
\textbf{CTP with tree-level propagators:} In this set-up, the rate is computed with Eq.~\eqref{eq:InRate_tree2} and accounts only for decays with thermal masses and quantum relativistic statistics but omits any scattering contribution. 
This rate is equivalent to the one obtained with HTL timelike propagators when approximating $(\Sigma_{F/f}^\mc{H})^2 = m_\text{th}^2$. 
Therefore, the interpretation closely follows what we have just discussed above for decays in the HTL approximation. \\
\newline
\textbf{Boltzmann approach with decays and scatterings using thermal masses:}
By using thermal masses, we underestimate scattering contributions, whereas the contributions coming from decays are generically larger. 
The first feature is a consequence of the use of Boltzmann statistics instead of  quantum statistics when calculating the collision term.
The overestimation for larger $z$, on the other hand, could originate from multiple reasons. 
Firstly, accounting for the appropriate Bose-Einstein and Fermi-Dirac distributions (red short-dashed lines) implies that final-state fermions are disfavoured by Pauli blocking so that simply using Boltzmann statistics (blue short-dashed lines) increases the decay contribution.
In addition, as indicated in Fig.~\ref{fig:Resummed_Fermion}, approximating the fermion dispersion relations only with in-vacuum and thermal masses tends to overestimate the total mass, increasing the decay contribution when decays are efficient. 
At the same time, a larger thermal mass also leads to an earlier closure of the kinematically-allowed window for decays.
Furthermore, we find that thermal mass-regulated scatterings provide a non-negligible contribution significantly longer than scattering processes captured by 1PI-resummed propagators. 
These differences explain why the height of the decay peaks is not the same for the solid blue and red curves. \\ \newline
\textbf{Boltzmann approach with vacuum decays:}
This method does not consider scattering contributions and thus strongly underestimates the production rate at small $z$.
For larger $z$, where decays dominate the interaction rate, it lies in between the rates obtained from decays with thermal masses and the CTP rates.
Given that this rate neglects the contribution from scatterings as well as thermal masses, the result is smaller than the one including the scattering contribution (solid blue).
The larger rate compared to the CTP-based methods (red lines) is in large parts caused by the use of Boltzmann statistics. 

\subsection{\label{sec:RelicDens}Relic density}
\begin{figure}[!th]
    \centering
    \begin{subfigure}{.49\textwidth}
        \centering
        \includegraphics[width=.99\textwidth]{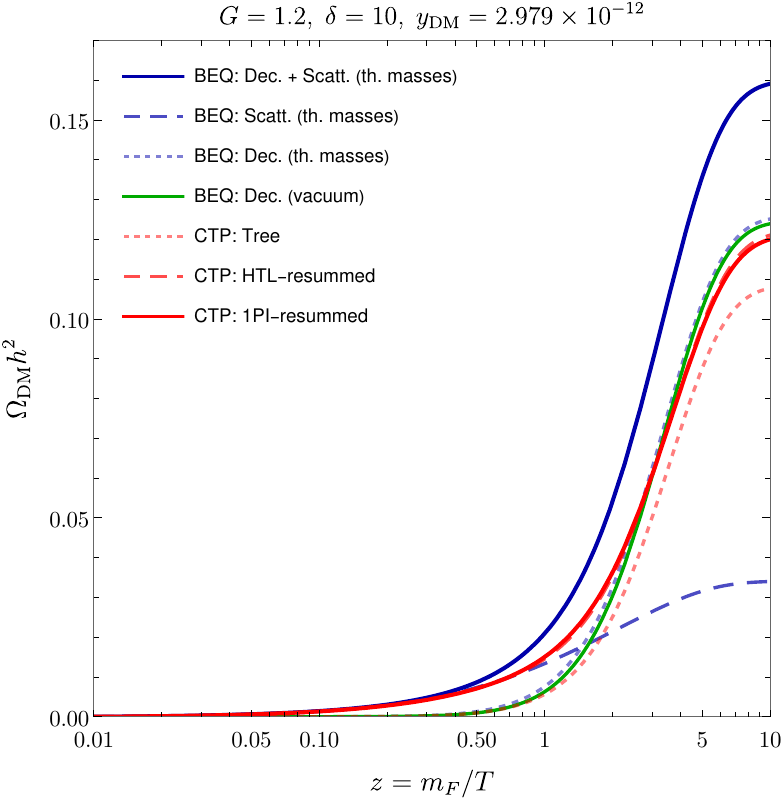}
    \end{subfigure}
    %
    %
    \begin{subfigure}{.49\textwidth}
        \centering
        \includegraphics[width=.99\textwidth]{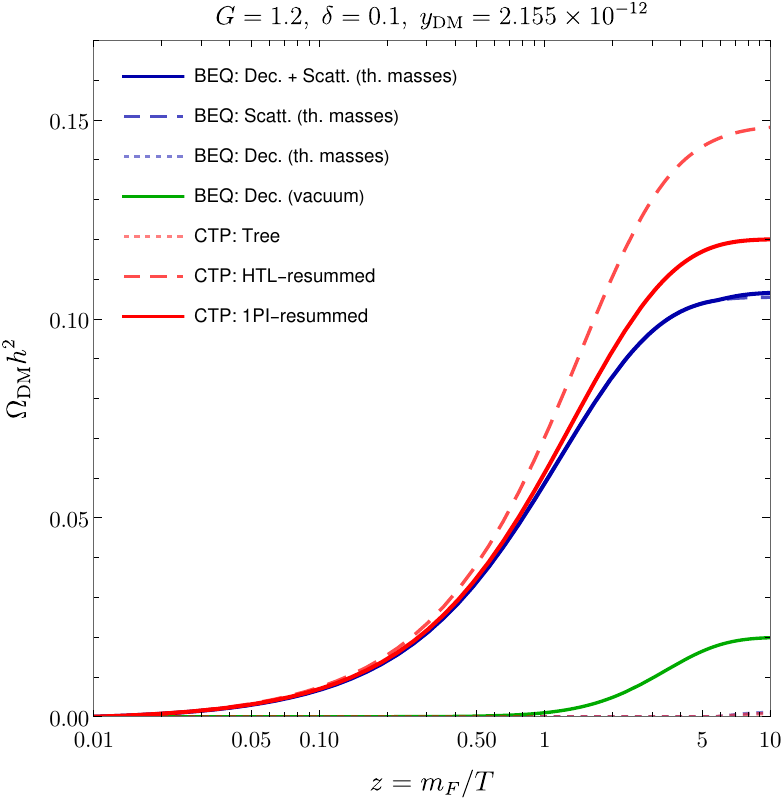}
    \end{subfigure}\\
    \vspace{3mm}
    \begin{subfigure}{.49\textwidth}
        \centering
        \includegraphics[width=.99\textwidth]{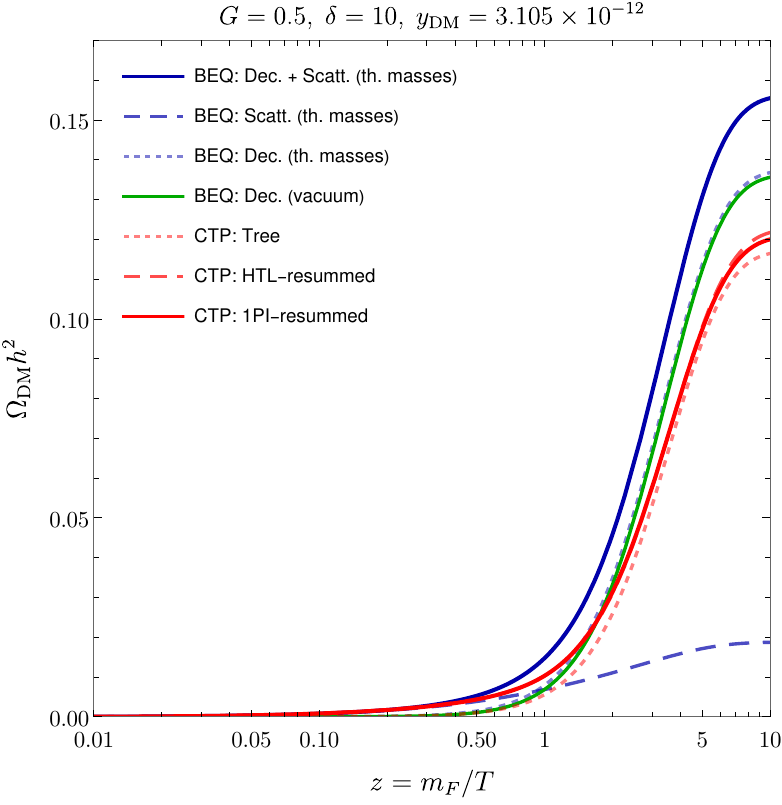}
    \end{subfigure}
    %
    %
    \begin{subfigure}{.49\textwidth}
        \centering
        \includegraphics[width=.99\textwidth]{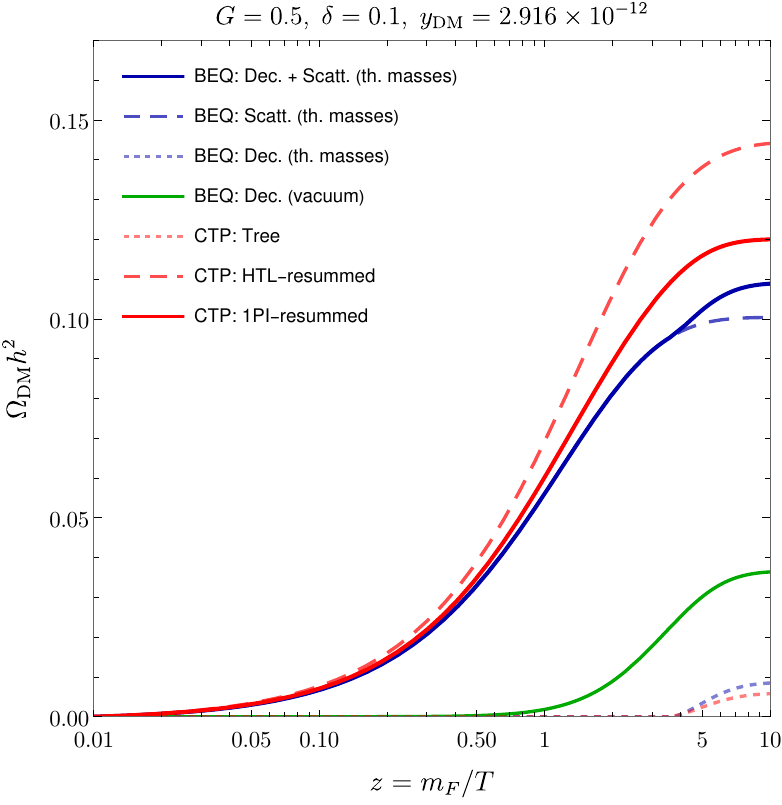}
    \end{subfigure}
    \caption{Evolution of the DM relic abundance. 
    The color scheme of the plotted lines and the values of $G$ and $\delta$ chosen in each panel are the same as in Fig.~\ref{fig:InteractionRateComp}.
    }
    \label{fig:OmegaDM_comparison}
\end{figure}
\begin{figure}[!t]
    \centering
    \captionsetup[subfigure]{labelformat=empty}
    \begin{subfigure}{.49\textwidth}
        \centering
        \includegraphics[width=.99\textwidth]{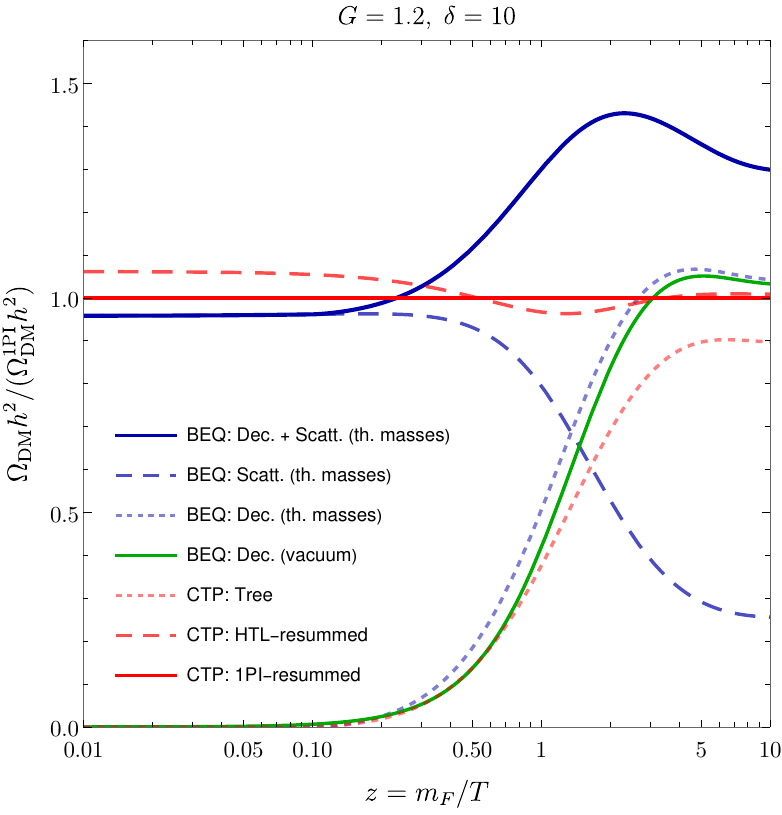}
    \end{subfigure}
    %
    %
    \begin{subfigure}{.49\textwidth}
        \centering
        \includegraphics[width=.99\textwidth]{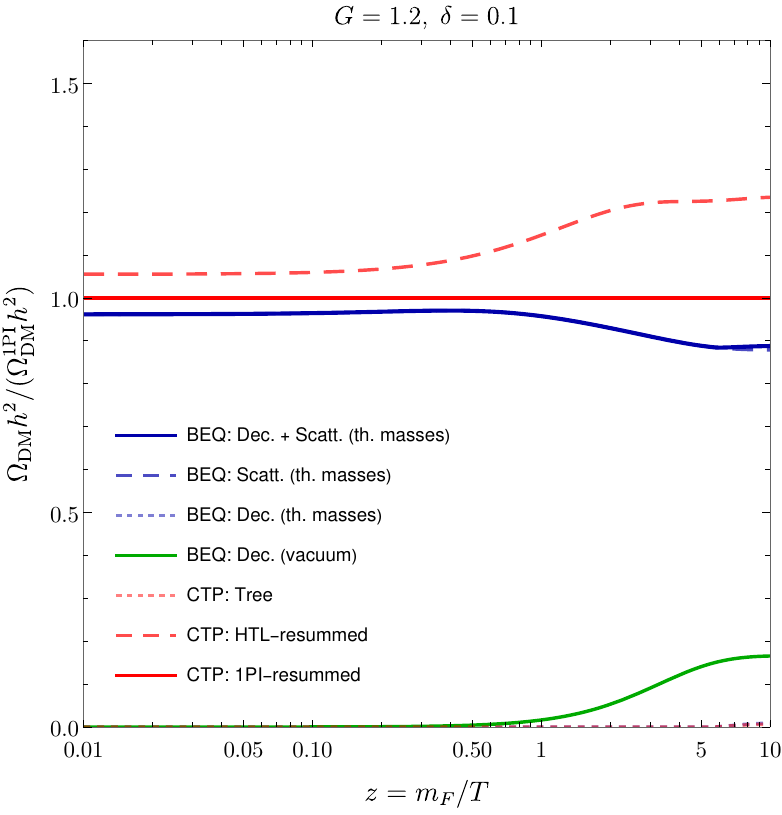}
    \end{subfigure}\\
    \vspace{3mm}
    \begin{subfigure}{.49\textwidth}
        \centering
        \includegraphics[width=.99\textwidth]{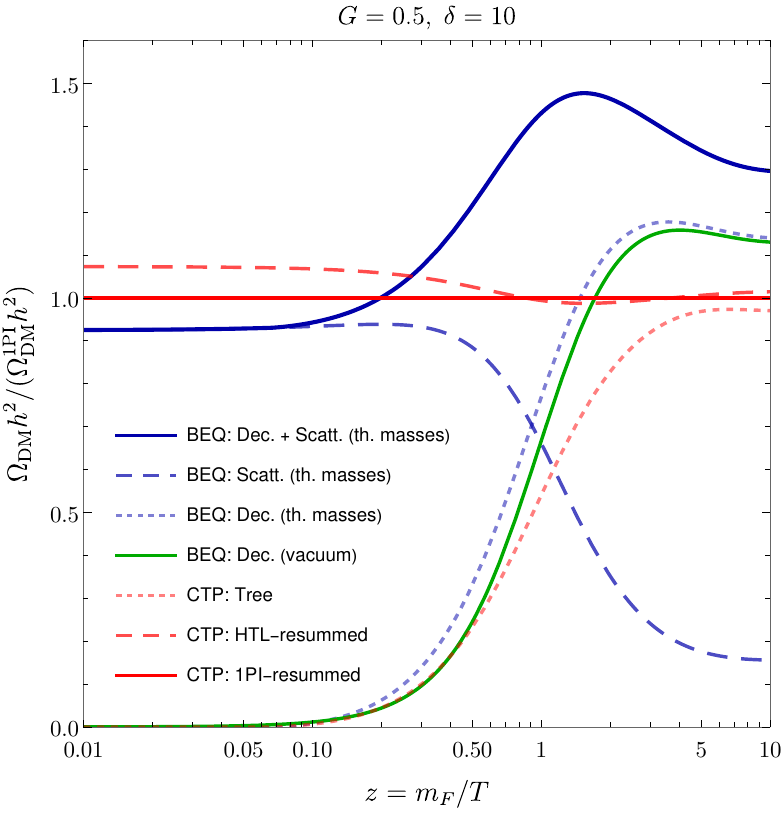}
    \end{subfigure}
    %
    %
    \begin{subfigure}{.49\textwidth}
        \centering
        \includegraphics[width=.99\textwidth]{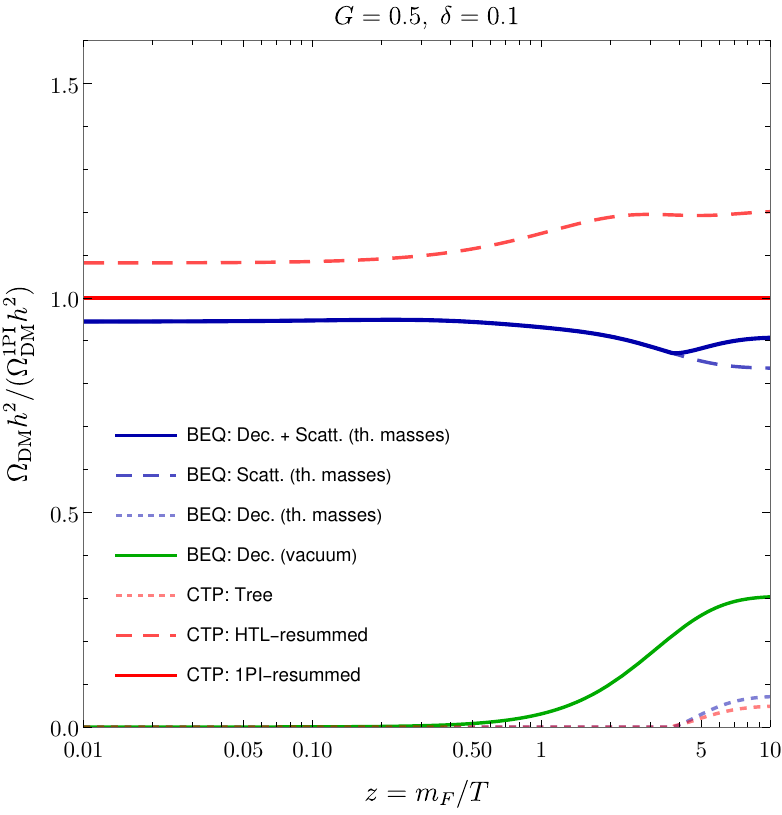}
    \end{subfigure}
    \caption{Evolution of the DM relic abundance normalized to the values obtained with 1PI-resummed propagators. The color scheme of the plotted lines and the values of $G$ and $\delta$ chosen in each panel are the same as in Fig.~\ref{fig:InteractionRateComp}.
    }
    \label{fig:OmegaDM_ratios}
\end{figure}

We now compare the different predictions to the DM relic density from the methods discussed in the previous section.
In Fig.~\ref{fig:OmegaDM_comparison}, we plot the time evolution of the relic density in Eq.~\eqref{eq:OmegaDM_final} based on the different evaluated interaction rate densities discussed above and depicted in Fig.~\ref{fig:InteractionRateComp}.
The results obtained in the CTP formalism from Eq.~\eqref{eq:numbdensEQ}, by employing 1PI-resummed propagators (cf. Sec.~\ref{sec:1PI}) are displayed with red solid lines and their HTL-resummed version (cf. Sec.~\ref{sec:HTL}) in red-dashed, while the solution using tree-level propagators with thermal masses is shown in red short-dashed lines. 
In addition, green solid lines represent the relic density evolution obtained in the Boltzmann equation approach using decays with vacuum masses only, see Eq.~\eqref{eq:DecayRate}. 
Moreover, we show the relic density including decays (Eq.~\eqref{eq:DecayRateTh}, small-dashed blue) and scattering processes (Eq.~\eqref{eq:Rate_scatdec}, dashed blue) each with in-vacuum and thermal masses. 
The combination of both is shown in solid blue.
This color scheme is in complete analogy to Fig.~\ref{fig:InteractionRateComp}.
We show the results for the four benchmark points $G=\left \lbrace 1.2, 0.5 \right \rbrace$ and $\delta=\left \lbrace 0.1, 10 \right \rbrace$.
Furthermore, Fig.~\ref{fig:OmegaDM_ratios} illustrates the evoution of the DM relic density normalized to the 1PI-resummed result for the same data points in the same color code.  

%

\begin{figure}[!t]
    \centering
    \includegraphics[width=0.6\textwidth]{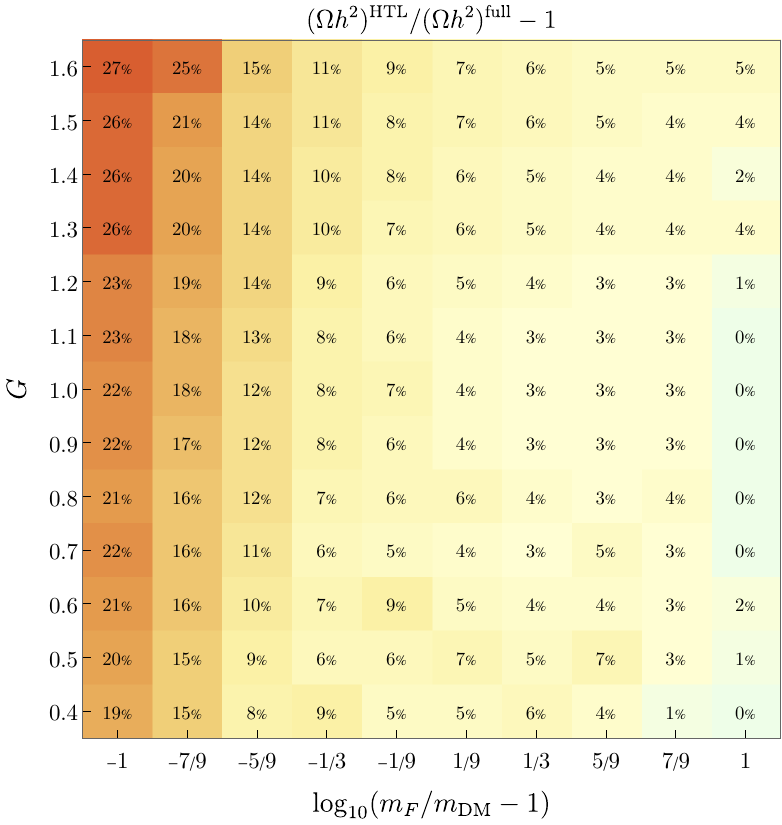}
    \caption{Relative correction on the relic abundance calculated with taking into account one-loop fully-resummed propagators with respect to the one calculated with HTL-resummed propagators in the real-time formalism. The percentage deviation is also shown in warm colors when positive, and with cold colors when negative.}
    \label{fig:OmegaDM_FullvsHTL}
\end{figure}

Finally, in the $(G,\delta)$-plane of Figs.~\ref{fig:OmegaDM_FullvsVac}-\ref{fig:OmegaDM_FullvsHTL} we show the relative deviation of the relic density obtained with the methods outlined in the beginning of this section with respect to the one obtained from 1PI-resummed propagators.
We also indicate with a color scheme if such methods lead to a larger (yellow to red colors) or smaller (azure to blue colors) DM abundance.
We state and briefly discuss the accuracy of each method in the following:
\begin{figure}[!t]
    \centering
    \includegraphics[width=0.6\textwidth]{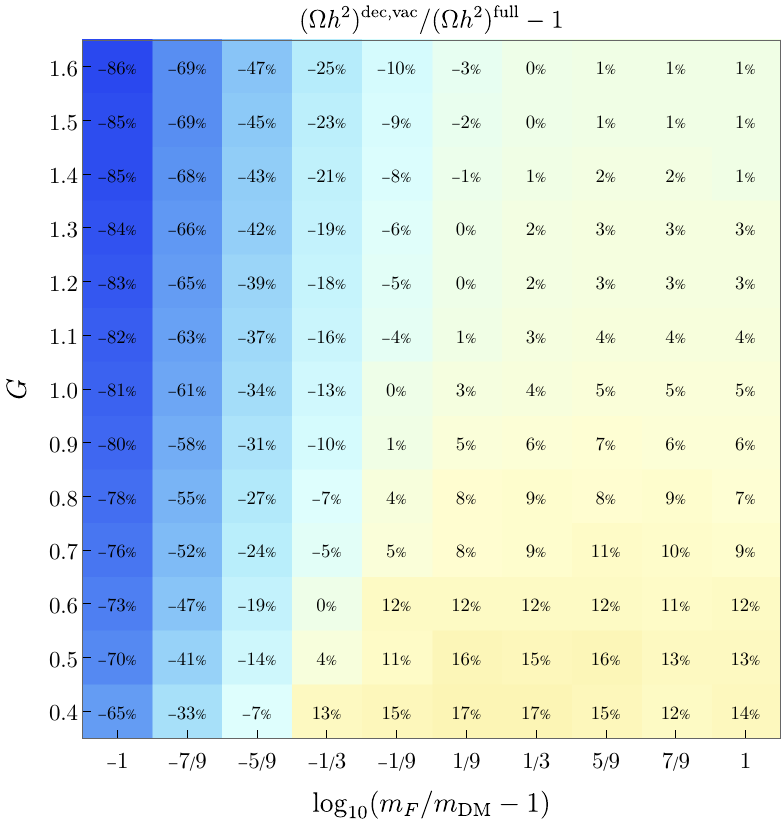}
    \caption{Relative deviation of the relic abundance calculated with the Boltzmann approach including DM production from decays in vacuum with respect to the result obtained with the CTP formalism with  one-loop fully-resummed propagators. The percentage deviation is also shown in warm colors when positive, and with cold colors when negative.}
    \label{fig:OmegaDM_FullvsVac}
\end{figure}
\\
\newline
\textbf{Boltzmann approach with decays (vacuum masses):} 
This approach does not include any DM production from scatterings, while it overestimates DM production from decays mainly due to the neglected quantum statistics.
As a consequence, as shown in Fig.~\ref{fig:OmegaDM_FullvsVac}, the rate tends to moderately overestimate the relic density for large $\delta$ (up to $14\%$), while it strongly underestimates it for small $\delta$ (up to $-86\%$), where decay contributions are suppressed. 
We notice that there exists a region in parameter space, where the overestimated decay contribution precisely compensates the neglected scatterings, corresponding to a diagonal across the plane in Fig.~\ref{fig:OmegaDM_FullvsVac}.
This explains why the overproduction (underproduction) of DM for large (small) $\delta$ becomes softer (stronger) for higher values of $G$. \\ \newline
\textbf{Boltzmann approach with decays (thermal masses):}
As previously elaborated, decays with thermal masses tend to overestimate DM production in the non-relativistic regime, while no DM particle is produced at high temperatures due to kinematics\footnote{This is different to the case of vacuum masses where decays of the parent particle are always kinematically allowed.}. 
As a consequence, as shown in Fig.~\ref{fig:OmegaDM_FullvsDecThM}, the rate also tends to moderately overestimate the relic density for large $\delta$ (up to $15\%$), while it even more strongly underestimates it for small $\delta$ (up to $-100\%$), where decay contributions are suppressed.  
This effect is exaggerated when including relativistic quantum statistics, where Pauli blocking further reduces the production rate, an effect captured by employing tree-level propagators in the CTP formalism (cf. Fig.~\ref{fig:OmegaDM_FullvsTree}).
Thus, remarkably, including only decays with thermal masses yields the largest deviation from the 1PI-resummed result.
Overall, neglecting DM production from scatterings leads to the largest deviations for the highest values of $G$ as the scattering contributions are directly proportional to powers of the gauge couplings (cf. Eq.~\eqref{eq:UVFit}).
As for vacuum decays, there exists a region in parameter space, where the overestimated decay contribution precisely compensates the neglected scatterings, which is evident in Fig.~\ref{fig:OmegaDM_FullvsDecThM}. \\ \newline
\textbf{Boltzmann approach considering decays and scatterings with thermal masses:} As the next step, we are combining the results from decays with thermal masses and scatterings in which contributions with $t$-channel propagators are regulated by thermal masses. 
The results are summarized in Fig.~\ref{fig:OmegaDM_FullvsThM}.
Again, DM is overproduced at large $\delta$ because of the overestimated production rate from decays, resulting in up to $\sim +35\%$ values of $\Omega_\text{DM}h^2$.
Additionally, for $\delta\lesssim 25\%$, deviations of the relic density are found to be negative up to $\sim -10\%$, because scatterings, dominating in this parameter region, yield a lower rate compared to the 1PI-resummed calculation. 
As in the previous scenario, the two effects can cancel each other for moderate mass splittings and gauge couplings, resulting in only $\mc{O}(1\%)$ deviations.
Furthermore, since the $G$ dependence of the scattering contribution is now taken into account, the relative deviation is only very mildly dependent on $G$ and almost entirely controlled by the mass-splitting. 
The accuracy of this method can be further enhanced if the correct quantum statistics of the bath particles are considered. 
While we have not performed this calculation for the scattering contribution, the effect for decays is exactly captured by the solution using tree-level propagators with thermal masses. 
Such a treatment reduces the deviation for large mass splittings to $\sim +15\%$. \\ \newline
\textbf{HTL propagators:}
As explained in the previous subsection, the HTL-approximated results provide a relatively accurate description of the decay contribution, whereas they tend to overestimate DM production from scatterings as summarized in Fig.~\ref{fig:OmegaDM_FullvsHTL}.
The former is caused by the zero-width of the timelike HTL-propagator, while the latter originates from the omission of the vacuum mass $m_F$. 
Consequently, using HTL propagators deviates from the results with 1PI-resummed up to only a few percent at large mass-splittings where decays dominate, while they lead to a larger DM abundance (up to $+27\%$) for smaller $\delta$.
Additionally, the deviation increases with $G$, 
an effect that was to be expected since the HTL expansion is effectively an expansion in $G/\pi$.\\ \newline
\begin{figure}[!t]
    \centering
    \includegraphics[width=0.6\textwidth]{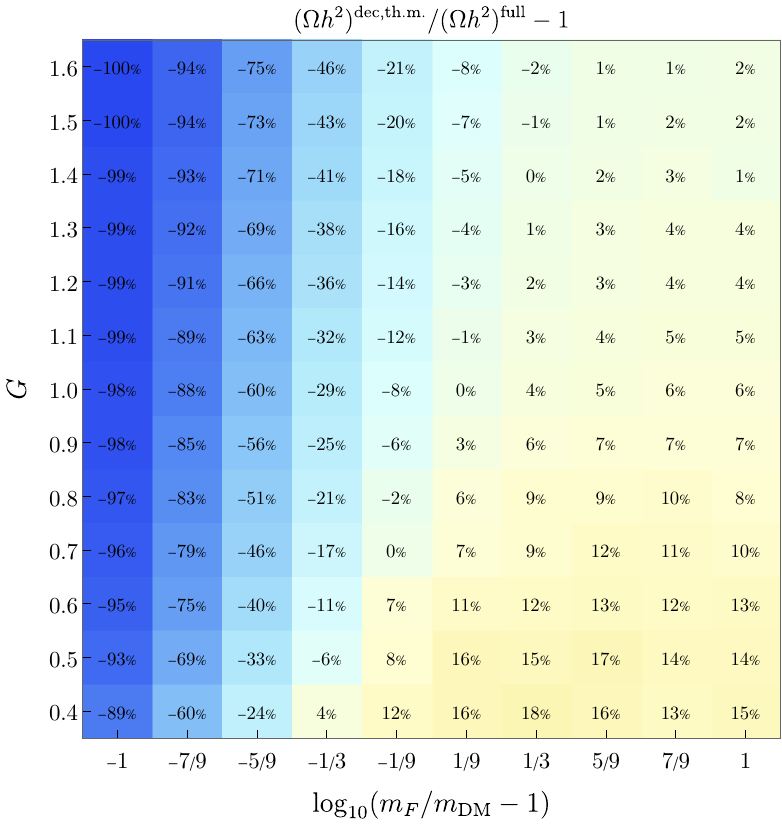}
    \caption{
    Same caption as Fig.~\ref{fig:OmegaDM_FullvsVac} but considering vacuum and thermal masses instead of only vacuum masses.}
    \label{fig:OmegaDM_FullvsDecThM}
\end{figure}
\begin{figure}[!t]
    \centering
    \includegraphics[width=0.6\textwidth]{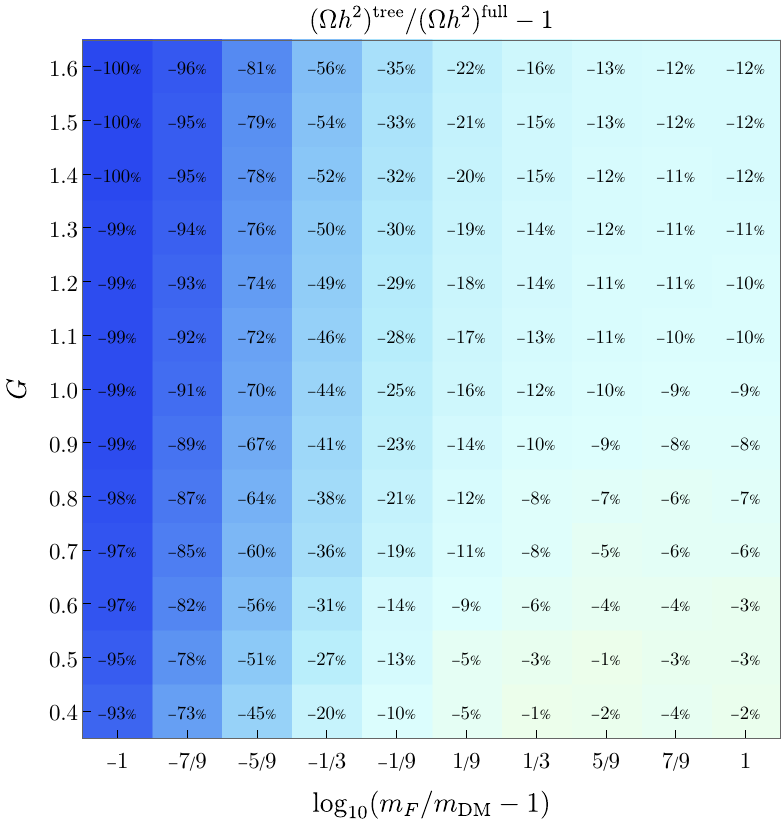}
    \caption{ Same caption as Fig.~\ref{fig:OmegaDM_FullvsVac} but considering vacuum and thermal masses as well as the correct quantum statistics instead of vacuum masses only and Boltzmann statistics.}
    \label{fig:OmegaDM_FullvsTree}
\end{figure}
\begin{figure}[!t]
    \centering
    \includegraphics[width=0.6\textwidth]{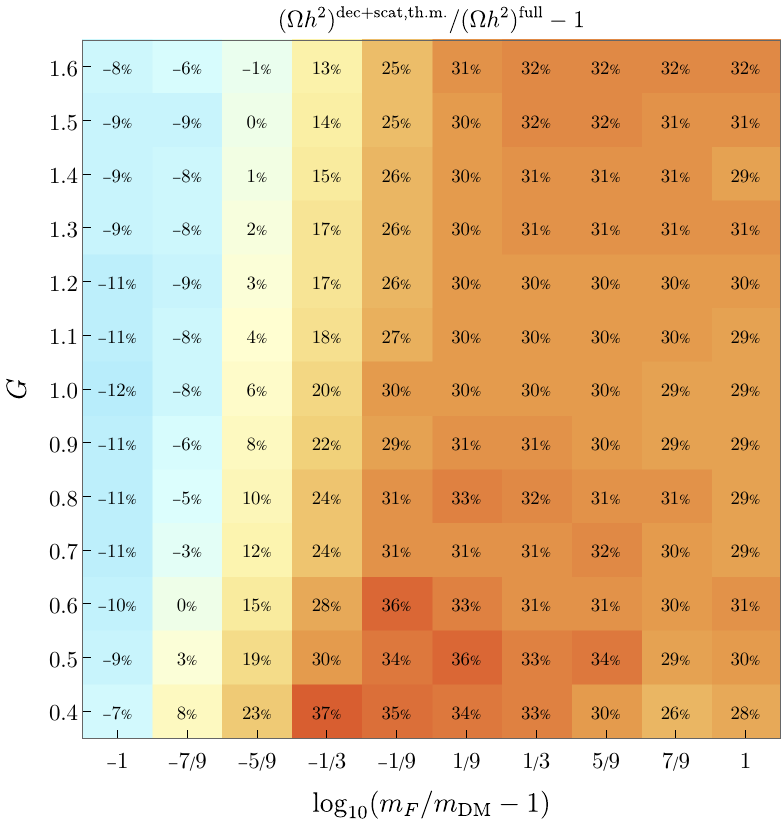}
    \caption{Same caption as in Fig.~\ref{fig:OmegaDM_FullvsVac}, but including thermal masses for decays as well as thermal mass regulated scatterings.}
    \label{fig:OmegaDM_FullvsThM}
\end{figure}

We want to stress that our findings apply to any model of scalar DM where the DM production vertex involves fermions that dominantly interact via gauge interactions. 
For models featuring top quarks, the Yukawa-induced corrections to the fermionic self-energies lead to a differene between the fermionic dispersion relations, with top quarks being more energetic due to the larger thermal masses.
Including these corrections would change the time window where DM production from decays efficiently contributes to the production rate and therefore would have significant implications for the comparison between the various methods here discussed. 

\begin{figure}[!t]
    \centering
    \includegraphics[width=0.6\textwidth]{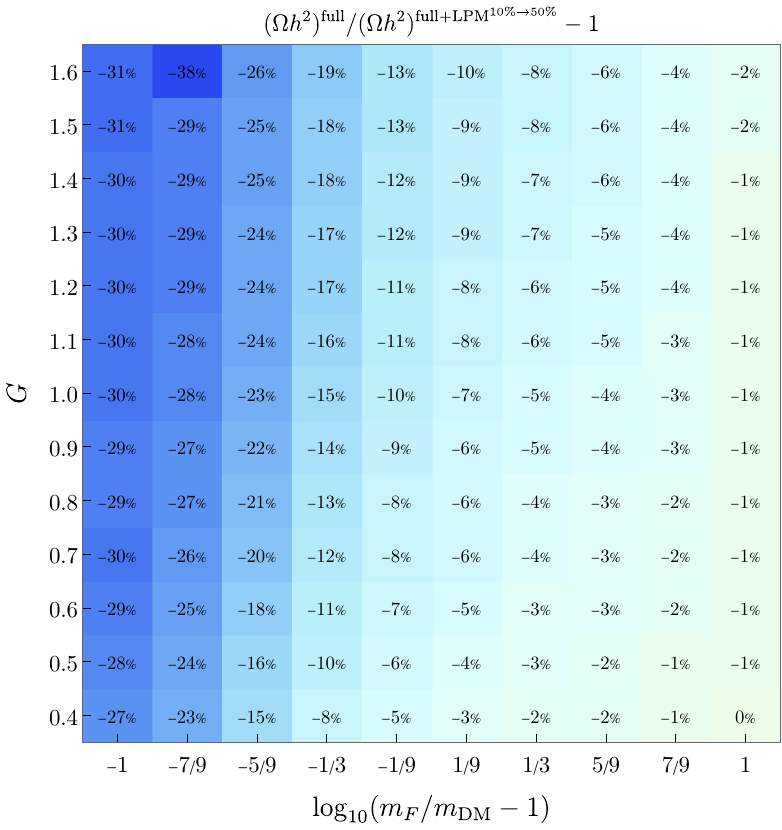}
    \caption{Relative correction to the relic density calculated with 1PI-resummed propagators when further considering a $50\%$ ($10\%$) enhancement of the scattering rate producing DM for $\delta=0.1$ ($\delta=10$), the estimated contribution of the LPM effect from Ref.~\cite{Biondini:2020ric}. For mass splittings in between we linearly interpolated the enhancement in $\delta$. The percentage deviation is shown in colder colors when more negative.}
    \label{fig:FullvsLPM}
\end{figure}

Finally, we would like to quantify the size of contributions to the relic density that we have omitted in this work.  
As discussed previously, we truncate the expansion of the 2PI effective action at leading order. 
This truncation results in the omission of two potentially relevant contributions. 
Firstly, interference terms between $t$-channel and $s$-channel scatterings only arise at NLO in the expansion of the 2PI effective action and are thus not included.
On the level of the Boltzmann equation, we find them to be sub-leading (around $\mc{O}(10\%)$) compared to squared $s$- and $t$-channel contributions and we expect similar contributions with our method. 
Secondly, we do not include contributions to DM production arising from the LPM effect. 
In Ref.~\cite{Biondini:2020ric}, the authors analyze a similar scenario to the one in this article, but with a scalar parent particle and fermionic dark matter candidate, and show that the LPM effect leads to an enhancement of the contribution from scatterings to the dark matter relic abundance by approximately $\mc{O}(50\%)$ for small mass splittings and around $\mc{O}(10\%)$ for large mass splittings.
These results, however, are obtained in the ultrarelativstic limit and interpolated to the non-relativistic regime.
The treatment of the LPM effect involving two in-vacuum mass scales, as in our scenario, is not yet discussed in the literature up to today and and we intend to address it in a follow-up work. 
To indicate the corrections that we expect, we estimate the maximal effect on our results in the following way: 
based on Ref.~\cite{Biondini:2020ric}, we assume that the LPM effect enhances the DM production rate from scatterings by $50\%$ for $\delta=0.1$ and by $10\%$ for $\delta=10$ and linearly interpolate the enhancement in $\delta$ in between\footnote{Practically speaking, we replace the coefficient $B \rightarrow (1.504-0.04 \delta) B$ in Eq.~\eqref{eq:FitFunction}.}. We then compare the resulting DM relic density to the one without this correction.
The results of such a comparison are presented in the $(G,\delta)$-plane in Fig.~\ref{fig:FullvsLPM}. 
We find that, when truncating the effective action at LO, the relic density can be underestimated by up to $\sim30\%$ for the smallest mass splittings, when scatterings dominate the DM interaction rate. 
For large mass splittings, the relic density is only mildly affected and differs by a few percent. 

Another source of theoretical uncertainty is related to our choice of the gauge fixing parameter.
In fact, truncating the loop-expansion of the 2PI-effective action, as done in this work, prevents the gauge-dependent parts of diagrams of higher-loop topologies to cancel each other, leading to gauge-dependent observables \cite{Arrizabalaga:2002hn,Carrington:2003ut,Mottola:2003vx,Reinosa:2009tc,Garbrecht:2015cla}.
In practice, however, a truncation and a gauge choice have to be performed, which in our case correspond to considering the LO (two-loop) expansion of the 2PI effective action and the Feynman gauge.
Thus, we expect any omitted gauge-dependent contribution to the 1PI-resummed propagators and to the DM relic abundance to fall within the same uncertainty band of the higher-order corrections that were not accounted for in our calculation.

\section{\label{sec:conclusions}Conclusions and Recommendations}
In this article, we have analyzed the impact of finite-temperature effects on the freeze-in production of DM.
We have considered a phenomenologically motivated model featuring a real scalar SM-singlet as our DM candidate, only feebly interacting with a SM fermion $f$ via a vectorlike BSM fermion $F$ with a coupling strength $\ydm$. 
In total, our model contains four parameters: the feeble coupling $\ydm$, the mass of the parent particle $m_F$, the mass splitting between DM and the parent particle $\delta = (m_F-\mdm)/m_F$ and the effective gauge coupling $G$ of the $f$ and $F$ particles.
In contrast to freeze-out, the dynamics of freeze-in are sensitive to the entire thermal history before the DM abundance stops growing at relatively large temperatures (around $T \sim m_F$) and are thus subject to non-negligible corrections from the thermal plasma environment.

At leading-order in vacuum, DM production proceeds via decays of the parent particle, $F \rightarrow f + s$. 
At next-to-leading order, scatterings due to gauge interactions of the fermions $F/f$ have to be included, although they can involve $t$-channel soft-collinear singularities.
A common approach to regularize these divergences is to introduce thermal masses $m_\text{th} \sim T$ for the exchanged virtual particles; however, this remains an ``ad-hoc'' prescription  and a more consistent quantum-field-theoretical treatment  is needed. 

For this reason, we have derived the freeze-in evolution equations from first principles of non-equilibrium quantum field theory by employing the closed-time-path (CTP) formalism (cf. Sec.~\ref{sec:CTP}), which yields a DM production rate directly related to the imaginary part of the DM self-energy, which can be expressed in terms of loop diagrams with dressed, exact propagators, cf. Fig.~\ref{fig:DMselfenergy}.
We approximate the exact fermionic propagators by resumming one-loop 1PI diagrams constructed with ordinary propagators (cf. Eqs.~\eqref{eq:SFresummed}, \eqref{eq:Sfresummed}). This approximation is able to capture the leading-order contributions to the DM interaction rate in both the large temperature limit, where they are proportional to $\sim G \log G$ and $\sim G$, and at zero temperature, they scale as $\sim G^0$.
The 1PI-resummed propagators exhibit an intricate momentum dependence,  particularly when accounting for the two different masses of the DM and the mediator. 
Therefore the DM production rate (cf. Eq.~\eqref{eq:RateDensityResummed}) was evaluated numerically with Monte Carlo techniques.

The resummed propagators can be simplified in appropriate limits. At high temperatures and for small couplings one can trade the full resummation for the Hard Thermal Loop (HTL) resummation, as we outlined in Sec.~\ref{sec:HTL}. Further simplifications of the HTL propagators allow us to bring them to a tree-level form and recover results that correspond to tree-level Boltzmann collision operators with full quantum relativistic statistics. These results themselves can be compared with the more traditional and widely-used Boltzmann approach of freeze-in (cf. Sec.~\ref{sec:Boltzmann}), in which the DM production rate is computed with S-Matrix elements without quantum statistics.

The HTL approximation and the tree-level-like propagators derived from it are expected to break down at the temperature regimes most relevant for freeze-in, when $T$ approaches the largest vacuum mass scale of the considered model (the mediator mass $m_F$ in our case). To assess in which temperature regime and how much the various analyzed methods disagree, we have computed 1PI-resummed rates for several values of $G$, $\delta$, and of the time variable $z=m_F/T$ (cf. Sec.~\ref{sec:results}). We have estimated the corresponding DM abundance, and compared the results with those obtained with the following variations of the aforementioned approximations: 1.~Boltzmann approach with DM production from decays and using vacuum masses. 2.~Same as before but including thermal masses. 3.~Boltzmann approach including decays plus scatterings and with thermally corrected masses. 4.~CTP calculation with tree-level-like simplified HTL propagators. 5.~CTP calculation with HTL-resummed propagators.

An important point concerning our most precise calculation on the CTP with 1PI-resummed propagators is that the truncation of the effective action at LO implies that our results do not include DM production stemming from multiple soft scatterings with the thermal bath, a phenomenon known as the LPM effect. 
While the inclusion of those effects is left for future work, based on results from Ref.~\cite{Biondini:2020ric}, we have estimated the theoretical uncertainty in our calculations stemming from the LPM effect, as summarized in Fig.~\ref{fig:FullvsLPM}.
We find that the LPM effect could increase the relic density up to $30 \%$ for a mass-degenerate dark sector ($\delta \rightarrow 0$), while we expect our results only to be affected on the percent level for large mass-splittings. These error estimates, however, stem from an assessment of the LPM effects that does not include  the effect of the DM and mediator mass scales, which could however play a relevant role.

Coming back to the comparison of different methods,  a summary of our findings follows:
\begin{itemize}
    \item The approach which relies on the semi-classical Boltzmann equation considering only decays with vacuum masses underestimates the relic density up to one order of magnitude for almost mass degenerate dark sectors, while for large mass splittings, the abundance is slightly overestimated by up to $+14\%$. The deviations are summarized in Fig.~\ref{fig:OmegaDM_FullvsVac} and are also sensitive to the size of the effective gauge coupling $G$.
    \item Remarkably, the approach that relies on using the Boltzmann equation considering only decays with thermal masses does not improve the accuracy of the method. It can completely miss DM production for small mass splittings (up to $-100\%$), while it also moderately overestimates the relic abundance for large mass splittings (up to $+15\%$), as summarized in Fig.~\ref{fig:OmegaDM_FullvsDecThM}. Considering the correct quantum statistics instead of Boltzmann statistics in this scenario further reduces the relic abundance such that it is underestimated for every choice of parameters, as illustrated in Fig.~\ref{fig:OmegaDM_FullvsTree}.   
    \item Considering the Boltzmann equation with decays and scatterings where both include thermal masses improves the agreement of the Boltzmann approach with our calculation, especially for small mass splittings. Here, the maximal deviations shrink down to $-10\%$. For large mass splittings, on the other hand, the deviations increase up to $35\%$. The results are summarized in Fig.~\ref{fig:OmegaDM_FullvsThM}. Note, however, that the mismatch for large mass splittings can be reduced to about $\sim 15\%$ when considering the correct quantum statistics for decay processes. Although we have not performed the calculation with the appropriate quantum statistics for scatterings in the Boltzmann approach, we expect a similar effect for small mass splittings. In general, the deviations of this approach are almost independent of the effective gauge coupling $G$. 
    \item The approach relying on the CTP formalism but with HTL approximated propagators provides percent-level accurate results for large mass splittings. However, it fails to accurately describe small mass splittings where it overestimates the rate up to $+27\%$. We find that the deviations of this approach increase with the effective gauge coupling $G$. The results are shown in Fig.~\ref{fig:OmegaDM_FullvsHTL}. Note that for a final verdict on the relic abundance for small mass splittings, contributions from the LPM effect should be added.  
\end{itemize}
Our findings apply to any model of scalar DM where the particles involved in the DM production vertex are dominantly interacting via gauge interactions.
With the above main results in mind, we would like to formulate the following recommendations:
\begin{enumerate}
    \item The choice of method to approximate the production of freeze-in DM should depend mainly on the mass splitting $\delta$ present in the dark sector.
    \item When only considering decays, the inclusion of thermal masses yields a less accurate result, especially for small mass splittings.
    \item For large mass splittings, Boltzmann equations only considering decays with vacuum masses and Boltzmann statistics are sufficient to arrive at $\mc{O}(15\%)$ accurate results. If a percent-level accurate result is required, a method based on a proper thermal field theory treatment has to be chosen. However, HTL approximated propagators are sufficient to arrive at a percent-level accuracy.
    \item For small mass splittings, considering thermal mass regulated scatterings on top of decays including thermal masses in the context of Boltzmann equations provides the best approximation to our results and is $\mc{O}(10\%)$ accurate. This statement might be altered as soon as DM production from the LPM effect is added to our results, which is expected to provide up to $30\%$ corrections to this regime.
\end{enumerate}
In summary, we can conclude that our calculation with full 1PI-resummed propagators is more accurate than all the alternative methods discussed, with the possible exception of the calculation with HTL-resummed propagators in the small $\delta$ regime.
We believe that this work provides a crucial step towards a consistent treatment of freeze-in dark matter production including thermal effects. 
We employed, for the first time in this context, the real-time formalism of thermal quantum-field theory, going beyond previous works by including the full dependence on the relevant mass scales as opposed to using (non)relativistic approximations. 
In particular, we used 1PI-resummed propagators without relying on the Hard-Thermal-Loop approximation and compared them to other approaches frequently used in the literature.
We hope that we can provide guidance to phenomenologists who wish to understand to what extent their models receive corrections from the non-trivial behavior at finite temperatures. 
Apart from being applicable to different simplified models of FIMP DM, our results are of generic relevance to scenarios involving feebly interacting particles, e.g. in the context of leptogenesis or the production of gravitational waves \cite{Ghiglieri:2015nfa,Ghiglieri:2020mhm,Ringwald:2020ist,Ringwald:2022xif,Drewes:2023oxg}.

\acknowledgments
E.~C. and M.~B. thank Martin A. Mojahed for useful discussions and Mar\'{\i}a Jos\'e Fern\'andez Lozano for partially proofreading the draft. M.~B., E.~C. and J.~H. acknowledge support from the Emmy Noether grant "Baryogenesis, Dark Matter and Neutrinos: Comprehensive analyses and accurate methods in particle cosmology" (HA 8555/1-1, Project No. 400234416) funded by the Deutsche Forschungsgemeinschaft (DFG, German Research Foundation). All authors acknowledge support by the Cluster of Excellence “Precision Physics, Fundamental Interactions, and
Structure of Matter” (PRISMA$^+$ EXC 2118/1) funded by the Deutsche Forschungsgemeinschaft (DFG, German Research
Foundation) within the German Excellence Strategy (Project No. 390831469).
The authors gratefully acknowledge the computing time granted on the supercomputer MOGON NHR at Johannes Gutenberg University Mainz (\href{hpc.uni-mainz.de}{hpc.uni-mainz.de}).

\appendix
\setcounter{equation}{0}
\section{\label{sec:AppA}The Closed-Time-Path formalism}
In this Appendix, we further expand and elaborate on the CTP formalism employed in Sec.~\ref{sec:CTP}.
First, we want to have a more rigorous definition of the path-integral definition of expectation values for operators.
Consider, for example, a time-dependent state $|S(t)\rangle$ evolving in time under the Hamiltonian $\hat H$, $|S(t)\rangle=e^{i\hat H(t-t_0)}|S(t_0)\rangle$. 
The average of an observable $\hat{\mc{O}}(t)$ can be written as
\begin{align}\label{eq:exp}
    \langle S(t)|O(t)|S(t) \rangle = \int {\cal D}\phi(t)\langle S(t_0)| e^{-i \hat H(t-t_0)}\hat{\mc{O}}(t) | \phi\rangle_t\, {}_t\langle \phi| e^{i \hat H(t-t_0)} |S(t_0) \rangle.
\end{align}
Above, we have introduced the spectral resolution of the identity in terms of eigenstates $|\phi\rangle_t$ of the field operators in the Heisenberg picture at time $t$. 
Each of the two factors inside the integral in Eq.~\eqref{eq:exp} is a transition amplitude which can be expressed as a path integral, with a time contour $\mc{C}_+$ going from $t_0$ to $t$ in the second factor, and a contour $\mc{C}_-$ from $t_0$ to $t$ in the first. 
The two path integrals can be joined into a single path integral along a combined time-contour  $\mc{C}=\mc{C}_+\oplus\mc{C}_-$, which goes from $t_0$ to itself passing through $t$, as we illustrated in Fig.~\ref{fig:CTP_contour}. 
This closed time-path justifies the name of the formalism. 
The choice of the initial state/statistical ensemble can be encoded in the path integral employing appropriate boundary conditions and choices of couplings in the Lagrangian density\footnote{
For example, if we consider an ensemble in thermal equilibrium with temperature $T=1/\beta$, this can be realized by appending a third branch to the time contour, going in the imaginary direction from $t=t_0$ to $t=-i\beta$ \cite{le_bellac_1996}.}.

From the path integral over the time-path $\mc{C}$, and introducing  sources $J,\bar J $  for the fields and their adjoints, one can construct a generating functional $Z_\mc{C}[J,\bar J]$.
A crucial simplification that can be made and that we employ in our paper is to stretch the time path to $t_0\rightarrow -\infty$ and $t\rightarrow+\infty$.
This is possible since we are mainly interested in the interactions of fields within a plasma in thermal equilibrium, where the information about initial conditions is lost.
In our scenario, all fields but the FIMP one are considered to be in thermal equilibrium and thus the only potentially problematic case for this simplification is the FIMP field itself.
However, as it was shown in \cite{Garbrecht:2011xw} and as we will recall in Sec.~\ref{sec:WignerGradient}, by performing a correct gradient expansion of the Schwinger-Dyson equations in Wigner space, one can still formally operate as if thermal equilibrium holds.
\subsection{\label{sec:CTPprop}CTP propagators}

One can define Green's functions in the usual manner by taking functional derivatives of the generating functional $Z_{\mc{C}}[J]$ with respect to the sources. 
In particular, one can define propagators (two-point functions), which are of special interest because they contain information about the number densities of states in the statistical ensemble, as well as the allowed momentum shells of the states. 
The propagators are defined as follows,
\begin{align}\label{eq:propagator_def}
    i G^{ab}(x,y)=-\dfrac{\delta^2}{\delta J(x^a)\delta \bar J(y^b)}\log Z[J, \bar J]\big|_{J_{\pm}=0}=\big\langle T_{\mc{C}}\phi(x^a)\bar\phi(y^b)\big\rangle.
\end{align}
Above, the contour $\mc{C}$ has been parameterized with coordinates $x^a$, where $a,b=\pm$ denotes the corresponding time branch. 
$T_\mc{C}$ indicates the time path ordering, and the angular parentheses $\big\langle\dots\big\rangle$ correspond to taking the expectation values of the quantum operators $\hat{\mc{O}}$ in the statistical ensemble, namely tracing the operators over all degrees of freedom weighted by the density matrix,
$\big\langle\hat{\mathcal{O}}\big\rangle=\Tr\hat{\rho}\,\hat{\mc{O}}$. 
The notation $\phi(x), \bar\phi(x)$ in Eq.~\eqref{eq:propagator_def} is meant to capture both scalar and fermion fields, and $G$ refers to a generic scalar of fermion two-point function. 
For a complex scalar $\varphi(x)$, one would have $\phi(x)=\varphi(x)$, $\bar\phi(x)=\varphi(x)^\dagger$,
while for a Dirac fermion $\Psi(x)$ with Dirac adjoint $\bar\Psi(x)$ one would have $\phi(x)=\Psi(x)$, $\bar\phi(x)=\bar\Psi(x)^\dagger$. 
In situations in which we want to distinguish between scalar and fermionic two-point functions, we will use $\Delta$ for scalar and $S$ for fermion propagators:
\begin{align}\label{eq:DeltaS}
i\Delta^{ab}(x,y)= &\,   \big\langle T_{\mc{C}}\,\varphi(x^a)\varphi(y^b)^\dagger\big\rangle, & iS^{ab}(x,y)= &\,   \big\langle T_{\mc{C}}\,\Psi(x^a)\bar\Psi(y^b)\big\rangle.
\end{align}
We can see that, for a given field, there are four possible CTP correlators: the time-ordered $G^{++}$ (or $G^T$) type, the anti-time-ordered $G^{--}$ (or $G^{\bar{T}}$) type, and the two so-called Wightman propagators with $G^{+-}$ (or $G^{<}$) and $G^{-+}$ (or $G^{>}$).
These two-point functions are not independent since they can be combined into retarded and advanced propagators as
\begin{align}\label{eq:ret_adv}
    G^\text{R}=G^T-G^{<}=G^{>}-G^{\bar{T}},\quad G^{\text{A}}=G^T-G^{>}=G^{<}-G^{\bar{T}}\,.
\end{align}
We also single out the Hermitian and anti-Hermitian parts of the retarded propagators as follows:
\begin{align}
    G^\mc{H}=\frac{1}{2}\left(G^\text{R}+G^\text{A}\right)=\frac{1}{2}\left(G^{++}-G^{--}\right),\quad G^\mc{A}=\frac{i}{2}\left(G^\text{R}-G^\text{A}\right)=\frac{i}{2}\left(G^{>}-G^{<}\right)\,.
    \label{eq:herm_antiherm}
\end{align}
$G^{\mc{A}}$ is of particular significance as it corresponds to the spectral density of states, a central object that encodes all the information regarding the spectrum of propagating single and multi-particle states.

As a last remark, we notice that, in thermal equilibrium with temperature $T\equiv 1/\beta$, the degeneracy between propagators further reduces to one single independent combination. 
This follows from the fact that, in thermal equilibrium, $\langle \dots\rangle = {\rm Tr}( e^{-\beta \hat H}\dots)$, and the cyclicity properties of the trace imply
\begin{align}
   { G^{\mathrm{eq},>}}(t)=\pm {G^{\mathrm{eq},<}}(t+i\beta).
    \label{eq:KMS}
\end{align}
where $\pm$ refers to bosons/fermions.
These are the so-called Kubo-Martin-Schwinger (KMS) relations and they will turn out to be extremely useful when written in momentum space.
\subsection{\label{sec:2PIEA}2PI effective action}
In order to track the DM density, we need to obtain equations for the time evolution of the DM two-point function, which includes information about the number density of DM particles. 
To obtain such an equation, it is useful to trade the generating functional $Z_\mc{C}$ for a functional that depends only on the one- and two-point functions of the theory, and whose equations of motion determine the evolution of the propagators. 
An appropriate functional is the so-called two-particle irreducible (2PI) effective action $\Gtpi_\mc{C}$, whose equations of motion correspond to the Schwinger-Dyson equations fixing the dynamics of the propagators.
This 2PI action can be obtained by generalizing $Z_\mc{C}[J]$ to include non-local sources $\sigma(x^a,y^b)$ for two-point functions, and performing Legendre transformations of $-i\log{Z_\mc{C}}[J, \sigma]$ with respect to both $J$ and $\sigma$ \cite{Cornwall:1974vz,Chou:1984es,Calzetta:1986cq,Berges:2004pu,Berges:2004yj}. 
The resulting functional $\Gtpi$ depends only on the averaged fields (i.e. the exact one-point functions) and on the exact (full) propagators of the theory.
It can be rewritten in the following form \cite{Cornwall:1974vz,Berges:2004pu,Prokopec:2003pj}
\begin{align}
    \Gtpi_\mc{C}[\Delta,S]=i\tr [\Delta^{0^{-1}}\Delta] - i\tr [S^{0^{-1}}S] +i\tr \ln \Delta^{-1} -i \tr \ln S^{-1} +\Gamma_2[\Delta,S]\,,
    \label{eq:2PIEA}
\end{align}
where $\Delta$ and $S$ are the exact two-point functions (full-propagators) for scalar and fermion fields, respectively, $\Gamma_2$ is the sum of all 2PI vacuum graphs constructed with the exact propagators, and $\Delta^{0}$ and $S^{0}$ are the tree-level propagators, i.e. the inverse of the corresponding kinetic operators in the classical action.

In what follows, we will not discuss the case for gauge bosons as their treatment closely follows the one for scalars, but with the obvious complications related to their vector nature and gauge fixing (see, for instance, Ref.~\cite{Berges:2004yj}).
For the DM model here considered, gauge bosons are simply part of the thermalized SM bath with which the dark sector particles may interact. 
Hence, only their spectral properties will be of relevance.

By varying the effective action and by imposing the stationary conditions in the absence of external sources, namely
\begin{align}
    \dfrac{\delta \Gtpi_\mc{C}[\Delta,S]}{\delta \Delta^{ba}(y,x)}=0, \qquad \dfrac{\delta \Gtpi_\mc{C}[\Delta, S]}{\delta S^{ba}(y,x)}=0\,,
    \label{eq:stationarity}
\end{align}
we can determine the equations of motion for the propagators, which give Eqs.~\eqref{eq:EoM_Delta}, \eqref{eq:EoM_S}.
%
%
In the latter, the free propagators are related to the exact ones by means of the proper self-energies, which are defined as
\begin{align}\label{eq:SelfEn_def}
    \Pi^{ab}(x,y)&=i\,ab\,\dfrac{\delta \Gamma_2[\Delta,S]}{i \delta \Delta^{ba}(y,x)}\,,\\
    \slashed{\Sigma}^{ab}(x,y)&=-i\,ab\,\dfrac{\delta \Gamma_2[\Delta,S]}{i \delta S^{ba}(y,x)}\,,
\end{align}
where, diagrammatically, the functional derivative corresponds to ``cutting'' one line from $\Gamma_2$, (which is the sum of all 2PI vacuum diagrams with full propagators).
In this notation, $-i$ times the self-energy corresponds to the sum of the corresponding 1PI diagrams with full propagators multiplied by the CTP branch indices $a,b=\pm$.
\subsection{\label{sec:SchwingerDyson}Schwinger-Dyson and Kadanoff-Baym equations}
We can now derive the Schwinger-Dyson equations for the dressed two-point functions.
To do this, we just need to convolute one full propagator with Eqs.~\eqref{eq:EoM_Delta} and \eqref{eq:EoM_S}, obtaining
\begin{align}
    \rp{-\partial^2-m^2}i\Delta^{ab}(x,y) = \,a\delta_{ab}i\delta^{(4)}(x-y)-i\sum_{c=\pm} c \int \dd^4 z\, i\Pi^{ac}(x,z)\,i\Delta^{cb}(z,y)\,,\label{eq:SchDys_scalar}\\
    \rp{i\dsl{\partial}-m}iS^{ab}(x,y) = \,a\delta_{ab}i\delta^{(4)}(x-y)-i\sum_{c=\pm} c \int \dd^4 z\, i\slashed{\Sigma}^{ac}(x,z)\,iS^{cb}(z,y)\,.\label{eq:SchDys_fer}
\end{align}
There are four equations for the four CTP propagators, but, since only two combinations are linearly independent, we can re-cast them into a set of equations for retarded/advanced propagators and another one for the Wightman $<,>$ functions.
In particular, for scalar propagators, one obtains
\begin{align}
    &(-\partial^2-m^2)\,i\Delta^{\text{R,A}}(x,y)+i\left(i\Pi^{\text{R,A}}\odot i\Delta^{\text{R,A}}\right)(x,y)=i\delta^{(4)}(x-y)\,,\label{eq:ReAd_scalar}\\
    &(-\partial^2-m^2)i\Delta^{<,>}(x,y)+i\left(i\Pi^{\text{R}}\odot i\Delta^{<,>}\right)(x,y)+i\left(i\Pi^{<,>}\odot i\Delta^{\text{A}}\right)(x,y)=0\,,\label{eq:<>_scalar}
\end{align}
and for fermion propagators one has
\begin{align}
    &(i\dsl{\partial}-m)\,i\dsl{S}^{\text{R,A}}(x,y)+i\left(i \slashed{\Sigma}^{\text{R,A}}\odot i\dsl{S}^{\text{R,A}}\right)(x,y)=i\delta^{(4)}(x-y)\,,\label{eq:ReAd_fermion}\\
    &(i\dsl{\partial}-m)i\dsl{S}^{<,>}(x,y)+i\left(i\slashed{\Sigma}^{\text{R}}\odot i\dsl{S}^{<,>}\right)(x,y)+i\left(i\slashed{\Sigma}^{<,>}\odot i\dsl{S}^{\text{A}}\right)(x,y)=0\,\label{eq:<>_fermion},
\end{align}
where ($A\odot B)(x,y)\equiv \int d^4z A(x,z) B(z,y)$. 
Here, we have also introduced the retarded, advanced, and Wightman self-energies in complete analogy to the definitions in Eq.~\eqref{eq:ret_adv}.

We notice that the Eqs.~\eqref{eq:ReAd_scalar} and \eqref{eq:ReAd_fermion} for the retarded and advanced propagators characterize the spectral properties of the fields in the system, as it will soon be manifest by moving from coordinate space into momentum space.
Eqs.~\eqref{eq:<>_scalar}, \eqref{eq:<>_fermion}, on the other hand, encode the statistical information, which will result in kinetic (fluid) equations for the distribution functions.
This can be already understood by replacing the retarded and advanced functions in Eqs.~\eqref{eq:<>_scalar}, \eqref{eq:<>_fermion} with spectral and Hermitian ones by using Eq.~\eqref{eq:herm_antiherm}, yielding the well-known Kadanoff-Baym (KB) equations \cite{kadanoff2018quantum}:
\begin{align}
    \rp{-\partial^2-m^2}\Delta^{<,>}-\Pi^{\mc{H}}\odot \Delta^{<,>}-\Pi^{<,>}\odot \Delta^{\mc{H}}&=\frac{1}{2}\rp{\Pi^{>}\odot\Delta^{<}-\Pi^{<}\odot\Delta^{>}},\label{eq:KB_scalar}\\
    \rp{i\dsl{\partial}-m}\dsl{S}^{<,>}-\slashed{\Sigma}^{\mc{H}}\odot \dsl{S}^{<,>}-\slashed{\Sigma}^{<,>}\odot \dsl{S}^{\mc{H}}&=\frac{1}{2}\rp{\slashed{\Sigma}^{>}\odot \dsl{S}^{<}-\slashed{\Sigma}^{<}\odot \dsl{S}^{>}}\,.
    \label{eq:KB_fermion}
\end{align}
The right-hand side can be viewed as the QFT equivalent of the collision operator in semiclassical Boltzmann equations, with the creation and destruction terms.
Therefore, as expected, it vanishes in thermodynamic equilibrium, as can be directly seen by employing the KMS relations in Eq.~\eqref{eq:KMS} to both Wightman propagators and self-energies.
\subsection{\label{sec:WignerGradient}Wigner transform and gradient expansion}
We want now to separate the microscopic physics, namely statistical fluctuations, from the macroscopic one, namely changes in the mean spacetime coordinates.
The latter is of less relevance to us, provided that the background on which the field dynamics occur varies slowly.
This is the case when, for example, the timescales of particle interactions are much shorter than those of the expansion of the Universe.
To see this, we first perform a Wigner transformation of the two-point Green's functions, namely a Fourier transform with respect to the relative coordinate $r$,
\begin{align}
 {G}(k,x)={\int {\rm d}^4 r}\,e^{ik\cdot r}G\left(x+\frac{r}{2},x-\frac{r}{2}\right)\,,
    \label{eq:Wigner_transform_app}
\end{align}
where $x$ is the mean coordinate.
In Wigner space, convolutions transform as
\begin{align}\begin{aligned}
A\odot B\rightarrow ({A \odot B})(k,x)=&\,\int\dd^4 r e^{ik\cdot r}\int \dd^4 z A\left(x+\frac{r}{2},z\right)B\left(z,x-\frac{r}{2}\right)\\
=&\,e^{-i\diamond}\{A(k,x)\}\{B(k,x)\}\,,
\end{aligned}
    \label{eq:MoyalWigner}
\end{align}
where the diamond operator (or Moyal product) is defined as
\begin{align}
    \diamond\{A\}\{B\}=\frac{1}{2}\left(\partial_{x}A\,\cdot\partial_k B-\partial_{k}A\,\cdot\partial_x B\right)\,,
    \label{eq:Moyal_product}
\end{align}
and the exponentiation has to be understood as a series expansion.
This series can be truncated to a certain order in the gradients if the convoluted operators have a weak dependence on the mean coordinate $x$.
What we are doing is effectively an expansion in the macroscopic correlation length, namely a measure for the variation of the expanded functions with respect to the macroscopic mean coordinate.
If the background fields change slowly, as in thermal equilibrium, their characteristic correlation length is much larger than the de Broglie length of particles in the plasma, typically of order $T^{-1}$, or, equivalently, their variation in spacetime is small compared to their momentum
\begin{align}
    \partial \ll k\sim \lambda_{\text{DB}}^{-1}\sim T\,.
\end{align}
In this case, the gradient expansion is well justified.

In Wigner space, the KMS relations in Eq.~\eqref{eq:KMS} become
\begin{align}
    { G}^{\mathrm{eq},>}(k)&=\pm e^{\beta k^0}{ G}^{\mathrm{eq},<}(k)\,,
    \label{eq:KMS_momentum}
\end{align}
where the upper (lower) sign is for bosonic (fermionic) species.
As in equilibrium one recovers translation invariance, the Green's functions $G^{\rm eq}$ only depend on the relative coordinate, and hence their Wigner transform reduces to a simple Fourier transform only depending on the momentum.
From Eqs.~\eqref{eq:KMS_momentum} and \eqref{eq:herm_antiherm} one can see that the Wightman propagators are unequivocally determined by the spectral density of states $G^{\rm eq,{\mc{A}}}$ in the following way
\begin{align}
    i{ G}^{\mathrm{eq}, <}(k)&=\pm 2 f_{\mp}(k^0){ G}^{\mathrm{eq},\mc{A}}(k)\,,  \label{eq:<_spectral_momentum}\\
    i{ G}^{\mathrm{eq},>}(k)&=\mp 2 f_{\mp}(-k^0){ G}^{\mathrm{eq},\mc{A}}(k)\,,
    \label{eq:>_spectral_momentum}
\end{align}
where the $\pm$ notation follows from above.
The $f_\mp$ are nothing but the thermal equilibrium Bose-Einstein and Fermi-Dirac distribution functions given by Eq.~\eqref{eq:BEFD_distr} with $T=1/\beta$.
\subsubsection{\label{sec:resum_prop}Resummed propagators}
In Wigner space and up to first order in the gradients, the Schwinger-Dyson equations for the retarded and advanced propagators, Eqs.~\eqref{eq:ReAd_scalar}-\eqref{eq:ReAd_fermion}, can be written as \footnote{Works addressing quantum kinetic equations without relying on 2PI effective action expansions or approximated Kadanoff-Baym equations have also been done for scalar field dynamics in the early Universe with applications to DM and reheating after inflation \cite{Boyanovsky:1994me,Boyanovsky:1998pg,Boyanovsky:1999cy,Boyanovsky:2004dj,Drewes:2010pf,Hamaguchi:2011jy,Drewes:2012qw,Drewes:2013bfa,Drewes:2013iaa,Drewes:2015eoa,Kainulainen:2021eki}.}
\begin{align}
    \left(k^2+ik\cdot\partial_x - m^2-{ \Pi}^{R/A}\right)i{\Delta}^{R/A}=i \label{eq:Wig_ReAd_sca}\,,\\
    \left(\dsl{k}+\frac{1}{2}\dsl{\partial}_x-m-\slashed{\Sigma}^{R/A}\right)i\dsl{S}^{R/A}=i\,.\label{eq:Wig_ReAd_fer}
\end{align}
By imposing spatial homogeneity and isotropy, we can algebraically solve these equations for the \emph{resummed propagators}.

For scalars, we have
\begin{align}
    i{\Delta}^{R/A}=\dfrac{i}{k^2-m^2-{\Pi}^{R/A}}=\dfrac{i}{k^2-m^2-{\Pi}^{\mc{H}}\pm i {\Pi}^{\mc{A}}}\,,\label{eq:Resum_Delta_ReAd}
\end{align}
where $+$ ($-$) is for the retarded (advanced) case.
The anti-Hermitian (spectral density) and Hermitian resummed scalar propagators therefore follow as 
\begin{align}
    {\Delta}^{\mc{A}}=-\Im {\Delta}^{R}=\dfrac{{\Pi}^{\mc{A}}}{\left(k^2-m^2-{\Pi}^{\mc{H}}\right)^2+\left({\Pi}^{\mc{A}}\right)^2}\,,\label{eq:Resum_Delta_AntiH}\\
    \Delta^{\mc{H}}=\Re \Delta^{R}=\dfrac{k^2-m^2-{\Pi}^{\mc{H}}}{\left(k^2-m^2-{\Pi}^{\mc{H}}\right)^2+\left({\Pi}^{\mc{A}}\right)^2}\,.\label{eq:Resum_Delta_H}
\end{align}
Similarly, for fermions, the retarded and advanced propagators read as
\begin{align}
    i\dsl{S}^{R/A}=\dfrac{i}{\dsl{k}-m-\slashed{\Sigma}^{R/A}}=\dfrac{i}{\dsl{k}-m-\slashed{\Sigma}^{\mc{H}}\pm i \slashed{\Sigma}^{\mc{A}}}\,.\label{eq:Resum_S_ReAd}
\end{align}
The resummed anti-Hermitian and Hermitian parts can be derived in an analogous way, with however the complication of a more intricate structure, due to the Dirac matrices. 
Assuming self-energies of the form
\begin{align}
    \slashed{\Sigma}^{\mc{H}/\mc{A}}=\gamma^\mu\, {\Sigma}^{\mc{H}/\mc{A}}_\mu,
\end{align}
a compact way to write the resummed propagators is:
\begin{align}
     \SsA=-\Im  \dsl{S}^R=\left(\dsl{k}+m-\slashed{\Sigma}^{\mc{H}}\right)\dfrac{\Gamma}{\Omega^2+\Gamma^2}-\slashed{\Sigma}^{\mc{A}}\dfrac{\Omega}{\Omega^2+\Gamma^2}\,,\label{eq:Resum_S_AntiH}\\
     \SsH=\Re  \dsl{S}^R=\left(\dsl{k}+m-\slashed{\Sigma}^{\mc{H}}\right)\dfrac{\Omega}{\Omega^2+\Gamma^2}+\slashed{\Sigma}^{\mc{A}}\dfrac{\Gamma}{\Omega^2+\Gamma^2}\,,\label{eq:Resum_S_H}
\end{align}
with
\begin{align}
    &\Gamma = 2 \left(k-\Sigma^{\mc{H}}\right)\cdot\Sigma^{\mc{A}},\label{eq:Gamma_Fer}\\
    &\Omega = \left(k-\Sigma^{\mc{H}}\right)^2-\rp{\Sigma^{\mc{A}}}^2-m^2\,.\label{eq:Omega_Fer}
\end{align}
When neglecting the interactions with the plasma, $\Gamma,\,\Omega\rightarrow 0$, the resummed propagators reduce to the tree-level ones. 
In particular, the spectral densities become Dirac delta functions
\begin{align}
\DeA(k) = \pi\,\sgn{k^0}\,\delta \left( k^2 - m^2 \right),\qquad \SsA(k) = \pi\, \sgn{k^0} \left( \slashed{k} + m \right) \delta \left( k^2 - m^2 \right),
\end{align}
which also determine the Wightman functions for scalar fields as
\begin{align}
i\Delta^{<}(k)&=2\Delta^{\mc{A}}(k)f_-(k) \, , \\
i\Delta^{>}(k)&=2\Delta^{\mc{A}}(k)\left(1+f_-(k)\right)\,,    
\end{align}
and for fermion fields as
\begin{align}
i \slashed{S}^{<}(k)&=- 2 \SsA(k) f_+(k) \, , \\
i \slashed{S}^{>}(k)&=2 \SsA(k)\left(1-f_+(k)\right)\,.    
\end{align}

\subsubsection{\label{sec:KineticEq}Kinetic equations}
Following our discussion in Sec.~\ref{sec:SchwingerDyson}, imposing the approximations there explained, and assuming spatial homogeneity and isotropy so that spatial derivatives can be neglected, we can re-cast the Kadanoff-Baym equations in Eqs~\eqref{eq:KB_scalar} in position space into a set of \emph{kinetic} equations in Wigner space.
Their imaginary parts lead to the following kinetic equation for the Wightman functions in Wigner space
\begin{align}
    k^0 \partial_t \left( i \Delta_s^{<,>} \right) &=- \frac{1}{2} \left( i {\Pi_s}^{>}\,i \Delta_s^{<} -  i {\Pi_s}^{<} \, i \Delta_s^{>} \right)+\mc{O}(\partial_t\Pi_s^{\mc{H};<,>}\Delta_s^{<,>;\mc{H}})\,.
    \label{eq:KinEq_scalar}
\end{align}
These equations will reduce to the fluid rate equations once we integrate over the energy, as shown in Appendix~\ref{sec:AppB}. 
The l.h.s of \eqref{eq:KinEq_scalar} is directly related to the time derivative of particle number, and the first term in the r.h.s. gives the dominant particle production rate. 
The discarded terms, involving a temporal derivative, represent gradient corrections to the production rate, which will be argued to be heavily suppressed in a freeze-in scenario.
\section{\label{sec:AppB}Derivation of the DM fluid equation for freeze-in}
In this section, we derive the DM rate equations for the scalar field $s$ directly from the kinetic equations for a scalar propagator in Eq.~\eqref{eq:KinEq_scalar}.
First of all, the DM self-energies ${\Pi}^{<,>}_{s}$ receive contributions from propagators of equilibrated particles, while corrections from internal FIMP propagators are suppressed by powers of the feeble DM coupling.
In this sense, we can turn the KMS relations valid for the bosonic correlators (cf. Eqs.~\eqref{eq:<_spectral_momentum}, \eqref{eq:>_spectral_momentum}) into KMS relations for the DM Wightman self-energies:
\begin{align}
    i{\Pi}^{<}_s(p)&=2{\Pi}_s^{\mc{A}}(p)f_-(p^0)\,,\\
    i{\Pi}^{>}_s(p)&=2{\Pi}_s^{\mc{A}}(p)\left(1+f_-(p^0)\right)\,.
\end{align}
Secondly, since finite-width effects are negligible due to the feeble coupling, we can simplify our calculation by employing a generalized fluctuation-dissipation relation, the Kadanoff-Baym \emph{ansatz}, for the two-point functions of the non-equilibrated DM particles, which allows us to bring them into the analogous form \cite{Greiner:1998vd,Berges:2004pu,Calzetta:2008iqa,Garbrecht:2011xw,Drewes:2015eoa}
\begin{align}
    i \Delta^{<}_s(p)&=2\Deltas^{\mc{A}}(p)f_s(p^0)\,,\\ \label{eq:KB-Ansatz-<}
    i\Delta^{>}_s(p)&=2\Deltas^{\mc{A}}(p)\left(1+f_s(p^0)\right)\,.
\end{align}
Above, we have used the notation $f_s$ for the DM distribution functions, which will differ from the equilibrium distribution $f_-$ in Eq.~\eqref{eq:BEFD_distr}.
The Wightman propagators $\Deltas^{<,>}$ are related to the momentum distribution functions $f_s$ and number densities $n_s$ of the associated particles  through the identity of Eq.~\eqref{eq:n_s},
%
%
as can be verified from Eq.~\eqref{eq:KB-Ansatz-<} and the spectral density corresponding to scalar particles with negligible width,
\begin{align}\label{eq:scalar_Pi_A}
    \DeA_s(p)=\pi\delta(p^2-m_s^2)\sgn{p^0},
\end{align}
together with the definition of $\omega_p$ in Eq.~\eqref{eq:omega_p}
Integrating Eq.~\eqref{eq:KinEq_scalar}  over $p^0>0$ and using \eqref{eq:n_s} leads to the following fluid equation
\begin{align}
 \partial_t f_s(t,\pvec)&=\int_0^\infty \frac{\dd p^0}{\pi}\,\frac{1}{2}\left[ \left( i {\Pi}_s^< \right) \left( i \Deltas^> \right) - \left( i {\Pi}_s^> \right) \left( i \Deltas^< \right)\right]\nonumber\\
 &=\int_0^\infty\frac{\dd p^0}{\pi}\DeA_s\PiA_s\left[f_-(p^0)-f_s(t,p)\right]\nonumber\\
 &=\frac{\PiA(\omega_p,|\vec{p}|)}{\omega_p}\left[f_-(\omega_p)-f_s(t,p)\right]\nonumber\\
 &\simeq \frac{\PiA(\omega_p,|\vec{p}|)}{\omega_p} f_-(\omega_p)\,.\label{eq:rateEQ}
\end{align}
Here, we have employed Eq.~\eqref{eq:scalar_Pi_A} and used $f_-(\omega_p)\gg f_s(t,p)$, valid under the freeze-in assumptions. 
Moreover, we have discarded the $\mc{O}(\partial_t \Pi \Delta)$ terms in Eq.~\eqref{eq:KinEq_scalar} because  as $\Pi_s$ is dominated by the effects of the particles in equilibrium, and as the smallness of $f_s$ in freeze-in production implies that  $\Delta_s$ can be approximated by its vacuum value, then neither $\Pi_s$ nor $\Delta_s$ are time-dependent.

We can now integrate the whole expression over the external momentum to arrive at the rate equation for the DM number density
\begin{align}
    \dfrac{\dd}{\dd t} n_s(t) = \int \dfrac{\dd^3 \vec{p}}{(2\pi)^3}\dfrac{{\Pi}^{\mc{A}}_s(\omega_p,|\vec{p}|)}{\omega_p}f_-(\omega_p)\,.
\end{align}
In order to quantify the relic abundance of DM today, we have to integrate the rates presented in the previous section.
However, in all the discussions carried out so far, we have not taken into account a fundamental ingredient to describe the evolution of particle production, which is the expansion of the Universe.
All the calculations have been made in a Minkowski background.
However, we can translate them in a Friedmann-Lem\^{a}itre-Robertson-Walker (FLRW) Universe through a conformal coordinate transformation \cite{Beneke:2010wd,Garbrecht:2018mrp}.
In detail, the conformal FLRW metric reads as
\begin{align}
    g_{\mu\nu}=a^2(\eta)\,\eta_{\mu\nu}=a^2(\eta)\,\mathrm{diag}\left(1,-1,-1,-1\right)\,,
    \label{eq:conf_FLRW_metric}
\end{align}
with $\eta$ being the conformal time and $a(\eta)$ the scale factor, which in radiation domination reads as $a(\eta)=a_R\,\eta$, with $a_R$ being a reference value.
We choose $a_R=T$ where $T=a\,T_\mathrm{ph}$ is the comoving temperature, while $T_\mathrm{ph}$ the physical one.
This implies that $T_\mathrm{ph}=1/\eta$ and for $a(t)=1$ we have $T=T_\mathrm{ph}$.
As a consequence, all the quantities we have described so far are correct in the expanding background provided that they are understood as comoving quantities.
Together with time and temperature, this includes the entropy density  $s=a^3\,s_\mathrm{ph}$, three-momenta $\mathbf{p}=a \mathbf{p}_\mathrm{ph}$, and all particle masses $m=a m_\mathrm{ph}$ and phase-space distributions $f=f(\mathbf{p})=f(a\mathbf{p}_\mathrm{ph})$.

Thus, the DM rate equation in Eq.\eqref{eq:rateEQ} is understood as in conformal coordinates.
It is straightforward to derive its version in physical coordinates.
For instance, the left-hand side reads as \cite{Beneke:2010wd}
\begin{align}
    \dfrac{\dd}{\dd t}f_s(\mathbf{p})&=
    \dfrac{1}{a(\eta)}\dfrac{\dd}{\dd \eta}f_s(\mathbf{p})=
    \dfrac{\partial}{\partial t}f(\mathbf{p}_\mathrm{ph})+\left(\dfrac{\partial}{\partial |\mathbf{p}_\mathrm{ph}|} f(\mathbf{p}_\mathrm{ph})\right)\dfrac{\partial |\mathbf{p}_\mathrm{ph}|}{\partial t}\\
    &=\dfrac{\partial}{\partial t}f(\mathbf{p}_\mathrm{ph})- H|\mathbf{p}_\mathrm{ph}|\dfrac{\partial}{\partial |\mathbf{p}_\mathrm{ph}|}f(\mathbf{p}_\mathrm{ph})\,.
\end{align}
where $H(t)=\dot{a}(t)/a(t)$ is the Hubble rate.
When integrated over the three-momentum, this expression transforms into
\begin{align}
    \dfrac{\partial}{\partial t}n_\mathrm{ph}+3H n_\mathrm{ph}\,,
\end{align}
the usual diffusion term in the FLRW background. The final equation for the rate of DM production is then
\begin{align}
    \dfrac{\dd}{\dd t} n_s(t) +3H n_\mathrm{ph}= \int \dfrac{\dd^3 \vec{p}}{(2\pi)^3}\dfrac{{\Pi}^{\mc{A}}_s(\omega_p,|\vec{p}|)}{\omega_p}f_-(\omega_p)\,.
\end{align}

\section{Integrals in the fermionic self-energies\label{sec:AppC}}
In this appendix, we collect the expressions entering the Hermitian and anti-Hermitian self-energies of a massless and massive gauge charged fermion needed in Sec.~\ref{sec:1PI} (see also Ref.~\cite{Petitgirard:1991mf}). As the SM fermions come in chiral representations of the gauge groups,  one has to distinguish the self-energies for left and right-handed components. Hence, in this appendix we will treat $\dsl\Sigma$ as a chiral object, $\dsl{\Sigma}\equiv P_{R/L} \gamma_\mu \Sigma^\mu$. To unclutter the notation, the chirality will be left unspecified, and the dependence on the representations of the $R/L$ fermions under the gauge groups is absorbed into the effective coupling $G$. In the SM, $G$  has to be assigned differently to $L/R$ fermions, c.f. table \ref{tab:G_val}, while for the exotic vectorlike fermion $F$ the coupling $G$ is independent of the chirality.

Essentially, we need to compute the scalar products $\mc{P}=p\cdot \Sigma$ and $\mc{U}=u\cdot \Sigma$.  
Thanks to the Dirac structure of the self-energy, these can be related to the traces
\begin{align}
    \Tr{\dsl{p}\dsl{\Sigma}}=2\,\left(p\cdot \Sigma\right)=2\mc{P},\quad  \Tr{\dsl{u}\dsl{\Sigma}}=2\,\left(u\cdot \Sigma\right)=2\mc{U}\,,
\end{align}
which are more straightforward to compute.\\
In particular, the contraction with the particle momentum $p^{\mu}$ yields
\begin{align}
\Tr{\dsl{p}\dsl{\Sigma}^{H}(p)}^{(T\neq 0)}&=\frac{G T^2}{6}+\frac{G}{\pi^2}\int_0^{\infty}\dd k \frac{k^2}{\omega_{F,k}}f_+(\omega_{F,k})\nonumber\\
&+\frac{G}{8\pi^2}\frac{p^2+m_F^2}{|\Vec{p}|}\int_0^{\infty}\dd k \,f_{-}(k)\sum_{\pm}\ln\left|\dfrac{p^2-m_F^2+2k(|\Vec{p}|\mp p_0)}{p^2-m_F^2-2k(|\Vec{p}|\pm p_0)}\right|\nonumber\\
&-\frac{G}{8\pi^2}\frac{p^2+m_F^2}{|\Vec{p}|}\int_0^{\infty}\dd k \,f_{+}\left(\omega_{F,k}\right)\sum_{\pm}\ln\left|\dfrac{p^2+m_F^2\mp2p_0\omega_{F,k}+2|\Vec{p}|k}{p^2+m_F^2\mp2p_0\omega_{F,k}-2|\Vec{p}|k}\right|,
\end{align}
where $k=|\Vec{k}|$, $\omega_{F,k} = \sqrt{k^2 + m_F^2}$, and $f_{\pm}$ are the Fermi-Dirac and Bose-Einstein distribution functions. 
For a massless fermion, this expression simplifies to
\begin{align}
\Tr{\dsl{p}\dsl{\Sigma}^{H}(p)}^{(T\neq 0)}_{m_F=0}=\dfrac{GT^2}{4}+\frac{G}{8\pi^2}\frac{p^2}{|\Vec{p}|}\int_0^{\infty}\dd k\,(f_-(k) - f_+(k))\sum_{\pm}\ln\left|\dfrac{p^2+2k(|\Vec{p}|\mp p_0)}{p^2-2k(|\Vec{p}|\pm p_0)}\right|.
\end{align}
For the contraction with $u^\mu$, we find
\begin{align}
\Tr{\dsl{u}\dsl{\Sigma}^{H}(p)}^{(T\neq 0)}&=\frac{G}{4\pi^2}\frac{1}{|\Vec{p}|}\int_0^{\infty}\dd k\,f_{-}(k)\sum_{\pm}(p_0\mp k)\ln\left|\dfrac{p^2-m_F^2+2k(|\Vec{p}|\mp p^0)}{p^2-m_F^2-2k(|\Vec{p}|\pm p^0)}\right|\nonumber\\
&+\frac{G}{4\pi^2}\frac{1}{|\Vec{p}|}\int_0^{\infty}\dd k\,k\,f_{+}(\omega_{F,k})\sum_{\pm}(\mp)\ln\left|\dfrac{p^2+m_F^2\mp2p_0\,\omega_{F,k}+2|\Vec{p}|k}{p^2+m_F^2\mp2p_0\,\omega_{F,k}-2|\Vec{p}|k}\right|,
\end{align}
for a massive fermion, while
\begin{align}
\Tr{\dsl{u}\dsl{\Sigma}^{H}(p)}^{(T\neq 0)}_{m_F=0}&=\frac{G}{4\pi^2}\frac{1}{|\Vec{p}|}\int_0^{\infty}\dd k\, k\,\left(f_{-}(k)+f_{+}(k)\right)\sum_{\pm}(\mp)\ln\left|\dfrac{p^2+2k(|\Vec{p}|\mp p^0)}{p^2-2k(|\Vec{p}|\pm p^0)}\right|\nonumber\\
&\quad+\frac{G}{4\pi^2}\frac{p_0}{|\Vec{p}|}\int_0^\infty\dd k\,f_-(k)\sum_\pm\ln\left|\dfrac{p^2+2k(|\Vec{p}|\mp p^0)}{p^2-2k(|\Vec{p}|\pm p^0)}\right|,
\end{align}
for a massless fermion.

As the anti-Hermitian self-energy is free of divergences, we can consider its complete form including vacuum and thermal contributions, yielding
\begin{align}
 \Tr{\dsl{p}\dsl{\Sigma}^{\mc{A}}(p)}=\frac{G}{8\pi}\sgn{p^2-m_F^2}\frac{p^2+m_F^2}{|\Vec{p}|}\left[\theta(-p^2)p^0+T\ln\left|\dfrac{\sinh(\frac{k^0_{+}-p^0}{2T})\cosh(\frac{k^0_{-}}{2T})}{\sinh(\frac{k^0_{-}-p^0}{2T})\cosh(\frac{k^0_{+}}{2T})}\right|\right],
\end{align}
and
\begin{align}
\Tr{\dsl{u}\dsl{\Sigma}^{\mc{A}}(p)}&=\frac{G}{4\pi}\frac{\sgn{p^2-m_F^2}}{|\Vec{p}|}\Bigg[\theta(-p^2)\left( -\frac{\pi^2 T^2}{2} +\frac{\left( p^0 \right)^2}{2}\right)\nonumber
\\ 
&+T\,\mathfrak{Re}\bigg\{k^0\ln\left|\frac{f_{-}(k^0-p^0)}{f_{+}(k^0)}\right|+T\Li{2}{-e^{k^0/T}}-T\Li{2}{e^{(k^0-p^0)/T}}\bigg\}\bigg|_{k^0_{-}}^{k^0_{+}}\Bigg],
\end{align}
where Li$_2$ is the dilogarithm function and where
\begin{align}
    k^0_{\pm}&=\dfrac{p^0}{2p^2}\rp{p^2+m_F^2-m_A^2}\pm\dfrac{|\Vec{p}|}{2p^2}\sqrt{\lambda\rp{p^2,m_F^2,m_A^2}}
\end{align}
which, for massless gauge bosons, reduces to
\begin{align}
    \dfrac{p^0}{2 p^2}\rp{p^2+m_F^2}\pm\dfrac{|\Vec{p}|}{2p^2}\left|p^2-m_F^2\right|,\quad \text{if $p^2\neq0$}.
\end{align}
When evaluating the scalar product of fermionic propagators in the context of the DM self-energy, the following relations prove to be useful:
\begin{align}
    &\dsl{\Sigma}(p)=\frac{1}{2|\Vec{p}|^2}[(p^0\gamma^0-\dsl{p})\Pc+(p^0\dsl{p}-p^2\gamma^0)\Uc],\\
    &\Sigma^H(p)\cdot\Sigma^\mc{A}(p)=-\frac{1}{2|\Vec{p}|^2}\left(\Pc^H \Pc^\mc{A}+p^2\Uc^H\Uc^\mc{A}-p^0(\Pc^H\Uc^\mc{A}-\Pc^\mc{A}\Uc^H)\right),\\
    &\Sigma^\mc{A}(p)\cdot\Sigma^\mc{A}(p)=-\frac{1}{2|\Vec{p}|^2}\left((\Pc^\mc{A})^2+p^2(\Uc^\mc{A})^2-2p^0(\Pc^\mc{A}U^\mc{A})\right),\\
    &\Sigma^H(p)\cdot\Sigma^H(p)=-\frac{1}{2|\Vec{p}|^2}\left((\Pc^H)^2+p^2(\Uc^H)^2-2p^0(\Pc^H\Uc^H)\right) \, ,\\
    &k\cdot\Sigma(p)=\frac{1}{2|\Vec{p}|^2}\left[(p^0 k^0-p\cdot k)\Pc+(p^0 p\cdot k-p^2 k^0)\Uc\right]\\
    &\hspace{13mm}=\frac{1}{2|\Vec{p}|^2}\left[|\Vec{p}||\Vec{k}|\cos\theta_{pk}\Pc-(p^0|\Vec{p}||\Vec{k}|\cos\theta_{pk}+|\Vec{p}|^2k^0)\Uc\right],\\
    &\Sigma^r(k)\cdot\Sigma^s(p)=\frac{1}{4}\Big[\cos\theta_{pk}\Big(k_0\,\Uc^r(k)\Pc^s-\Kc^r\Pc^s+p_0\,\Kc^r\Uc^s(p)\Big)\nonumber\\
    &\hspace{42mm}-\left(\cos\theta_{pk}k_0 p_0-|\Vec{k}||\Vec{p}|\right)\Uc^r(k)\,\Uc^s(p)\Big] \, .
\end{align}
In the last line, the indices $r$ and $s$ refer to either $\mc{A}$ or $\mc{H}$. 
\section{Collection of results for HTL propagators\label{sec:AppD}}
\subsection{HTL spectral densities}
{
For massless fermions, we can split the spectral function in the HTL approximation into positive and negative helicity over chirality ratios, yielding \cite{Weldon:1982bn,le_bellac_1996}
\begin{align}
    \SsA_{F/f} (k) = \dfrac{1}{2}\left[(\gamma^0-\hat{k}\cdot\hat{\gamma})\rho_{+}(k)  +(\gamma^0+\hat{k}\cdot\hat{\gamma})\rho_{-}(k)\right]\,,
    \label{eq:HTL_rho}
\end{align}
where $\hat{k}=\vec{k}/|\vec{k}|$.
The two terms correspond to
\begin{align}
    \rho_{\pm}(k)=2\pi\left[Z_\pm (|\vec{k}|)\,\delta\rp{k_0-\omega_{\pm}(|\vec{k}|)}+Z_\mp (|\vec{k}|)\,\delta\rp{k_0-\omega_{\mp}(|\vec{k}|)}\right]+\rho_{\pm}^{\mathrm{cont}}(k)\,,
    \label{eq:rho_pm}
\end{align}
where the residues at the poles $\omega_{\pm}(|\vec{k}|)$ of the propagator are given by
\begin{align}
    Z_\pm(|\vec{k}|)=\dfrac{\omega_\pm^2(|\vec{k}|)-|\vec{k}|^2}{4y^2 |\vec{k}|^2}\,,
    \label{eq:Z_pm}
\end{align}
with $y=(GT^2) / (4|\vec{k}|^2)$, while the \emph{continuum} term reads as
\begin{align}
     \rho_{\pm}^{\mathrm{cont}}(k^0,|\vec{k}|)&=\frac{y^2}{|\vec{k}|}(1\mp x)\theta(1-x^2)\times\\
     &\,\times\left[\left((1\mp x)\pm y^2\left[(1\mp x)\ln\bigg|\frac{x+1}{x-1}\bigg|\pm 2\right]\right)^2+\pi^2 y^4(1\mp x)^2\right]^{-1}\,,
     \label{eq:rho_c}
\end{align}
and corresponds to the anti-Hermitian part of the HTL self-energy.
}
\subsection{HTL self-energies}
In the following, we collect the expressions for the DM self-energy in the three different integration regimes of the DM interaction rate given in Eq.~\eqref{eq:rateEQ}. 
We split the integration domain into four different regions that distinguish between time- and spacelike momenta $q$ and $k$. 
The DM self-energies in Eq.~\eqref{eq:SelfEnergy_HTLTOT} are made explicit in the following.
\subsubsection*{Double timelike momenta ($k^2 >0$ and $q^2 > 0$)}
We use use Eq.~\eqref{eq:HTL_prop} to evaluate the DM self-energy in Eq.~\eqref{eq:PiA_master} and obtain
\begin{align}
    \Pi^{\mc{A}}_{s,TT}(p)=\frac{y_{DM}^2}{ 2 \pi }\int \dd |\vec{k}| &\sum_i \sum_j \left[ k_i q_j - q_j \Sigma_F^\mathcal{H} \left( k_i \right)  - k_i \Sigma_f^\mathcal{H} \left( q_j \right) + \Sigma_f^\mathcal{H} \left( q_j \right) \Sigma_F^\mathcal{H} \left( k_i \right) \right] \nonumber \\ &\times \left[1-f_+(k^0_i)-f_+(-q^0_j)\right] \frac{\text{sign} \left( k^0_i q^0_j \right) \kvec^2}{| g'_F \left( k^0_i \right)| | g'_f \left( \cos \theta_j \right)|} \nonumber \\ &\times \theta \left( \qvec_j - \left| \pvec - \kvec \right| \right) \theta \left( -\qvec_j + \left| \pvec + \kvec \right| \right) \label{eq:DM_selfenergy_TT} \, ,
\end{align}
where 
\begin{align}
    g'_F \left( ( k^0_i \right) &= \frac{\dd}{\dd k^0} \left[ \left( k_\mu - \Sigma_{F,\mu}^{\mathcal{H},HTL} \left( k \right) \right)^2- m_F^2 \right]\bigg|_{k^0=k^0_i} \, , \label{eq:Delta_HTL_F}
\end{align}
with 
\begin{align}
    k^0 = k^0_i \left( |\vec{k}| \right) = \left\{\begin{array}{ll} \omega_{+,m_F} \left( |\vec{k}| \right), & i=1 \\
         \omega_{-,m_F} \left( |\vec{k}| \right), & i=2 \\ -\omega_{+,m_F} \left( |\vec{k}| \right), & i=3 \\ -\omega_{-,m_F} \left( |\vec{k}| \right), & i=4\end{array}  \right. , \label{eq:Dispersion_Massive}
\end{align}
being the dispersion relations for an in-vacuum massive fermion and with
\begin{align}
    g'_f \left(  q \right) &= \frac{\dd}{\dd \cos \theta} \left( q_\mu - \Sigma_{f,\mu}^{\mathcal{H},HTL} \left( q \right) \right)^2\bigg|_{\cos \theta=\cos \theta_j} \, .
\end{align}
The Heaviside step functions in the integral ensure that the kinematic constraints imposed by $q = k - p$ are respected. 
Note that the dispersion relations of an in-vacuum massive fermion cannot be obtained analytically and must be solved numerically.
\subsubsection*{Timelike-spacelike: $k^2>0$ and $q^2<0$}
For this momentum assignment, only the $F$ propagator has a vanishing width and takes the form of Eq~\eqref{eq:HTL_prop}.
We choose to eliminate the $k^0$ integration with the on-shell delta of $S_F^\mc{A}(K)$, which fixes $k^0$ according to Eq.~\eqref{eq:Dispersion_Massive}.
Then, the DM self-energy is given by 
\begin{align}
    \Pi^{\mc{A}}_{s,TS}(p)&=2\frac{y_{DM}^2}{\left( 2 \pi \right)^2}\int \dd \cos \theta \,\dd |\vec{k}| \sum_i \frac{|\vec{k}|^2}{|g'_F \left( k^0_i \right)|} \left(A - B\right)\,\text{sign} \left( k^0_i \right) \left[1-f_+(k^0_i)-f_+(p^0-k^0_i)\right]\,, \label{eq:HTL_ST}\\
    A & = \frac{\Gamma_f \left( q  \right) \left( k_i q - k_i \Sigma_f^\mathcal{H} \left( q \right)  - q \Sigma_F^\mathcal{H} \left( k_i \right) + \Sigma_F^\mathcal{H} \left( k_i \right) \Sigma_f^\mathcal{H} \left( q \right)  \right)}{\Omega_f^2(q)+\Gamma_f^2(q)}\,,\\
    B & = \frac{\Omega_f \left( q \right) \left( k_i \Sigma_f^\mathcal{A} \left( q \right) - \Sigma_F^\mathcal{H} \left( k_i \right) \Sigma_f^\mathcal{A} \left( q \right) \right)}{\Omega_f^2(q)+\Gamma_f^2(q)}\,,
\end{align}
where $g'_F$ is given by Eq.~\eqref{eq:Delta_HTL_F}.

After $k^0=k^0_i$ is fixed, we set $q^0=k^0_i - p^0$ and $\qvec = \sqrt{\pvec^2 + \kvec^2 - 2 \pvec \kvec \cos \theta}$ and perform the integrations over $\kvec$, $\cos \theta$ and $\pvec$ numerically. 
The integration for the case $k^2<0$ and $q^2>0$ is conducted in the same way and can be obtained from Eq.~\eqref{eq:HTL_ST} by interchanging $k \leftrightarrow q$ and $f \leftrightarrow F$ everywhere.
\subsubsection*{Double spacelike momenta ($k^2<0$ and $q^2<0$)}
If both momenta are spacelike, the expression for the DM self-energy cannot be further simplified and all integrations have to be carried out numerically. 
As in the case of the 1PI resummed propagator, the DM self-energy $\Pi^\mc{A}_{s,SS}$ is given by Eq.~\eqref{eq:PiA_master}, where the integration is restricted to space-like momenta for both $k$ and $p$. %

\bibliographystyle{JHEP}
\bibliography{biblio}
\end{document}